\documentclass{aastex631}
\usepackage[T1]{fontenc}

\usepackage{multirow}
\usepackage{amsmath}
\usepackage{hhline}

\newcounter{tr}
\ifnum \value{tr}>5
\newcommand{\deletedD}[1]{{\color{red} Damien - Deleted: } \sout{#1}}

\newcommand{\authorcommentD}[1]{{\color{purple} Damien - Comment :} {\color{blue} #1}}

\newcommand{\deletedA}[1]{{\color{red} Asaf - Deleted: } \sout{#1}}

\newcommand{\authorcommentA}[1]{{\color{purple} Asaf - Comment :} {\color{cyan} #1}}

\else
\newcommand{\deletedD}[1]{}

\newcommand{\authorcommentD}[1]{}

\newcommand{\deletedA}[1]{}

\newcommand{\authorcommentA}[1]{}

\fi

\begin{document}

%\title{cuHARM: MAD}
\title{A study of the MAD accretion state across black hole spins for radiatively inefficient accretion flows.}

\shorttitle{cuHARM}
\shortauthors{G.-Q.  Zhang et al.}

\author[0000-0001-6545-4802]{G.-Q.  Zhang}
\affiliation{School of Astronomy and Space Science, Nanjing University, Nanjing 210093, China}
\affiliation{Bar Ilan University, Ramat Gan, Israel}

\author[0000-0003-4477-1846]{Damien B\'egu\'e}
\affiliation{Bar Ilan University, Ramat Gan, Israel}

\author[0000-0001-8667-0889]{A. Pe{'}er}
\affiliation{Bar Ilan University, Ramat Gan, Israel}

\author[0000-0003-4111-5958]{B.-B. Zhang}
\affiliation{School of Astronomy and Space Science, Nanjing University, Nanjing 210093, China}
\affiliation{Key Laboratory of Modern Astronomy and Astrophysics (Nanjing University), Ministry of Education, China}

\begin{abstract}
The study of Magnetically Arrested Disks (MAD) attract strong interest in recent years, as these disk configurations were found to generate strong jets as observed in many accreting systems. 
Here, we present the results of 14 general relativistic magnetohydrodynamic
(GRMHD) simulations of advection dominated accretion flow in the MAD state across black hole spins, carried with cuHARM. 
Our main findings are as follows.
(i) The jets transport a significant amount of angular momentum to infinity in the form of Maxwell stresses. For positive, high  spin, the rate of angular momentum transport is about 5 times larger than for negative spin. This contribution is nearly absent for a non-rotating black hole. (ii) The mass accretion rate and the MAD parameter, both calculated at the horizon, are not correlated. However, their time derivatives are anti-correlated for every spin. (iii) 
For zero spin, the contribution of the toroidal component of the magnetic field to the magnetic pressure is negligible, while for fast spinning black hole, it is in the same order as the contribution of the radial magnetic component. For high positive spin, the toroidal component even dominates. (iv) For negative spins, the jets are narrower than their positive spin counterparts, while their fluctuations are larger. The weak jet from the non-rotating black hole is the widest with the smallest fluctuations. Our results highlight the complex, non-linear connection between the black hole spin and the resulting disk and jet properties in the MAD regime. 
\end{abstract}

\keywords{Accretion -- Magnetohydrodynamics -- Black hole physics -- Computational methods}
\section{Introduction}

Accretion disks are ubiquitous in many astronomical objects, such as active
galactic nuclei (AGNs) and X-ray binaries. 
The structure of the accretion disk mainly depends on the accretion rate.
At high accretion rate, close to the Eddington limit, the disks are
typically geometrically thin and optically thick, and the models
from \citet{Novikov1973blho.conf..343N}
and \citet{Shakura1973A&A....24..337S} are thought to accurately describe their physics. When the accretion rate is
much lower than the Eddington accretion rate,
the cooling time becomes longer than the accretion time, leading
to a radiatively inefficient accretion flow (RIAF), and the disk becomes
geometrically thick and optically thin. In this regime, there are several theoretical disk models, such as the Advection-dominated accretion flow
\citep[ADAF,][]{Narayan1994ApJ...428L..13N,Narayan1995ApJ...452..710N,
Abramowicz1995ApJ...438L..37A,Yuan2014ARA&A..52..529Y}.
The estimated low
luminosity of Sagittarius A$^\star$ as well as the black hole at the center of the M87 galaxy compared to the Eddington luminosity suggests that
these black holes accrete in the form of an ADAFs \citep{Yuan2002A&A...383..854Y}.

It is widely believed that the structure of an accretion flow consists of
a turbulent accretion disk, a bipolar jet and a magnetized
wind \citep[see e.g.][]{McKinney2004ApJ...611..977M,DHK03}.
The details of this structure strongly depend on the configuration and the strength
of the magnetic field inside and outside
the disk. The magnetic
fields in the disk, either advected from large distances or created in situ by the dynamo effect, are amplified by the magnetorotational instability
\citep[MRI,][]{Balbus1991ApJ...376..214B,Balbus1998RvMP...70....1B}.
They ultimately drive angular momentum transport, regulate
accretion and produce a bipolar, strongly magnetized jet.
There are two distinct modes of accretion, depending on
the magnetic fields surrounding the black hole, which ultimately lead to two
different disk configurations. In the Standard And Normal Evolution
\citep[hereinafter SANE, ][]{Narayan2012MNRAS.426.3241N,
Sadowski2013MNRAS.436.3856S}, the magnetic field pressure is not
strong and the accretion process is smooth.
The accretion disk, although turbulent, extends nearly evenly
up to the horizon. In this accretion mode, angular
momentum is transported mostly radially inside the disk by MRI \citep{Chatterjee2022ApJ...941...30C}.

The second type of disk is termed Magnetically Arrested Disk
\citep[MAD,][]{BR74, Narayan2003PASJ...55L..69N, Igu08}. In this model,
the magnetic flux accumulates near the  horizon until it saturates. In fact, the
accumulated magnetic field becomes so strong close to the black hole
that it can change the dynamics of the in-falling matter, thereby regulating the accretion. It was found
in 2D simulations that the accretion can be nearly fully stopped by
the magnetic pressure, and then resumes following the reconnection
of the magnetic field lines at the equator \citep[see, e.g.][]{CBL21}. This
picture however only partially holds in 3D simulations: accretion continuously
proceeds, via the development of non-axisymmetric instabilities 
\citep{SSP95,Begelman2022MNRAS.511.2040B}, with the in-falling gas being shaped into
filaments by the strong magnetic field (see e.g. Figure 1 of \cite{XZ19}, and \cite{Wong2021ApJ...914...55W}).

The MAD state attracted increasingly more attention in the past few years,
following the observation of the closest region to the super-massive black hole
in M87 and Sagittarius A$^\star$ by the Event Horizon Telescope (EHT) collaboration
\citep{EHT19,EHT22a}. By comparing the images taken in the radio band to
post-processed GRMHD simulations, it was determined that the accretion should
operate in the MAD state for those two black holes \citep{EHT_21b, EHT22e}. \citet{YWY22} independently arrived at
the same conclusion for M87 by studying rotation measurements.
Moreover, several studies, including \citet{DTJ20, PMY21, RLC22, SDB22},
proposed a model in which the flares observed in Sagittarius A$^*$ by
the GRAVITY experiment \citep{Gra18, Gra20} have their origin in the
magnetic flux eruptions, characteristics of the MAD state \citep{Igu08}. It is
clear that the MAD state is ubiquitous at low accretion rate, and better
understanding its properties will shed light on key observations of black holes
and their physics.

An important characteristics of the MAD state is the saturated
magnetic flux at the horizon. Since numerical studies of disk evolution depend on assumed initial conditions, in order to numerically study these systems one has to use an appropriate initial magnetic field configuration. \citet{Tchekhovskoy2011MNRAS.418L..79T} found such a configuration that results in the transport of a large amount of magnetic flux
%used a special initial magnetic field to ensure that large amount of magnetic flux are transported 
to the horizon. Using normalized black hole spin parameter $a=0.99$, they found that the
disk would be in the MAD state when the MAD parameter
reaches $\Phi_B \sim$50. Here and below, the MAD parameter is defined as the ratio of the magnetic flux to the square root of the mass accretion rate at the horizon. 
Later, \citet{Tchekhovskoy2012MNRAS.423L..55T} performed
two simulations with $a = 0.9$ and $a = -0.9$ and found that the
MAD parameter of the retrograde disk is about 30, smaller than that of
prograde disk. \citet{Narayan2022MNRAS.511.3795N} studied the
dependence on the spin of the MAD parameter, confirmed the results from \citet{Tchekhovskoy2012MNRAS.423L..55T}
and also found that the maximal value of the MAD parameter (for a given initial magnetic configuration) is, in fact,
reached for a black hole spin $a \simeq 0.5$.

The MAD parameter $\Phi_B$ is an important quantity in quantifying both the disk structure and the emerging jet. 
It was found that disks in
the MAD state around rotating black holes launch
strong and powerful jets 
via the \citet{BZ77} mechanism \citep[see e.g.][]{Tchekhovskoy2011MNRAS.418L..79T}. The emerging power of the magnetized  
jet is 
%the black hole efficiency to create jet power is 
proportional to the square of the MAD parameter multiplied by the mass accretion rate \citep{Tchekhovskoy2011MNRAS.418L..79T}.  It is therefore important to constrain the
accretion parameters and mechanisms which determine and regulate the value and the duty cycle of the MAD parameter.

The strong magnetic field during the MAD state pushes out the gas and stops, or at least regulates the dynamics of
the in-falling of matter. 
Therefore, one may naively expect that
an increase of the magnetic flux should result in
the decrease of the mass accretion rate, namely, that they
are anti-correlated. An anti-correlation is also expected if accretion
proceeds by interchange instability, as matter replaces a highly magnetized region closer to the black hole resulting in a drop in the MAD parameter
\citep{PMY21}. However, \citet{PMY21} did not find a correlation or an anti-correlation between the mass accretion rate
$\dot M $ and $\Phi_B$. Here, we expand their results for all black hole
spins: as we show below, none of our simulations show a correlation or an anti-correlation between $\dot M$ and $\Phi_B$. On the other hand, we do find an anti-correlation between their time derivatives.

A complete understanding of the interplay between accretion and saturated
magnetic field close to the horizon and how it impacts the structure and
power of the disk, jet and wind, as well as their evolution is still missing.
\citet{Narayan2022MNRAS.511.3795N} found that prograde disks have wider
jets and that the shape itself depends on the spin of the black hole. The
structure of the jet and the disk mainly depends on the magnetic field
pressure and the gas pressure inside the disks and jets.
\citet{Begelman2022MNRAS.511.2040B}, using the simulation in the MAD regime
around a rotating BH with $a = 0.9375$ from \citet{DJR20,DTJ20}, proposed
that the disk properties in this regime are due to a dynamically important
toroidal field in the close vicinity of the black hole. However,
\citet{Chatterjee2022ApJ...941...30C} studying accretion in the case of a
non-rotating black hole, showed that the radial magnetic field $b^r$ is
actually stronger than the toroidal magnetic field at small radii. This
apparent inconsistency needs to be resolved, even though its origin, namely
the black hole spin, is trivially identified. Here, we find that, close to the BH horizon, the toroidal component of the magnetic field, $b^\phi$ is increasing with the absolute value of the BH spin, $|a|$, being dynamically unimportant for $a=0$, but dominant for $|a| \rightarrow 1$.

It is generally thought that the strong magnetic field close to the black hole suppresses the development of MRI, as the disk height is smaller than the wavelength of the most unstable vertical mode \citep{MTB12, MAM18, White2019ApJ...874..168W}. This, in turn, affects the rate of angular momentum transport inside the disk, which details are still poorly understood in the MAD state. 
Recently, \citet{Begelman2022MNRAS.511.2040B} argued that MRI is actually not suppressed closest to the black-hole by considering non-axisymmetric modes \citep{DBL18}. On the other hand, \citet{Chatterjee2022ApJ...941...30C} demonstrated that a for non-rotating black hole, angular momentum is transported predominantly by magnetic flux eruptions, characteristics of a disk in the MAD state. It remains to understand if the contribution of this process still dominates the transport of angular momentum over MRI in the case of a rotating black hole. In addition, angular momentum from the disk is transported by the emerging winds above and below it. 

An additional source of angular momentum from an accretion disk system is the jet, in the form of Maxwell stresses. This contribution to the angular momentum originates mainly from the black hole, rather than the disk, and as such acts to spin down the black hole \citep{Chatterjee2022ApJ...941...30C}. As we show here, this transport can significantly contribute to the amount of angular momentum deposited to the external medium. Since the power of the jet depends on the BH spin, this also constrains the cosmic period over which the system is active. The net rate of angular momentum has a strong dependence on the spin and is the largest for prograde disks.

The shape of the jet itself also depends on the BH spin. It is determined by the balance between the internal and external stresses. As explained above, the main source of stresses are the magnetic field components, which, in turn depend on the BH spin. Furthermore, the BH spin not only affects the time-average jet shape, but also fluctuations around it. As we show here, retrograde disks produce the narrowest jets with the largest fluctuations, while non-spinning BHs produce the widest jets with the narrowest fluctuations.   

Given the complexity of accreting systems, many previous works used general
relativistic (eventually radiative and two temperatures) magnetohydrodynamic
(GRMHD) simulations to investigate the properties and evolutionary process of
MAD disks \citep{Narayan2012MNRAS.426.3241N, Sadowski2013MNRAS.436.3856S,
Porth2017ComAC...4....1P, White2020ApJ...891...63W, Porth2021MNRAS.502.2023P,
Begelman2022MNRAS.511.2040B, Narayan2022MNRAS.511.3795N,Chatterjee2022ApJ...941...30C}.
In the past two decades, with the rapid increase in compute capability, these
simulations have become increasingly popular and practical
\citep{Gammie2003ApJ...589..444G, Anninos2005ApJ...635..723A, Stone2008ApJS..178..137S,
Noble2006ApJ...641..626N, Porth2017ComAC...4....1P, Tchekhovskoy2019ascl.soft12014T, Liska2022ApJS..263...26L,
Begue2023ApJS..264...32B}.
In addition, general-purpose graphics processing units (GPU) started
to be used to accelerate fluid simulations in recent years, as they are particularly
well-suited to be ran on GPUs.  As a result, several GRMHD codes can now use
GPU accelerators, see e.g. \citet{Chandra2017ApJ...837...92C,Liska2020MNRAS.494.3656L,
Begue2023ApJS..264...32B, Shankar2022arXiv221017509S}. \textit{grim}
uses the library ArrowFire to achieve GPU compatibility
\citep{Chandra2017ApJ...837...92C}. \citet{Liska2020MNRAS.494.3656L}
developed H-AMR with openMP, MPI and CUDA. Building on HARMPI, our group developed a new GPU-accelerated
GRMHD code, cuHARM, which uses openMP and cuda \citep{Begue2023ApJS..264...32B}. This code is thoroughly optimized for maximal harness of the power available in NVidia GPUs, with more than 50\% computation efficiency on NVidia A100 cards. For the results presented here, the simulations are made on a single multi-GPU workstation.

In this paper, using our GRMHD code cuHARM, we study the
role played by the magnetic field in the MAD state for different black hole spins, and its effect on the structure of the disk and the jet.
For this, we present several simulations with different initial magnetic field
strengths and black hole spins. 
This paper is organized as follows. In Section \ref{sec:simulations},
we present the setup of our simulations and introduce the numerical diagnostics
used in our analysis. We discuss the dynamics of the accretion disk system in
our simulations in Section \ref{sec:general}. In particular, after specifying the inflow and outflow equilibrium radius and the time evolution of $\dot M$ and $\phi_B$ in section \ref{subsec:inflowoutflow} and \ref{subsec:mass_accretion} respectively, we study (i) the absence of correlation between the mass accretion rate and the MAD parameter, and introduce the anti-correlation between their time derivative in section \ref{subsec:time_characteristics}; (ii) the shape of the jet as a function of spin in section \ref{subsec:disk-jet}, finding that the retro-grade disks are narrower than their corresponding prograde disks; and (iii) the component-wise contributions of the magnetic pressure to underline the differences between a spinning and a non-spinning black-hole in section \ref{subsec:pressure}, where the toroidal component is found to be sub-dominant for $a = 0$ but similar to the radial component for large |a|. In section \ref{sec:angular}, we discuss the transport of angular momentum for our simulations with spin $a = -0.94$, $a = 0$ and $a = 0.94$, underlying the differences between each black hole spins. The summary and conclusions of the paper are given in Section \ref{sec:conclusion}.

\section{Simulations}

\label{sec:simulations}

We perform several simulations with cuHARM \citep{Begue2023ApJS..264...32B}, which uses the finite volume method to numerically solve the conservative GRMHD equations \citep[for reviews, see \textit{e.g.}][]{Marti2003LRR.....6....7M, Font2008LRR....11....7F,Rezzolla2013rehy.book.....R}. The code is written in CUDA-C and openMP, and all calculations of cuHARM are accelerated by GPU (only the data transfer and exports are powered by CPU). To perform the simulations which results are presented in this article, we use an Nvidia DGX-V100 server with 8 Nvidia V100 GPUs .

%\subsection{cuHARM and Simulation Setup}
\subsection{Initial setup}

In this paper, we study on the accretion flows in the MAD state around a
spinning black hole. Our simulations begin with the stationary axisymetric torus
described by \citet{Fishbone1976ApJ...207..962F}. We set the gas adiabatic index to
$\Gamma = 14/9$, and consider an initially large disk with $r_{\mathrm{in}} = 20 r_g$
and $r_{\mathrm{max}} = 41 r_g$, where $r_{\mathrm{in}}$ is the inner boundary of the
disk, $r_{\mathrm{max}}$ is the radius at which the pressure reaches its maximum, and
$r_g$ is the gravitational radius. The matter and internal energy densities are
normalized such that, for the initial disk, the maximum matter density $\rho$ in the
entire disk is normalized to $\rho_{\mathrm{max}} = 1$. The internal energy density is scaled
accordingly. Since the initial torus is in equilibrium, it does not spontaneously 
evolve. We therefore add small random perturbations (set to $4\%$) to the internal
energy density $u$ as the seed of instabilities, which will promote accretion.

This initial torus is in full hydrodynamic equilibrium and, as such, it does not
contain any magnetic field. We introduce a purely poloidal subdominant magnetic
field defined by the vector potential ${\bf A}$, such that $A_r = A_\theta = 0$ and 
\begin{equation}
    A_\phi = \mathrm{max}\left[0, \left(\frac{\rho}{\rho_{\mathrm{max}}}\right) \left(\frac{r}{r_\mathrm{in}} \sin\theta\right)^3 \exp \left (-\frac{r}{400} \right ) - 0.2\right],
\end{equation}
which has been previously employed in, \textit{e.g.}, \citet{Wong2021ApJ...914...55W,
Narayan2022MNRAS.511.3795N}. Here $r$, $\theta$ and $\phi$ are the horizon
penetrating spherical Kerr-Schild coordinates. The corresponding  magnetic field is initially
 a single loop confined to the disk. The magnitude of the magnetic field  is further normalised by the parameter
$\beta_0 = p_{\mathrm{gas,max}} / p_{\mathrm{b,max}} \gg 1$, where
$p_{\mathrm{gas, max}}$ is the maximum gas pressure,
$p_{\mathrm{b, max}} = b^2 / 2$ is the maximum of the
magnetic field pressure, and $b = b^\mu b_\mu$ is the norm of the 4-vector magnetic field, see Section \ref{subsec:diagnostics} below. This expression of the magnetic field is designed to ensure
that  enough magnetic flux can be transported
to the black hole throughout the course of the simulation and
"saturates" its magnetosphere; see further discussion in  \citet{Tchekhovskoy2011MNRAS.418L..79T}. % \authorcommentD{I think you should cite Tchekovskoy here.} 

We conduct a series of simulations with different black hole spins,
$a \in \{-0.985, -0.94, -0.85, -0.5,
0, 0.5, 0.85, 0.94, 0.985 \} $ and an initial magnetization $\beta_0 = 100$.
In the case of the retrograde disk with $a= -0.94$, we also varied the initial magnetic
field strengths, with $\beta_0 \in \{100, 200, 400, 800 \}$. We evolve most of the
simulations until $t = 2\times 10^4 t_g$, where $t_g = r_g / c$, apart for aM94b800,
which is evolved to $t = 2.5 \times 10^4 t_g$ due to the weak initial magnetic field
and the longer time required to reach the MAD state for this setup. Additionally,
simulation aM94b100h is evolved until $t = 5 \times 10^4 t_g$ in order to study the
long time behavior of our accretion disk system.
% and aM94b100h, which is evolved until $t = 50,000 t_g$ in order to
% assert the long time behaviour of our simulations. We
We use the spin and the initial $\beta_0$ to name the simulations: "a" stands for spin,
"M" indicates a negative value, and "b" represents the initial $\beta_0$,
such that, for instance, aM94b100 stands for a simulation with the negative ("M") spin
$a = -0.94$ and an initial $\beta_0 = 100$. A summary of all the simulations used in this work is given in Table \ref{tab:runs}.

\subsection{Numerical aspects}

Since we are studying accretion around rotating black hole, we
use the Kerr metric for our simulations. The Kerr-Schild (KS) coordinate system
$(t, r, \theta, \phi)$ is used as the physical coordinates, while
to both enhance the robustness of the calculation and
focus the computation in the region of interest, namely close to the black hole and
at the equator, the modified Kerr-Schild
\citep[MKS, see \textit{e.g.}][]{McKinney2004ApJ...611..977M} coordinates
$(t, q^1, q^2, \phi)$ are used in the numerical calculation. The relation between these coordinates, as implemented in cuHARM, can be found in section 4.1 of \citet{Begue2023ApJS..264...32B}.

We use the inflow and  outflow boundary conditions in the radial
direction at small and large radii, respectively. 
In the $\theta$ direction we use 
the reflective boundary condition, and the periodic boundary condition is used in the $\phi$ direction.
To address the potential numerical errors in empty or strongly magnetized region,
we adopt the same flooring model as in \citet{Begue2023ApJS..264...32B}, which is used in many other papers, e.g. \citet{PCN19}. The density $\rho$
and the internal energy $u$ are limited using 
\begin{equation}
    \rho = \max \left ( \rho , 10^{-20}, 10^{-5} r^{-\frac{3}{2}} \right ),
\end{equation}
\begin{equation}
	u = \max \left( u, 10^{-20}, \frac{10^{-5}}{3} r^{-\frac{5}{2}}\right ).
\end{equation}
Matter and energy are added when needed to preserve the conditions $b^2/\rho < 50$ and
$b^2/u < 2.5\times 10^3$.

The reference resolution of most simulations is ($N_r\times N_\theta \times N_\phi$)
= (192, 96, 96), except for aM94b100h and a0b100h, which have a slightly higher resolution
($N_r\times N_\theta \times N_\phi$) = (256, 128, 128). Here, the "h" appended to
the name stands for "high resolution". The resolution of the simulations presented
in this paper is somewhat lower than that used in some of the simulations presented in recent works. For example,
\citet{Narayan2022MNRAS.511.3795N} performed fairly similar simulations with a
resolution of 288x192x144. \citet{White2020ApJ...891...63W}
examined the impact of different resolutions, and they argued that the accretion rate and the general disk structure agree across
the simulations with different resolution. We use our higher
resolution simulations, aM94b100h and a0b100h, to check the solidity
of our results to a change in the resolution. We did not find any
significant difference between the low and high resolution simulations.

\begin{table}[]
\centering
\begin{tabular}{|c||c|c|c||c|c|c|}
\hline
Name      & $\beta_0$ & spin   & Resolution $N_r \times N_\theta \times N_\phi$ & MAD parameter        & Accretion Rate            & Jet Efficiency             \\ \hhline{|=|=|=|=|=|=|=|}
aM985b100 & 100       & -0.985 & 192   $\times$ 96         $\times$ 96       & $12.26^{+2.58}_{-2.02}$ & $23.54^{+9.54}_{-6.52}$   & $0.28^{+0.15}_{-0.11}$ \\ \hline
aM94b100  & 100       & -0.94  & 192   $\times$ 96         $\times$ 96       & $14.46^{+3.99}_{-2.23}$ & $24.30^{+10.19}_{-8.60}$  & $0.26^{+0.15}_{-0.09}$ \\ \hline
aM85b100  & 100       & -0.85  & 192   $\times$ 96         $\times$ 96       & $17.59^{+3.52}_{-2.97}$ & $22.62^{+10.76}_{-6.55}$  & $0.26^{+0.12}_{-0.10}$ \\ \hline
aM5b100   & 100       & -0.5   & 192   $\times$ 96         $\times$ 96       & $26.24^{+2.28}_{-3.34}$ & $31.55^{+9.58}_{-9.15}$   & $0.11^{+0.02}_{-0.02}$ \\ \hline
a0b100    & 100       & 0      & 192   $\times$ 96         $\times$ 96       & $30.94^{+2.25}_{-5.08}$ & $15.12^{+2.44}_{-2.46}$   & $0.06^{+0.01}_{-0.01}$ \\ \hline
a5b100    & 100       & 0.5    & 192   $\times$ 96         $\times$ 96       & $32.73^{+7.02}_{-6.56}$ & $45.16^{+15.28}_{-12.90}$ & $0.22^{+0.07}_{-0.06}$ \\ \hline
a85b100   & 100       & 0.85   & 192   $\times$ 96         $\times$ 96       & $29.94^{+5.10}_{-4.79}$ & $34.23^{+14.34}_{-11.36}$ & $0.90^{+0.29}_{-0.25}$ \\ \hline
a94b100   & 100       & 0.94   & 192   $\times$ 96         $\times$ 96       & $25.37^{+4.27}_{-4.20}$ & $32.15^{+13.39}_{-10.26}$ & $1.05^{+0.36}_{-0.38}$ \\ \hline
a985b100  & 100       & 0.985  & 192   $\times$ 96         $\times$ 96       & $22.69^{+3.93}_{-3.69}$ & $32.63^{+15.74}_{-10.80}$ & $1.22^{+0.47}_{-0.32}$ \\ \hhline{|=|=|=|=|=|=|=|}
aM94b200  & 200       & -0.94  & 192   $\times$ 96         $\times$ 96       & $15.09^{+2.66}_{-2.89}$ & $20.75^{+6.46}_{-6.04}$   & $0.28^{+0.12}_{-0.11}$ \\ \hline
aM94b400  & 400       & -0.94  & 192   $\times$ 96         $\times$ 96       & $16.37^{+2.40}_{-2.21}$ & $20.41^{+5.85}_{-4.92}$   & $0.34^{+0.10}_{-0.08}$ \\ \hline
aM94b800  & 800       & -0.94  & 192   $\times$ 96         $\times$ 96       & $16.22^{+1.80}_{-1.67}$ & $15.36^{+5.78}_{-4.60}$   & $0.33^{+0.07}_{-0.06}$ \\ \hhline{|=|=|=|=|=|=|=|}
a0b100h   & 100       & 0      & 256   $\times$ 128         $\times$ 128     & $32.84^{+4.23}_{-4.10}$ &                  $43.12^{+17.25}_{-11.14}$ &  $0.07^{+0.01}_{-0.01}$                 \\ \hline
aM94b100h & 100       & -0.94  & 256   $\times$ 128        $\times$ 128      & $15.13^{+1.81}_{-3.18}$ & $23.94^{+10.65}_{-8.57}$  & $0.27^{+0.07}_{-0.12}$ \\ \hline
\end{tabular}
\caption{List of the simulations presented in this paper with their initial magnetization $\beta_0$, spin $a$ and resolution. The three last columns give the time-average value of the MAD parameter $\Phi_B$, of the accretion rate $\dot M$ and of the jet efficiency $\eta$ for $10^4 t_g< t < 2\times 10^4 t_g$ ($1.5 \times 10^4 t_g < t < 2.5 \times 10^4 t_g$ for aM94b800, $10^4 t_g < t < 5 \times 10^4 t_g$ for aM94b100h).}
\label{tab:runs}
\end{table}
%\authorcommentGQ{I don't understand why the mass accretion rate of a0b100h is so high. I guess maybe you use different adiabatic index?}

\subsection{Diagnostics}\label{subsec:diagnostics}
Following \citet{Kom99}, let $b^\mu \equiv (\star F)^{\mu \nu} u_\nu$ represent the 4-vector magnetic field and $u^\mu$ be the 4-velocity, which is orthogonal to $b^\mu$. In the ideal MHD limit, the dual to the Faraday tensor is given by 
\begin{equation}
    (\star F)^{\mu \nu} = b^\mu u^\nu - b^\nu u^\mu.
\end{equation}
In this limit of a perfect magnetized fluid, the stress energy tensor $T^{\mu \nu}$ is given by
\begin{equation}
    T^{\mu\nu} = (h + b^2) u^\mu u^\nu + \left (p_g + \frac{b^2}{2} \right )g^{\mu \nu} - b^\mu b^\nu.
\end{equation}
Here, $ h = \rho + u + p_g$ is the enthalpy, $p_g$ is the gas pressure, 
$ b^2 = b^\mu b_\mu$ and $g^{\mu \nu}$ is the metric tensor with determinant noted $\sqrt{-g}$. Using cuHARM, we solve the general relativistic magneto-hydrodynamic equations, namely 
\begin{equation}    
    \nabla_\mu \left ( \rho u^\mu \right ) = 0   \label{eq:mass_conservation}
\end{equation}
\begin{equation}    
    \nabla_\mu \left ( T^{\mu\nu}\right ) = 0  \label{eq:energy_momentum_conservation}
\end{equation}
\begin{equation} 
    \nabla_\mu \left ( \star F^{\mu \nu}\right ) = 0  \label{eq:maxwell} 
\end{equation}
which respectively are the equations of mass conservation, the equation of energy and momentum conservation, and the homogeneous Maxwell's equations.

The MAD state mainly depends on the magnetic flux through the horizon,
Therefore, we define the  following radial diagnostics, which can
eventually be evaluated at the horizon:
\begin{enumerate}
    \item The mass accretion rate:
        \begin{equation}
            \dot{M}(r) = \int_{\theta = 0}^{\theta = \pi} \int_{\phi = 0}^{\phi = 2\pi} \sqrt{-g} \rho u^r d\theta d\phi.
            \label{eq:mdot}
        \end{equation}
    \item The magnetic flux crossing the horizon (through one hemisphere):
        \begin{equation}
            \phi_B(r = r_H) = \frac{1}{2} \int_{\theta = 0}^{\theta = \pi} \int_{\phi = 0}^{\phi = 2\pi} \sqrt{-g} | \star F^{rt}| d\theta d\phi.
            \label{eq:phiB}
        \end{equation}
    \item The energy flux through the horizon towards the black hole:
        \begin{equation}
            \dot{E}(r) = - \int_{\theta = 0}^{\theta = \pi} \int_{\phi = 0}^{\phi = 2\pi} \sqrt{-g}   T^{r}_{~t} d\theta d\phi.
            \label{eq:dotE}
        \end{equation}
    \item The angular momentum flux in the radial direction:
        \begin{equation}
             \dot{J}_r (r) = \int_{\theta = 0}^{\theta = \pi} \int_{\phi = 0}^{\phi = 2\pi} \sqrt{-g} T^{r}_{~\phi}  d\theta d\phi.
            \label{eq:dotL}
        \end{equation}
    \item The MAD parameter:
        \begin{equation}
            \Phi_B = \frac{\phi_B}{\sqrt{\dot M (r = r_H)}} 
            \label{eq:MAD_param}
        \end{equation}
    \item The jet efficiency at the horizon:
        \begin{equation}
            \eta(r = r_H) = 1 + \frac{\dot{E} (r = r_H)}{\dot{M} (r = r_H)}.
            \label{eq:jet-efficiency}
        \end{equation}
\end{enumerate}

\begin{figure}
    \centering
    \includegraphics[width=\textwidth]{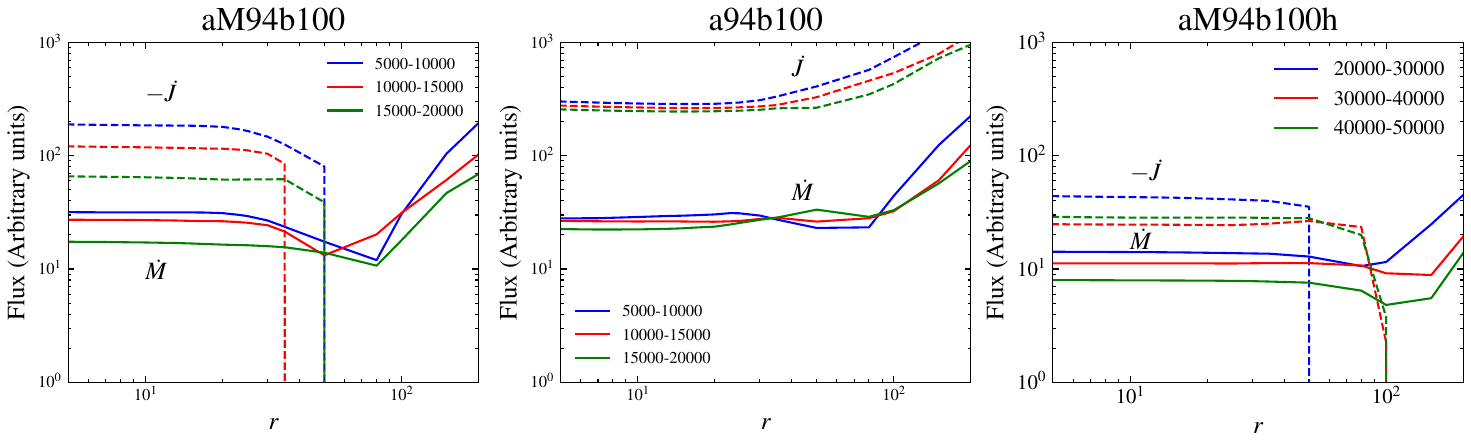}
    \caption{ The radial profile of the mass accretion rate $\dot{M}$ (solid line) and
    of the angular momentum flux $\dot{J}$ (dashed line) for aM94b100
    (left panel), a94b100 (middle panel) and for the long term evolution aM94b100h
    (right panel) at different time periods (see
    legend). As the time increases, the steady region extends to larger radius.
    This suggests that aM94b100 and a94b100 have established inflow outflow
    equilibrium at $r < 30 r_g$ for $1.5\times 10^4 t_g <t < 2\times 10^4 t_g$, while
    it is about $70r_g$ at $ 4\times 10^4 < t_g < 5 \times 10^4 $ for aM94b100h.
    For the simulations with the negative spin on the left and right panels,
    the angular momentum fluxes change sign at $r = 30-50 r_g$ and $r = 50-100 r_g$, respectively.}
    \label{fig:flux-r}
\end{figure}
%\deletedD{Left panel: The radial profile of aM94b100. The solid lines are the mass accretion rate and the dashed lines are the angular momentum flux. The different colors denote different averaged time period.} 

Note that our definition of angular momentum flux $\dot J$ and energy flux
$\dot E$ are opposite to those employed by \citet{Narayan2022MNRAS.511.3795N},
but are in agreement with the definitions of \textit{e.g.} \citet{PCN19}.
We are also interested in the structure of the disk and of the jet. Therefore,
we define the additional following diagnostics.
\begin{enumerate}
    \item The disk height, denoted by $(h/r)$:
        \begin{equation}
            (h/r)~(t,r) = \frac{\int_{0}^{2\pi} \int_0^\pi | \frac{\pi}{2} - \theta | \rho \sqrt{-g} d\theta d\phi}{\int_{0}^{2\pi} \int_0^\pi \rho \sqrt{-g} d\theta d\phi}.
            \label{eq:hr}
        \end{equation}
    \item The $\phi$-average of a quantity $q$:
        \begin{equation}
            \langle q \rangle_{\phi} (t, r, \theta) = \frac{\int_0^{2\pi} q \sqrt{-g} d\phi}{\int_0^{2\pi} \sqrt{-g} d\phi}.
            \label{eq:phi-average}
        \end{equation}
    \item The disk-average of a quantity $q$:
        \begin{equation}
            \langle q \rangle_{\theta, \phi}(t,r) = \frac{\int_0^{2\pi} \int_0^{\pi} q \rho \sqrt{-g} d\theta d\phi}{\int_0^{2\pi} \int_0^{\pi} \rho \sqrt{-g} d\theta d\phi}.
        \end{equation}
    \item The disk-average of a quantity $q$ but within a narrow $\theta$ range  (used below in calculating the pressure):
        \begin{equation}
            \langle q \rangle_{\theta, \phi}(t,r) = \frac{\int_0^{2\pi} \int_{\theta = \pi / 8}^{\theta = 7\pi / 8} q \rho \sqrt{-g} d\theta d\phi}{\int_0^{2\pi} \int_{\theta = \pi / 8}^{\theta = 7\pi / 8} \rho \sqrt{-g} d\theta d\phi}.
            \label{eq:pressure-average}
        \end{equation}
\end{enumerate}
We will also present time-averaged quantities and time- and $\phi-$averaged maps,
which are computed for $ 10^4 t_g < t < 2 \times 10^4 t_g$, unless specified otherwise,
and the full $2\pi$ range for $\phi$.

\section{General accretion dynamics}

\begin{figure}
    \centering
    \gridline{\fig{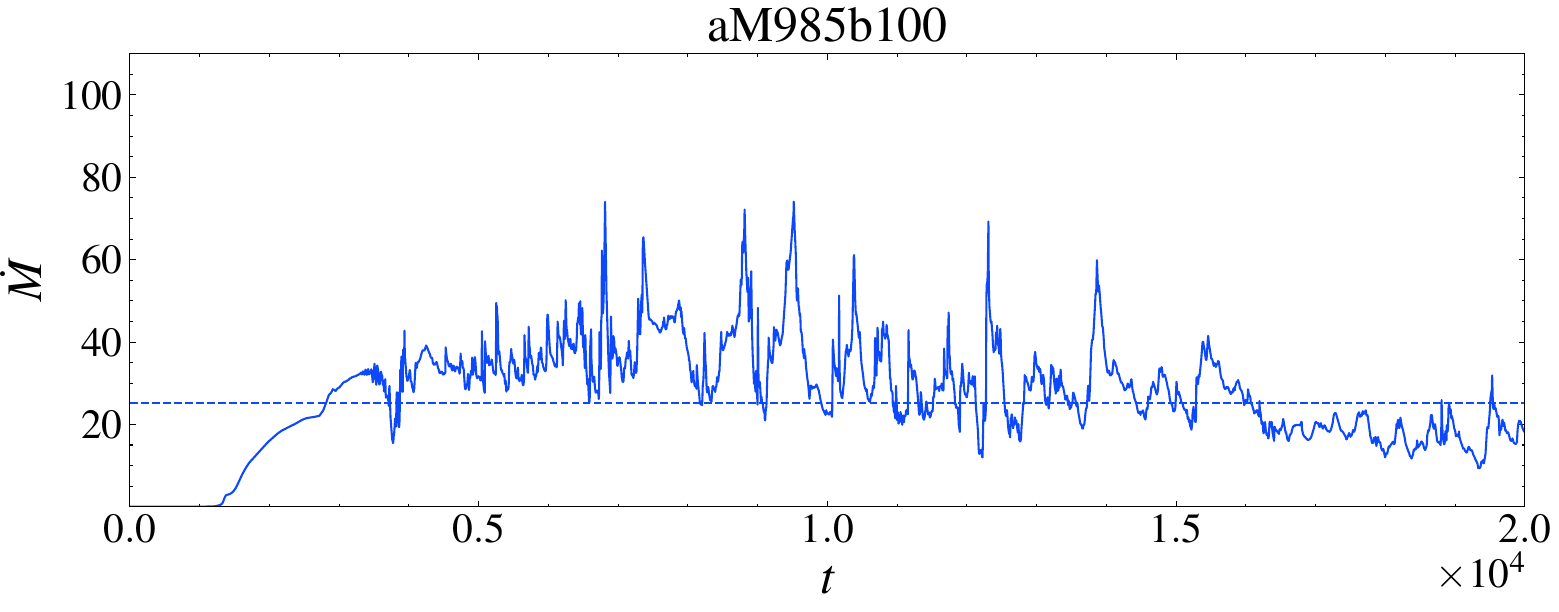}{0.48\textwidth}{}
              \fig{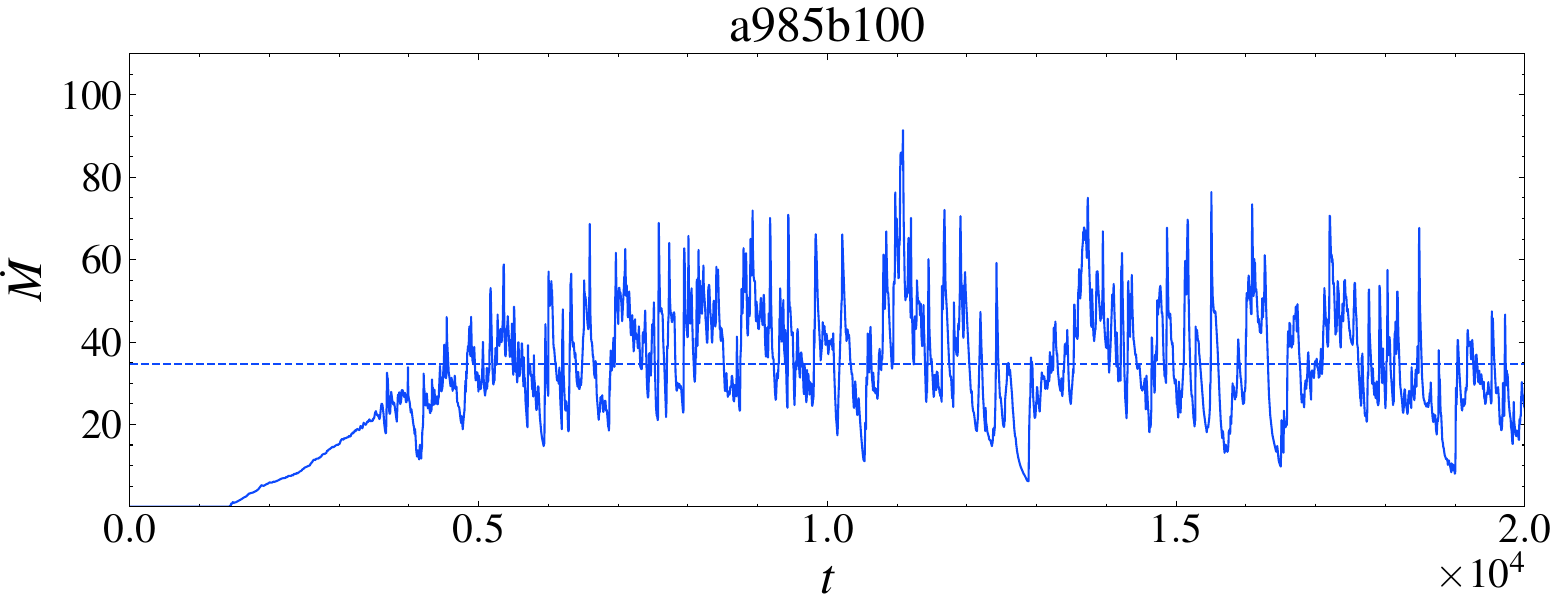}{0.48\textwidth}{}}
                \vspace*{-8mm}
    \gridline{\fig{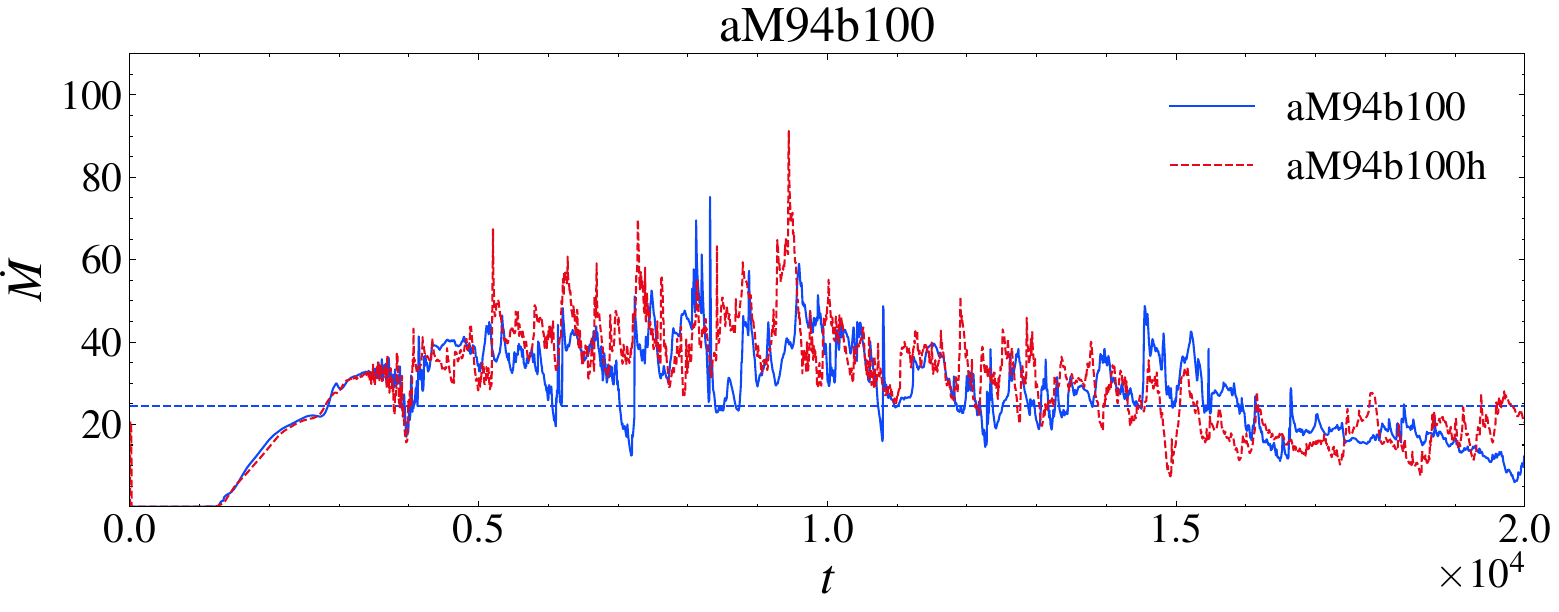}{0.48\textwidth}{}
              \fig{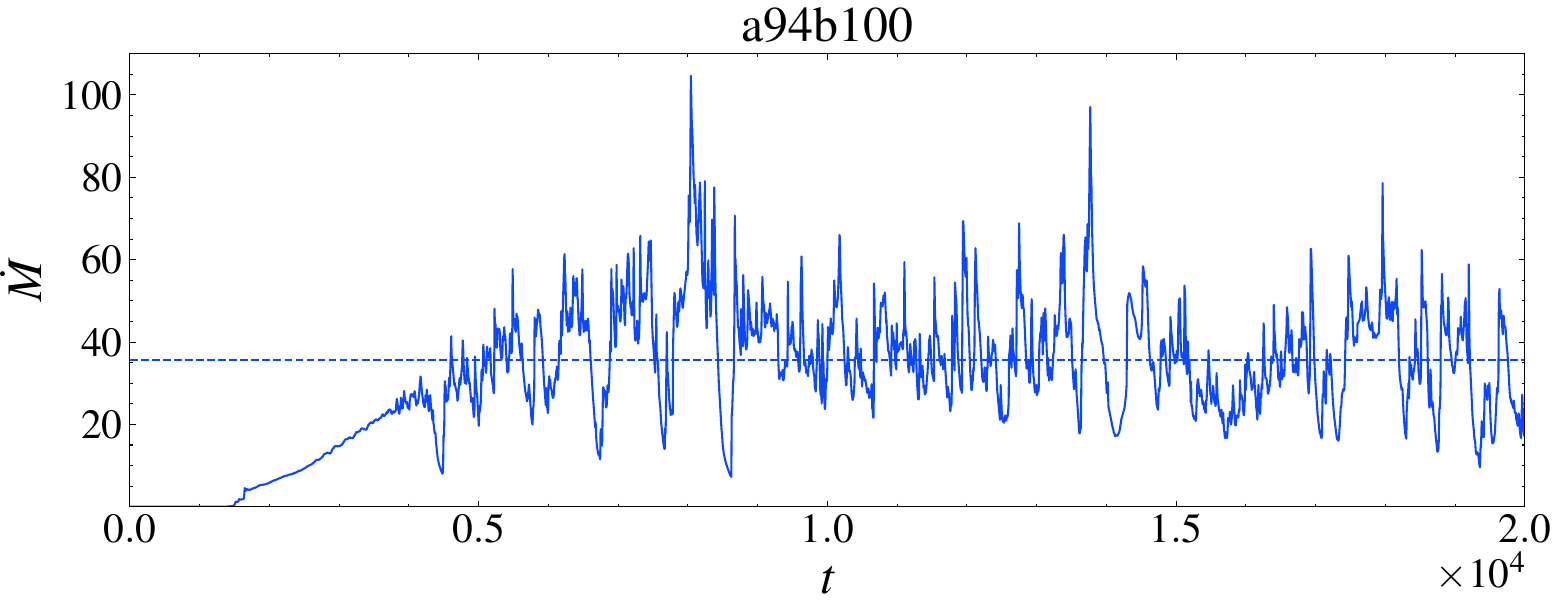}{0.48\textwidth}{}}
              \vspace*{-8mm}
    \gridline{\fig{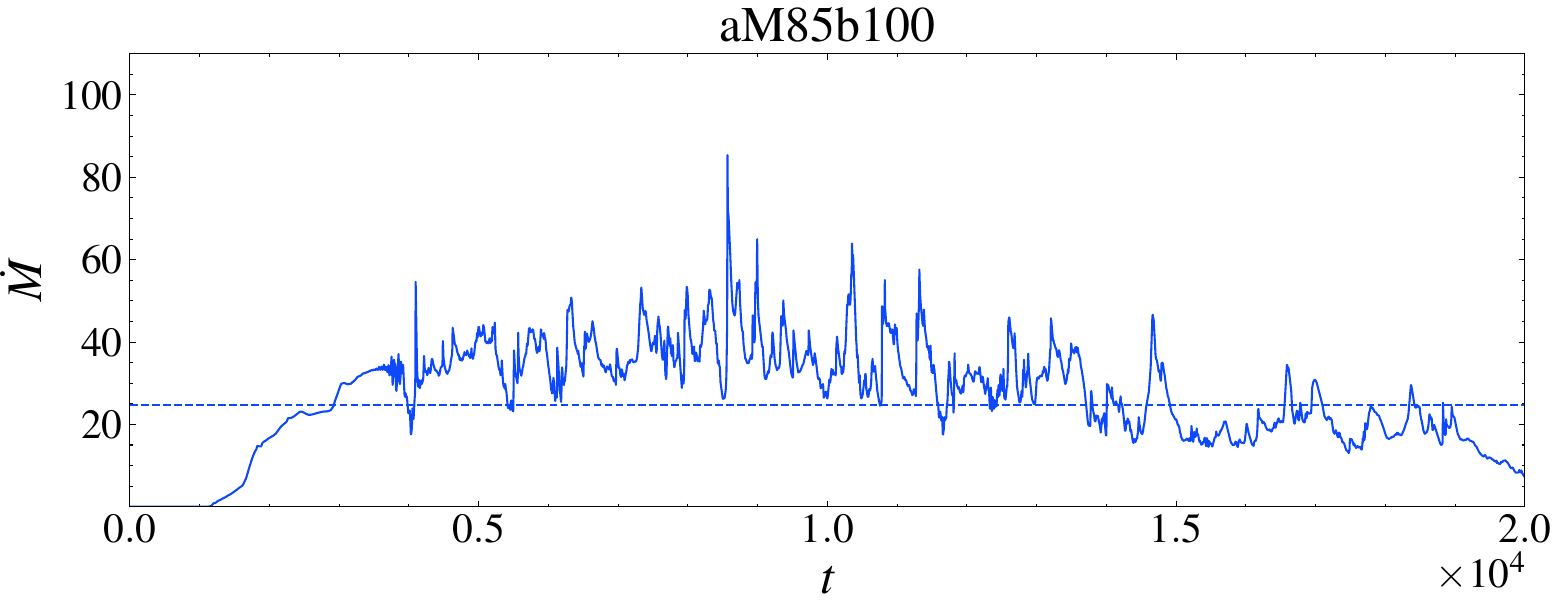}{0.48\textwidth}{}
              \fig{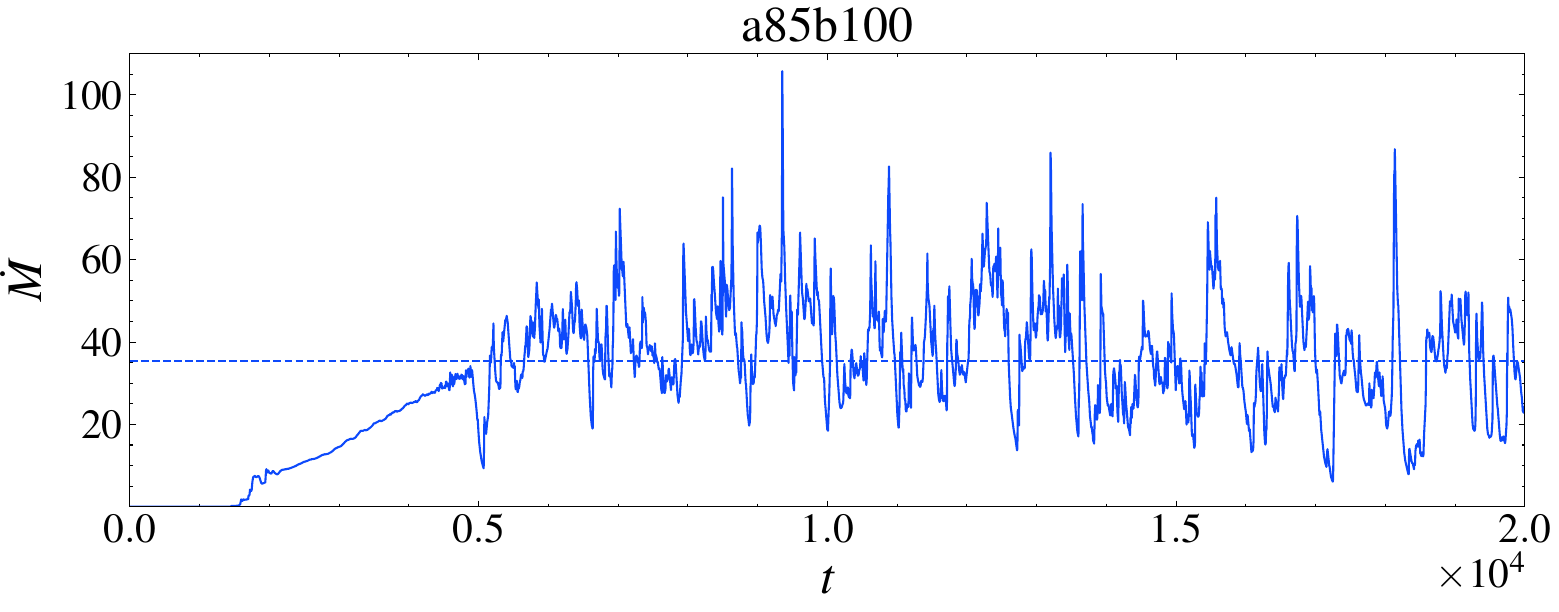}{0.48\textwidth}{}}
              \vspace*{-8mm}
    \gridline{\fig{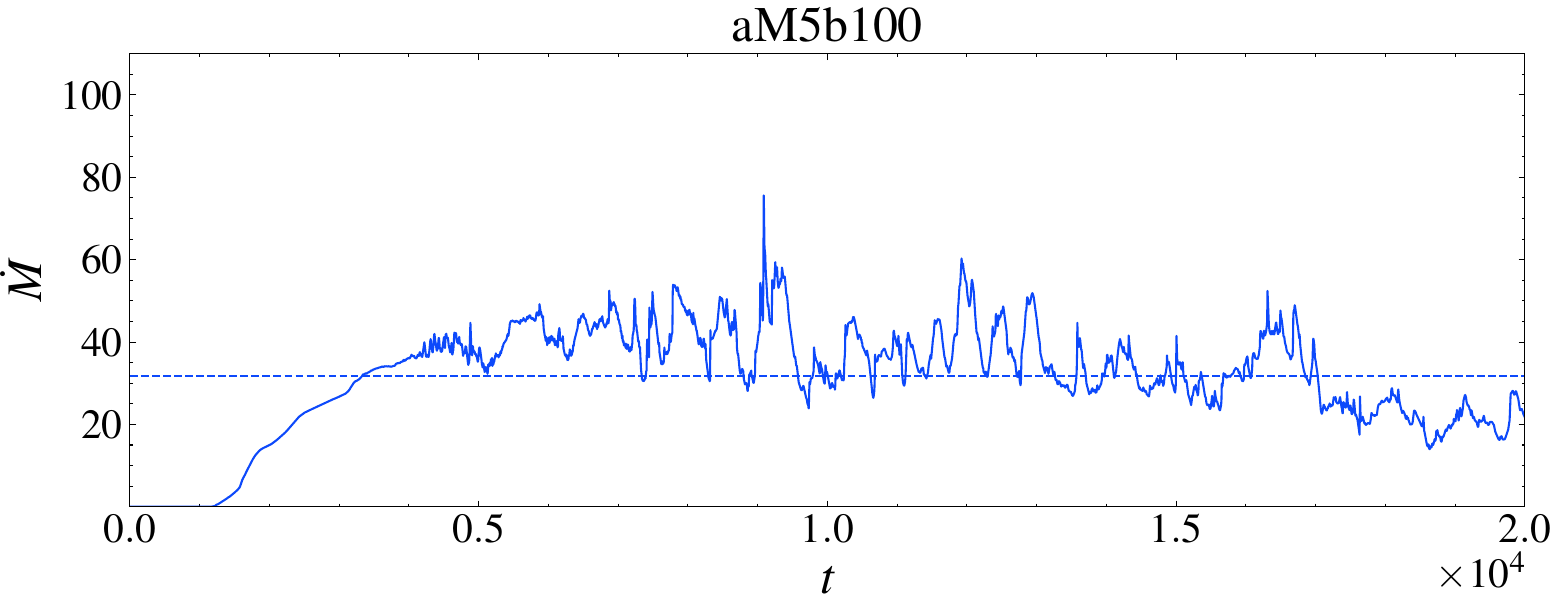}{0.48\textwidth}{}
              \fig{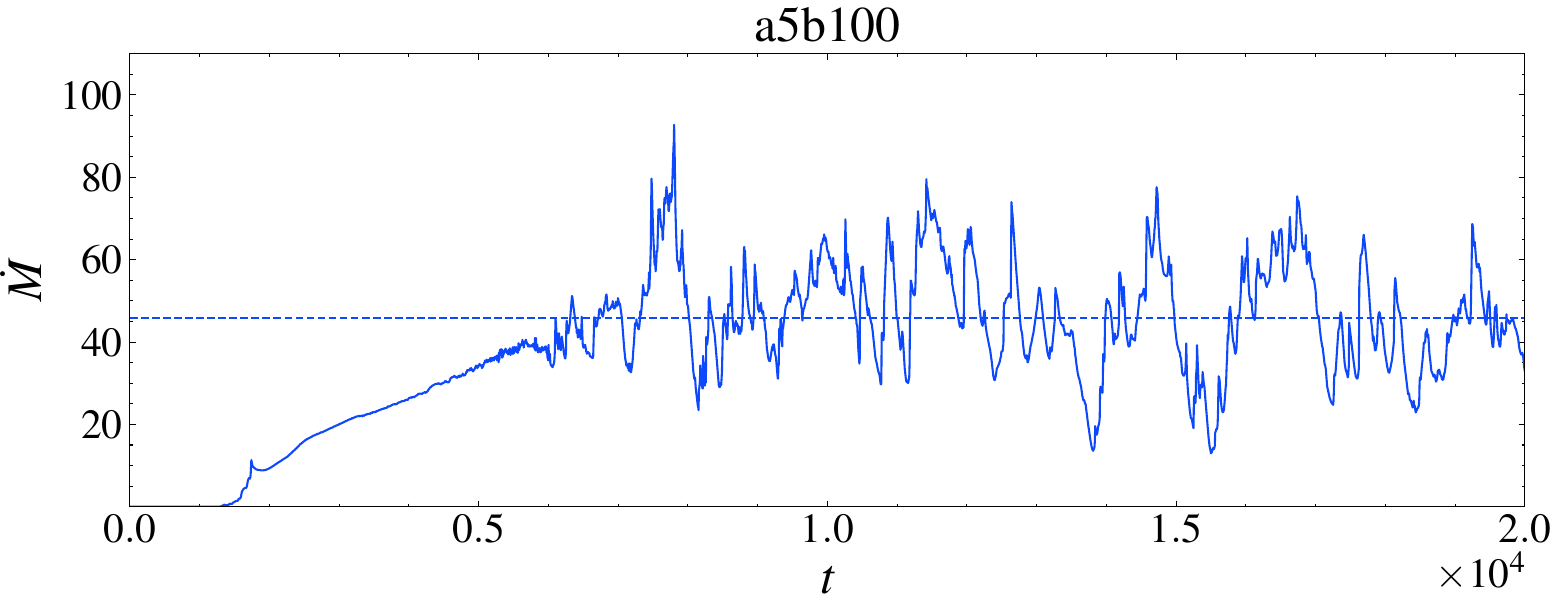}{0.48\textwidth}{}}
              \vspace*{-8mm}
    \gridline{\fig{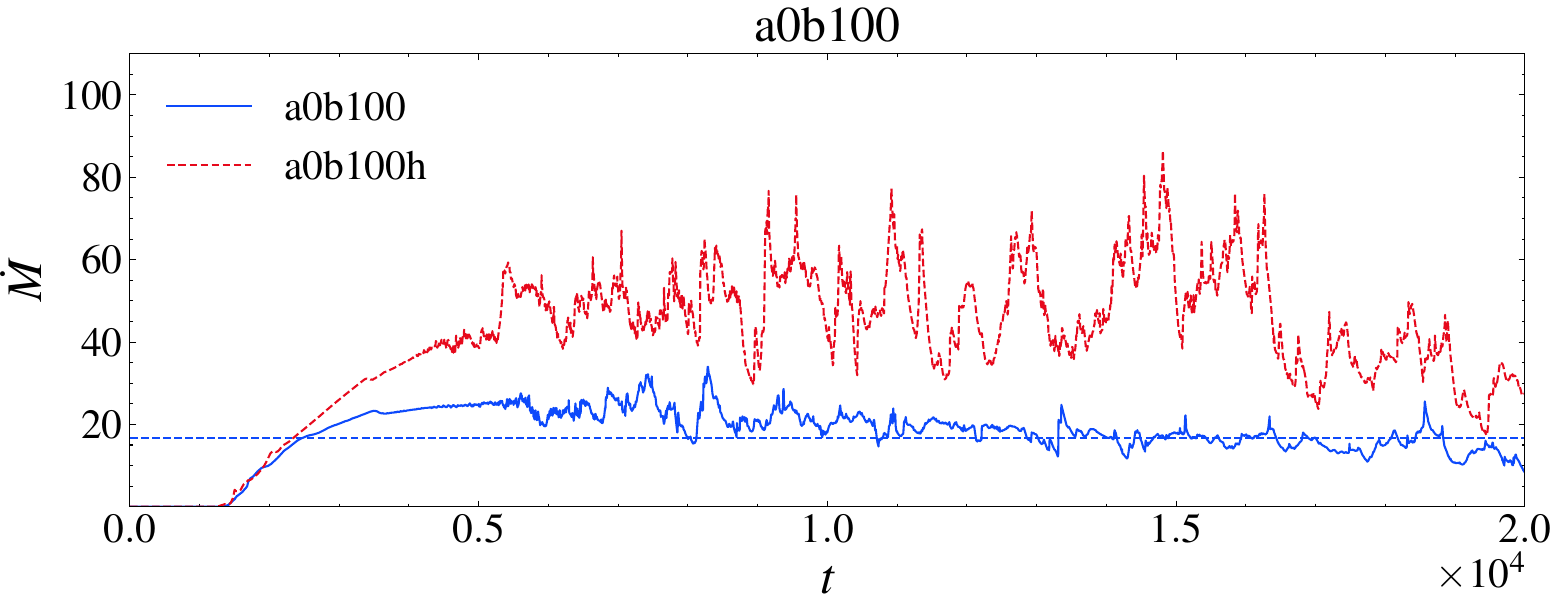}{0.48\textwidth}{}}
    \caption{Time evolution of the mass accretion rate $\dot M$. The
    left column pertains to retrograde disks, while the right columns
    corresponds to prograde disks. For convenience, we set the y-axes
    to identical scales to facilitate direct
    comparison. The horizontal lines are the time-averaged mass accretion
    rate $\dot{M}$, where the averaged is taken from $10^4 t_g$ to $2\times 10^4 t_g$.}
    %\authorcommentGQ{There are something wrong for a0b100h. I don't understand why the mass accretion rate of a0b100h is so high.}
    \label{fig:dotM}
\end{figure}

\label{sec:general}

\subsection{Inflow equilibrium radius and radial limit of our analysis}\label{subsec:flux-r}

\label{subsec:inflowoutflow}

Most simulations are evolved until $t = 2\times 10^4 t_g$, which is sufficient
for the disk to be in the MAD state. We assume this state to be established
when the MAD parameter reaches its average value at late time. We find that
in all cases, the averaged MAD parameter is comparable to or greater
than 15, the limiting value proposed by  \citet{Tchekhovskoy2011MNRAS.418L..79T}
to define the MAD state. Only aM94b800 barely reaches the MAD state at
$t = 2\times 10^4M$ because of the initial small magnetic field normalisation.
So this simulation is evolved further until $t = 2.5\times 10^4M$ at which time the MAD state for this initial setup is well-established. %Simulation aM94b100h evolves until  $t = 5\times 10^4M$ for comparison with previous works.  

We first study the radial profile of the mass accretion rate $\dot{M}$
and of the angular momentum flux $\dot{J}$, as given by Equations
\eqref{eq:mdot} and \eqref{eq:dotL}, respectively. The results are
displayed on the left and middle panel of Figure \ref{fig:flux-r} for simulations aM94b100 and a94b100, serving as examples. The results are similar for
all other simulations. For aM94b100 and a94b100, the radial profiles
of $\dot M$ and $\dot{J}$ are averaged over
3 different time intervals, namely from $5\times 10^3$ to $10^4 t_g$,
from $10^4$ to $1.5\times 10^4t_g$ and from $1.5\times 10^4$ to
$2\times 10^4t_g$. In the last time interval, the mass accretion
rates of these two runs are independent of the radius $r$ for $r < 30 r_g$. The angular
momentum fluxes also exhibit a similar pattern, remaining constant at
$r < 30 r_g$. This means that the inner region with $r < 30 r_g$
is in inflow equilibrium state.

There are two key differences between the prograde and retrograde
cases. Firstly, the sign of the angular momentum flux is different. The
prograde black hole has a positive angular momentum flux, namely,
the black hole is losing angular momentum, as was previously reported in
\citet{Narayan2022MNRAS.511.3795N}.
%\authorcommentD{meaning the there is a net flux of angular momentum away from the BH}.
Conversely, the angular momentum flux of the retrogade black hole
is negative, meaning that it accumulates positive angular momentum and spins down.  The second difference
is the existence of a radius at which the angular momentum flux changes sign for a retrograde black hole.
This indicates that there is a net flux of angular momentum 
away from the black hole at large distance. This radius is also observed in the simulation
presented in \citet{Narayan2022MNRAS.511.3795N}, see their figure 3 with 
the sharp drop of angular momentum flux at $r \sim 10^2 r_g$. We
discuss in more details the differences for the angular momentum transport
between prograde and retrograde disks in section \ref{sec:angular}.

The duration of most of our simulations is shorter than recent long-time
simulations. For example, \citet{Narayan2022MNRAS.511.3795N}
evolved the accretion system until $t = 10^5 t_g$,  such that they obtained
an inflow equilibrium radius at $\sim 100 r_g$, larger than ours by a factor
3 to 4. To verify the reliability of our results over a longer time span,
we extend the simulation aM94b100h to $t = 5 \times 10^4 t_g$. In the right panel of Figure \ref{fig:flux-r},
we present the radial profile of
the mass accretion rate $\dot{M}$ and the angular momentum flux $\dot{J}$ for this extended period.  
The time-averaged quantities
for $2 \times 10^4 t_g < t < 3\times 10^4 t_g$,
$3\times 10^4 t_g < t <  4\times 10^4 t_g$ and
$4\times 10^4 t_g < t <  5\times 10^4 t_g$ are displayed.
These profiles show the same behavior as aM94b100 but the
inflow equilibrium radius now extends to $r > 50 r_g$.
Therefore, the shorter simulation time does not affect the establishment of the
equilibrium state but only limit the inflow equilibrium to smaller radius. In the following,
we will focus on the inner region characterized by $r \leq 30_g$
and demonstrate, where relevant, that we obtain similar results to those presented in 
\citet{Narayan2022MNRAS.511.3795N}. Considering that we are using (i) a
different numerical scheme, (ii) a different numerical resolution, and (iii)
analysing the results earlier in the simulation, this shows the solidity of these
results.

\subsection{Mass accretion rate and magnetic flux at the horizon}
\label{subsec:mass_accretion}

\begin{figure}
    \centering
    \gridline{\fig{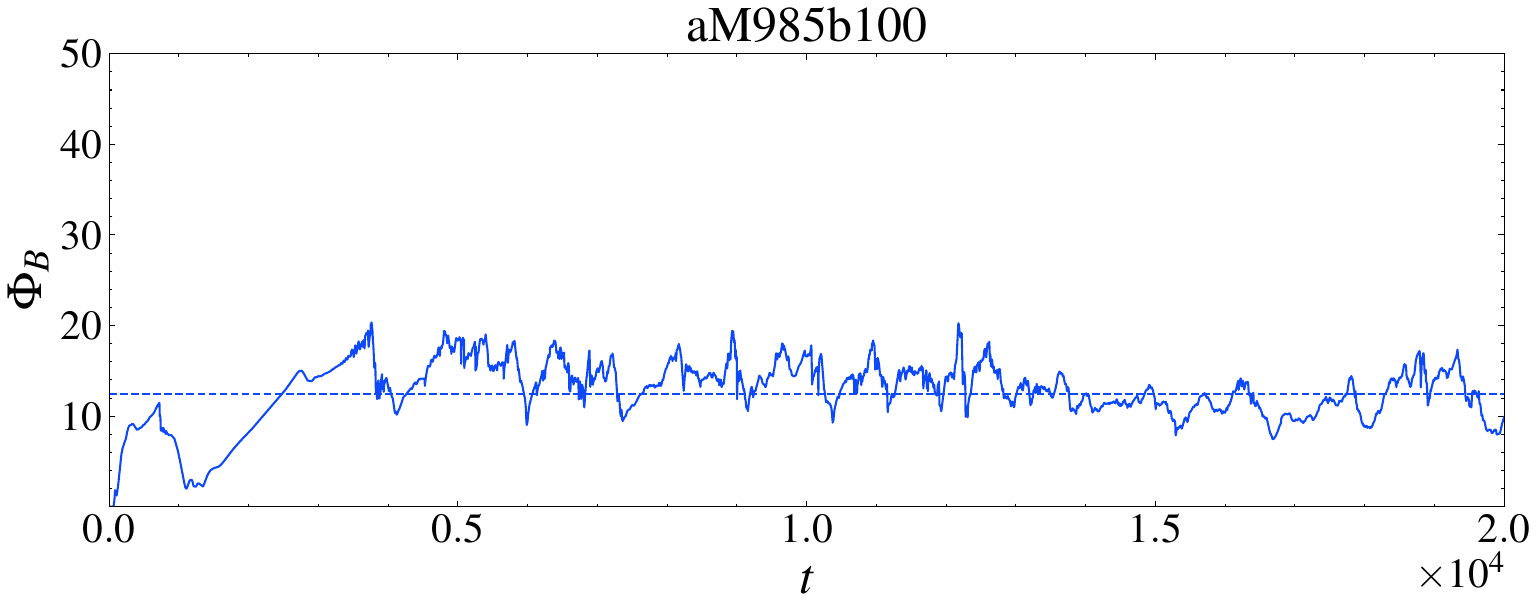}{0.48\textwidth}{}
              \fig{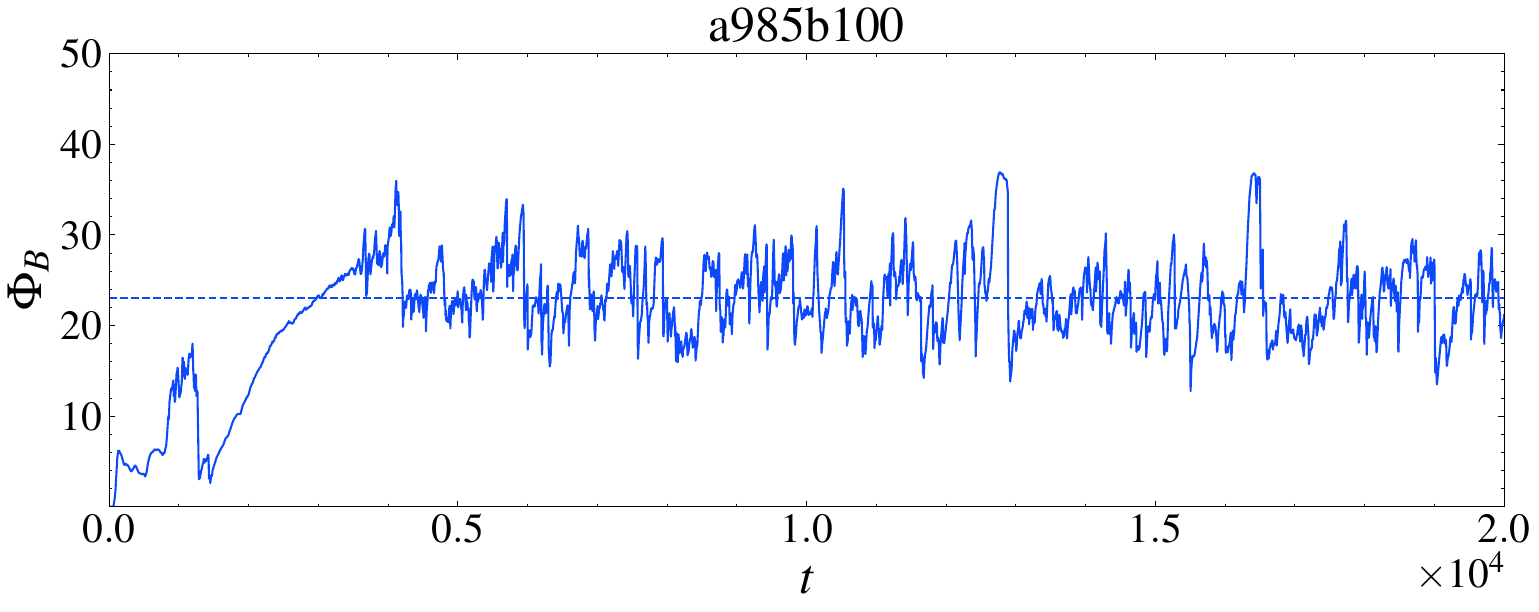}{0.48\textwidth}{}}
              \vspace*{-8mm}
    \gridline{\fig{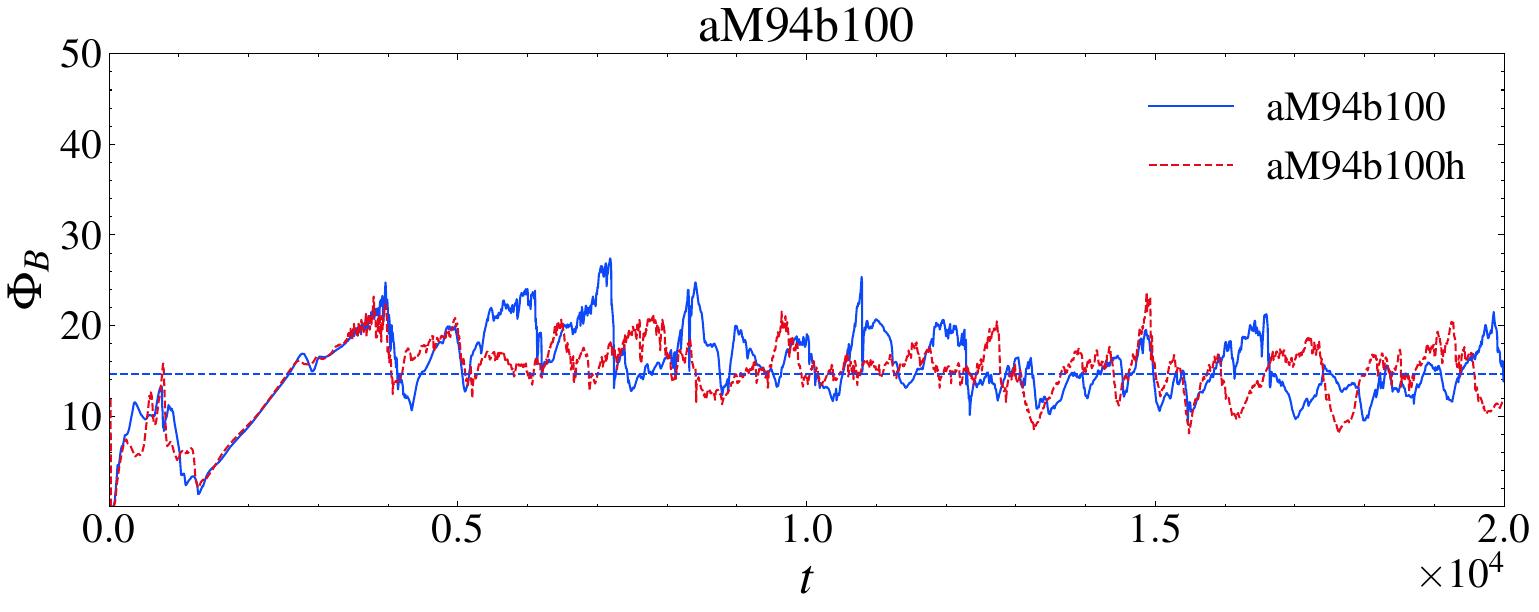}{0.48\textwidth}{}
              \fig{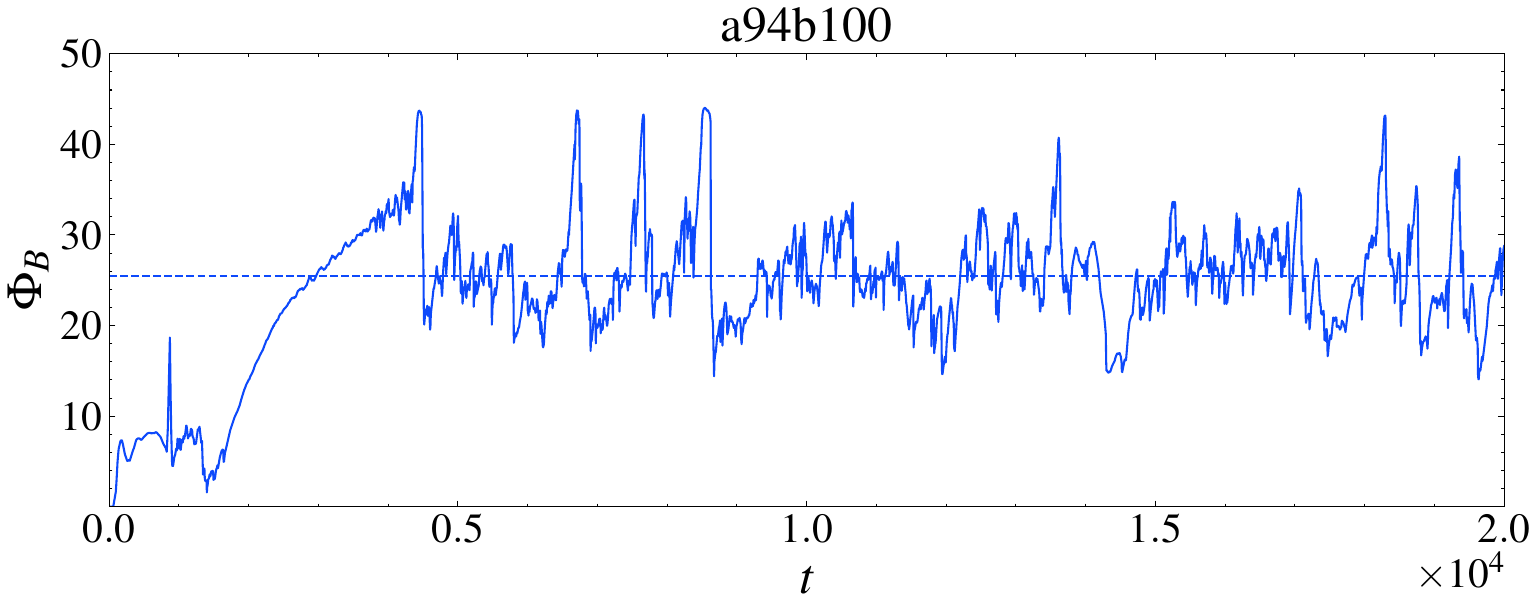}{0.48\textwidth}{}}
              \vspace*{-8mm}
    \gridline{\fig{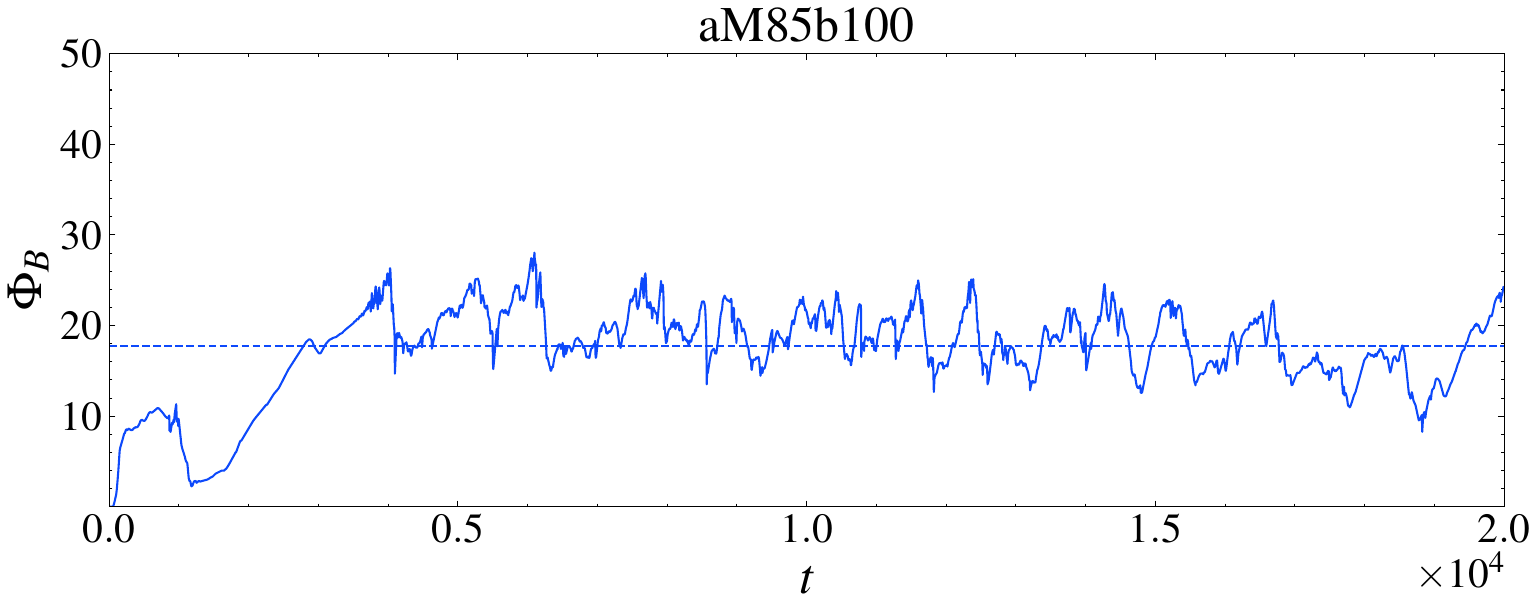}{0.48\textwidth}{}
              \fig{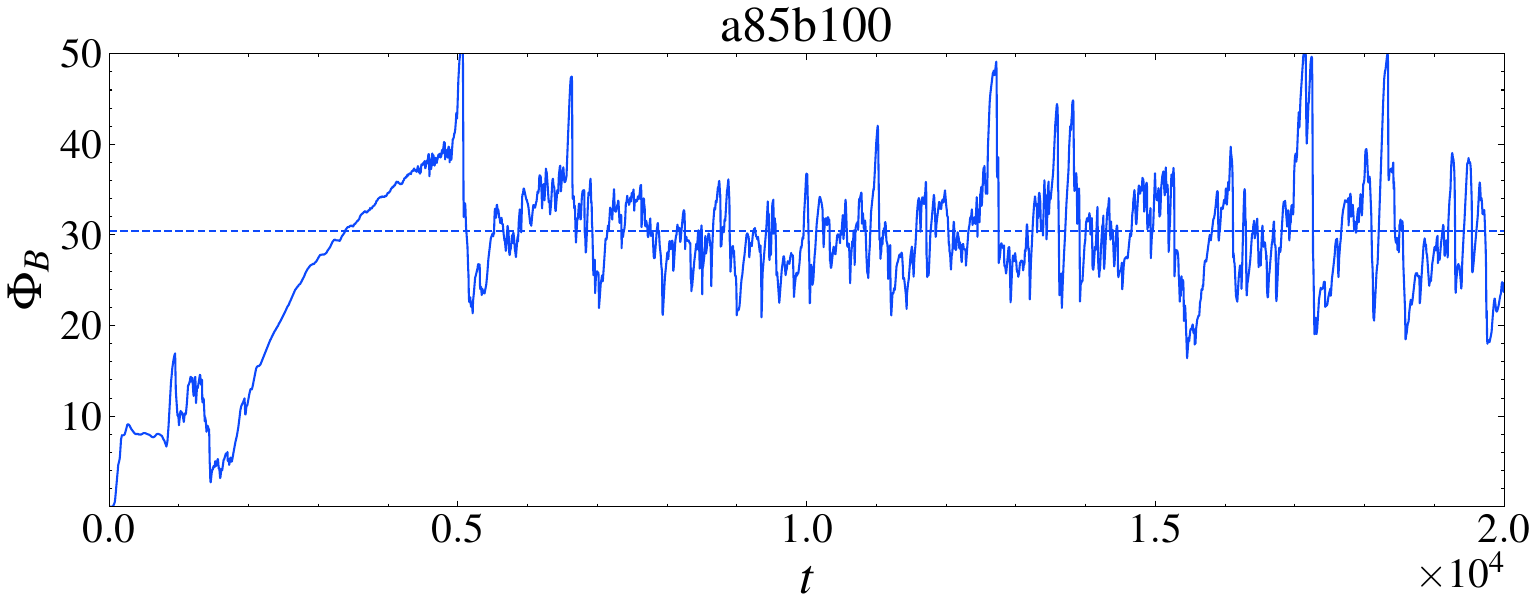}{0.48\textwidth}{}}
              \vspace*{-8mm}
    \gridline{\fig{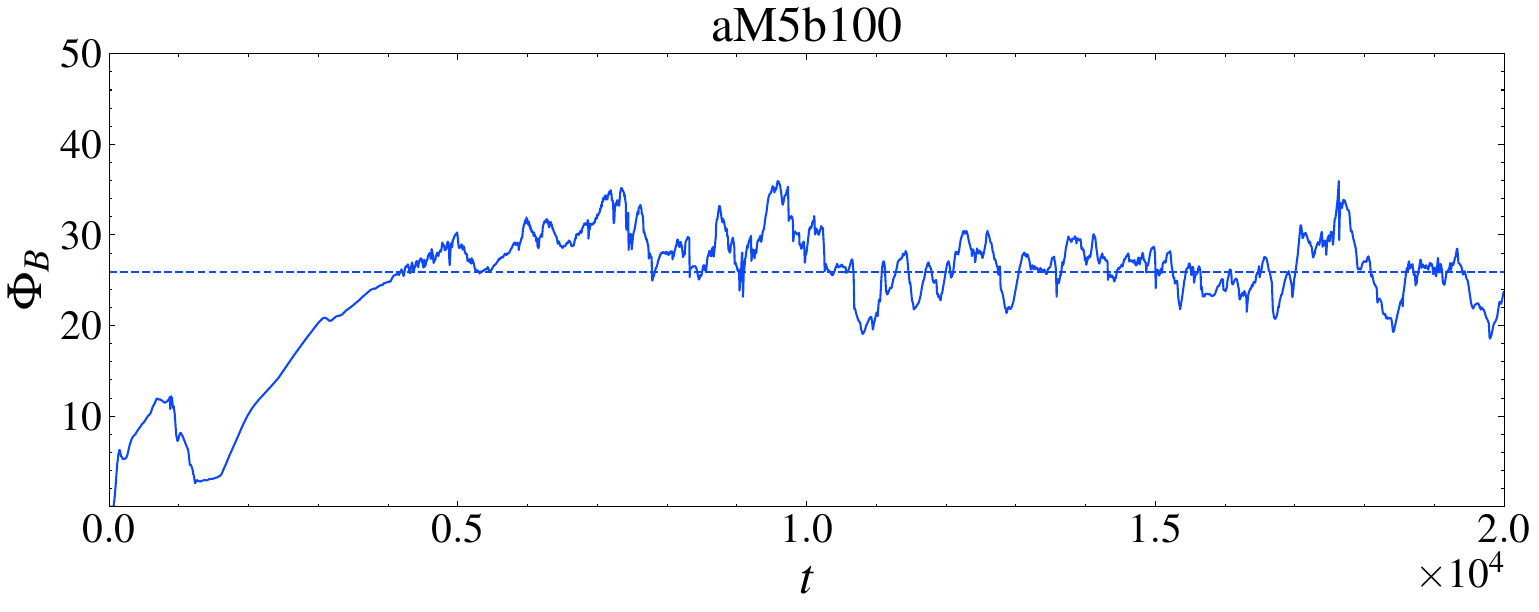}{0.48\textwidth}{}
              \fig{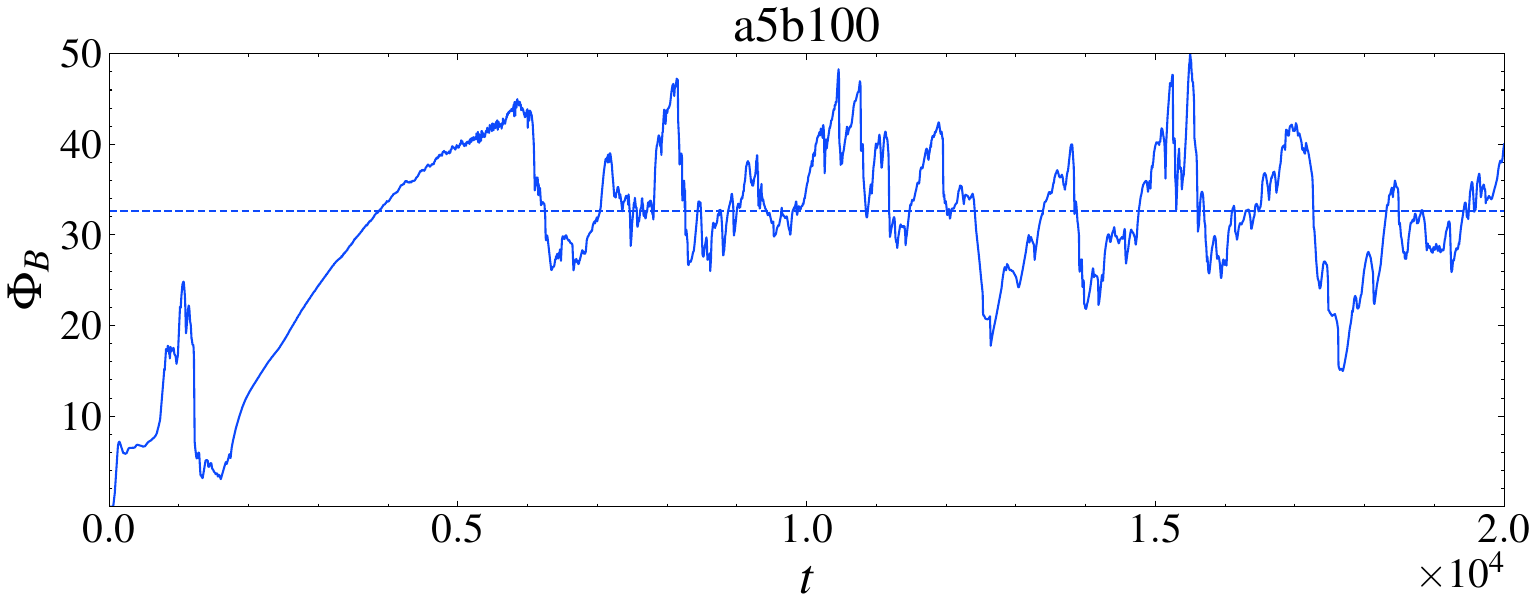}{0.48\textwidth}{}}
              \vspace*{-8mm}
    \gridline{\fig{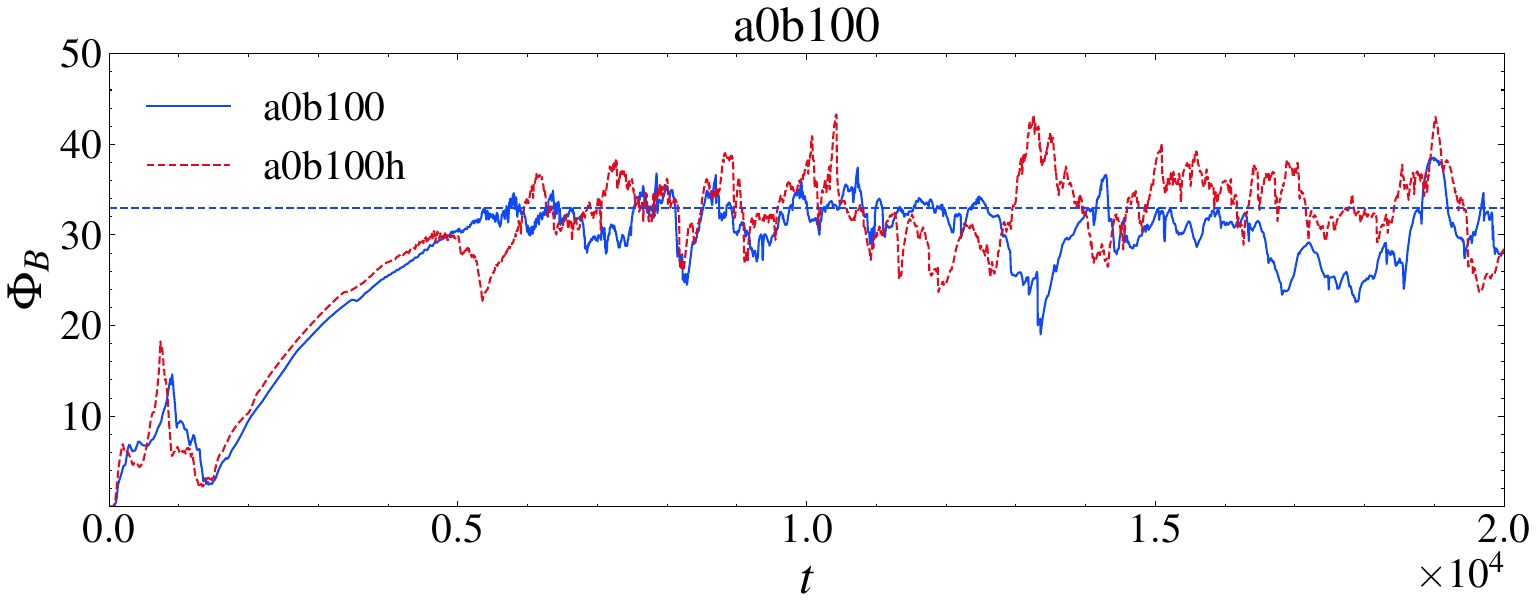}{0.48\textwidth}{}}
    \caption{The time evolution of MAD parameters $\Phi_B$. The left column
    represents retrograde disks, while the right column is for
    prograde disks. We use the horizontal lines to denote the time-averaged
    $\Phi_B$ from $10^4 t_g$ to $2 \times 10^4 t_g$, which is referred to as the
    typical MAD parameter during the MAD state.}
    \label{fig:mad-factor}
\end{figure}

The time evolution of the mass accretion rate $\dot M$ at the horizon, derived
from Equation \ref{eq:mdot}, are shown in Figure \ref{fig:dotM} for all
simulations with different spins. The same rough behaviour is observed in all cases.
After an initial quiet period, the mass accretion rate steadily increases
to reach its late time average around $t\sim 5\times 10^3 t_g$. The
left column of Figure \ref{fig:dotM} shows the time evolution of
$\dot M$ for retrograde disks, while the corresponding prograde
disks are displayed in the right column. The y-axis scale is
identical to facilitate comparison. There is a clear dependence
of the mass accretion rate on the black hole spin. First,
as displayed by the dashed line, representing the time average of the
mass accretion rate for $10^4 t_g < t < 2\times 10^4 t_g$, prograde disks have a higher mass accretion
rate than retrograde ones. The time-averaged accretion rate
from $ t = 10^4 t_g$ to $t = 2\times 10^4 t_g$ is listed in Table
\ref{tab:runs} alongside the 1$\sigma$ temporal variation.
Second, the variation of the mass accretion rate is more pronounced
in prograde disks than in retrograde disks.

\begin{figure}
    \centering
    \includegraphics[width=\linewidth]{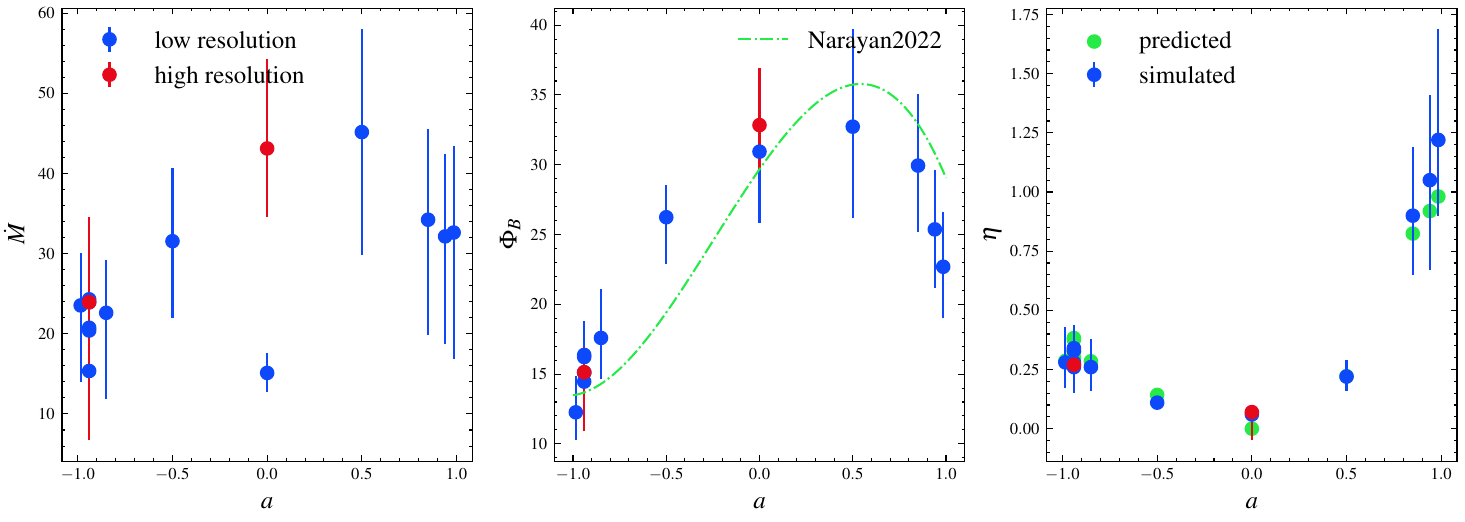} \\
    \begin{tabular}{cc}
    \includegraphics[width=0.48\textwidth]{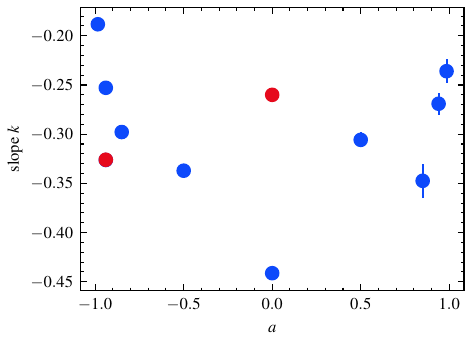}     &  
    \includegraphics[width=0.48\textwidth]{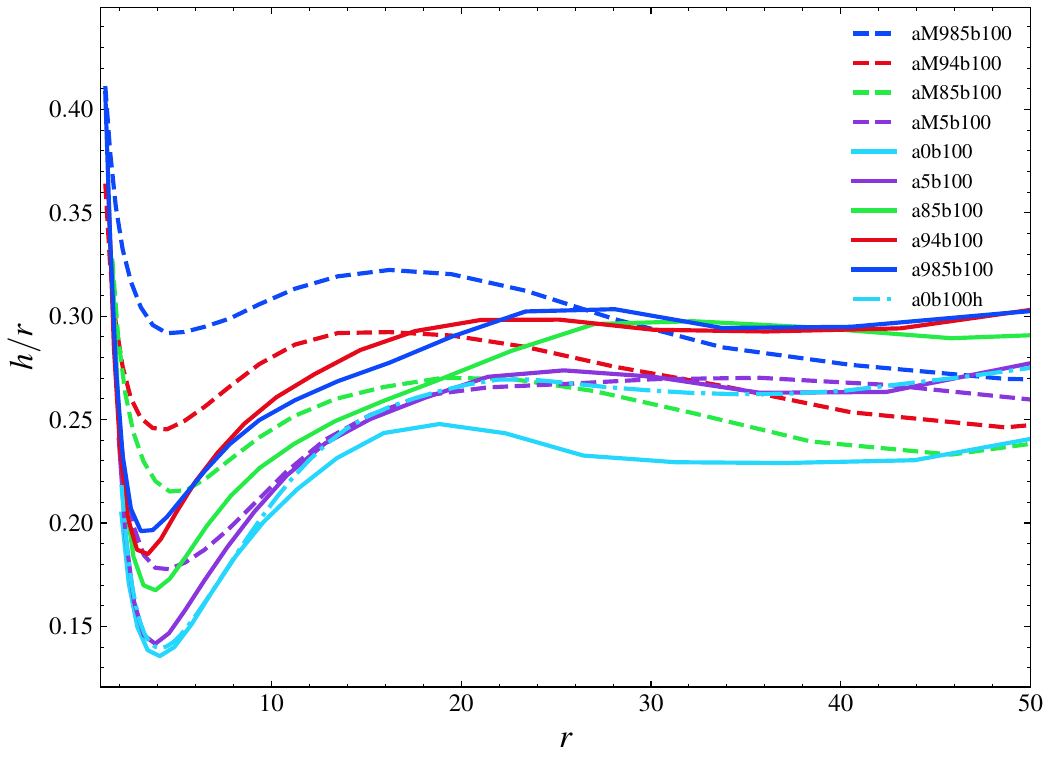}
    \end{tabular}
    \caption{Spin-dependence of several disk and jet properties. Top left panel:
    mass accretion rate $\dot{M}$ at the horizon for different spins. The blue points are the results of low resolution simulations, while the red points are the high resolution simulations. 
    Top middle panel:
    the normalized MAD parameters for different spins. Plotted on top (green dashed line) the
    fit results of \citet{Narayan2022MNRAS.511.3795N}; Their formula is divided by $\sqrt{4\pi}/2$ to be
    consistent with our definition of MAD parameter, see Section
    \ref{subsec:mass_accretion} for more details. Top right panel: the jet
    efficiency as a function of spin. The green points are
    the predicted efficiencies and the blue (red) points are the efficiencies derived
    from our simulation (red- higher resolution). The error bars represent the 1 sigma temporal variation
    of each quantity. $\eta > 1$ represents more than 100\% efficiency. Bottom left: the dependence of the slope $k \equiv (d \Phi_B/dt)/(d \dot M/dt)$ on the spin $a$. The slope $k$ decreases as the spin $a$
    increases until $a = 0$, at which point the slope $k$ begin to increase as
    $a$ continues to increases. However, $a = 0.85$ slightly deviates from this
    trend. Bottom right panel: the time averaged disk height $h/r$ given by
    Equation \eqref{eq:hr} as a function of radius.  }
    \label{fig:spin-factor}
\end{figure}

The time evolution of the MAD parameter can be described as follows.
As the accretion proceeds, the magnetic flux at the horizon accumulates until
it saturates. At this point, the MAD parameter, as well as
the magnetic flux threading the horizon and the mass accretion rate,
are regulated by eruptions of magnetic flux. These eruptions expel the magnetic field to far
distances from the black hole. In the MAD state, the  pressure
of the saturated magnetic field is balanced with the 
gas pressure \citep{BR74, NIA03}.
Although the flux eruptions cause the fluctuation
in both the magnetic flux \citep{Begelman2022MNRAS.511.2040B} and the mass accretion rate, the MAD parameters
generally remains stable around their time-averaged value.

To explore the differences in the MAD state for different spins,
we present the time evolution of the MAD
factors in Figure \ref{fig:mad-factor}, where the left column is for negative spin simulations and the right column is for positive spin simulations, similar to Figure \ref{fig:dotM}. The horizontal dashed lines 
represent the time-averaged MAD parameters from $t = 10^4 t_g$ to $t = 2\times 10^4 t_g$,
at which interval the MAD state is well established. The
time-averaged MAD parameters during this period
are also listed in Table \ref{tab:runs}. We find that the MAD parameters of prograde disks
are higher compared to the corresponding retrograde disks. Additionally, similarly to
the mass accretion rate, we find that the
temporal fluctuations of the MAD parameters are greater for prograde disk.
This is in agreement with the findings of \citet{PMY21}, who also found
that their co-rotating simulation has weaker flux expulsions than the
counter-rotating case. We also observe many more small flux
eruptions for the co-rotating case, but they are accompanied by several
strong eruptions, such as the one at $\sim t = 9 \times 10^3$ for a94b100, as seen on Figure \ref{fig:mad-factor}.

Both the mass accretion rate and the MAD parameter are strongly dependent 
on the black hole spin, $a$. This relation  is 
displayed in the top left panel of Figure \ref{fig:spin-factor}.
We find that the mass accretion rate is highest for low, positive spin, and drops for higher values of the spin, both for prograde and retrograde disks, with retrograde disks show lower values than prograde disks. For the $a=0$ spin we show two results, one with the standard resolution and one with a higher resolution, which we believe is more accurate (see discussion below).
We also show in Figure \ref{fig:spin-factor} the $1\sigma$ temporal fluctuation of the mass accretion rate as error bars. Clearly, prograde disks have larger fluctuations (represented by larger error bars on Figure \ref{fig:spin-factor}) than retrograde disks.

We present the relationship
between the MAD parameters during the MAD state and the spin of black hole
in the top middle panel of Figure \ref{fig:spin-factor}. The MAD parameter
increases as the spin increases from $a = -0.985$ to  $a = 0.5$, and then decreases
as the spin increases from $a = 0.5$ to $a = 0.985$. 
\citet{Narayan2022MNRAS.511.3795N} reported a similar trend and 
used a third order polynomial to fit the relationship between $a$
and the MAD parameter $\Phi_B$. We include their fit result
(displayed by the red dashed line) in the top middle panel
of Figure \ref{fig:spin-factor} to compare with our results.
Their definition of MAD parameters slightly differs from ours,
so we renormalised their fit formula by a factor of $\sqrt{4\pi}/2$
to account for this discrepancy.
Our results are in good agreement with theirs, which enhances credibility of the results
considering that we use a different
code, different resolutions, and shorter simulation duration.

Our simulations have a somewhat lower resolution (by a factor of $\approx 4$) compared to
\citet{Narayan2022MNRAS.511.3795N,Chatterjee2022ApJ...941...30C}. To address this limitation,
we conducted two simulations, aM94b100h and a0b100h, which have identical initial
conditions as aM94b100 and a0b100, respectively, but differ by having a higher resolution.
We show the mass accretion rates and the MAD parameters
of these two simulations in Figures \ref{fig:dotM} and
\ref{fig:mad-factor}, alongside their counterpart with lower resolution.
In the case of aM94b100 and aM94b100h, there is no significant difference in
terms of mass accretion rates and MAD parameters. However, for a0b100 and a0b100h,
the mass accretion rates are different by a factor of nearly three.
According to \citet{White2019ApJ...874..168W}, this indicates a too low resolution for our simulation with spin $a = 0$ (but not for our simulations with rotating black-holes as is suggested by aM94b100h). They demonstrate that resolving the mass accretion rate demands a better resolution than to resolve the MAD parameter, as is also shown here. Indeed,
the saturation values of the MAD parameters are nearly identical. This indicates that the
properties of the MAD state should be similar.

\begin{figure}
    \centering
    \includegraphics{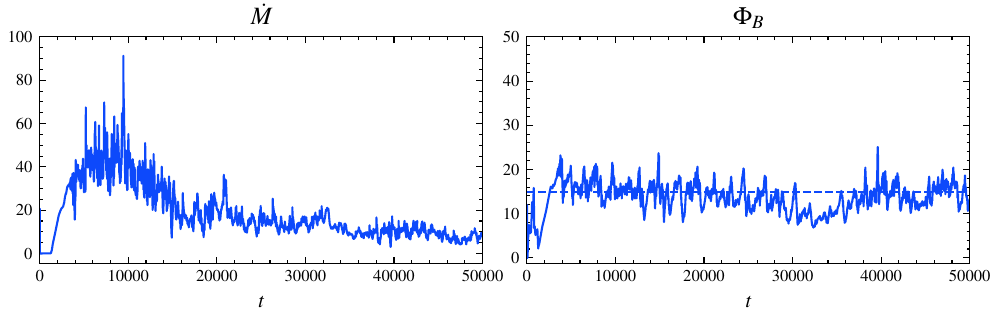}
    \caption{The long time evolution of the mass accretion rate and MAD parameters for simulation aM94b100h. The MAD parameter remains steady with time although the mass accretion drops.}
    \label{fig:aM94b100h}
\end{figure}

Most of our simulations end at $t = 2  \times 10^4 t_g$. As discussed
in sub-section \ref{subsec:flux-r}, we extend aM94b100h to
$t = 5\times 10^4 t_g$. The time evolution of the mass accretion rate
$\dot{M}$ and of the MAD parameter $\Phi_B$ is shown in Figure
\ref{fig:aM94b100h}. The MAD parameter remains relatively stable, oscillating
around its average value, once the MAD state is established. However,
the mass accretion rate $\dot{M}$ gradually decreases until
$t   = 5\times 10^4  t_g$, which is caused by the spread of the
disk as the simulation advances.

\subsection{Characteristic dynamical timescale}

\label{subsec:time_characteristics}

\begin{figure}
    \centering
    \gridline{\fig{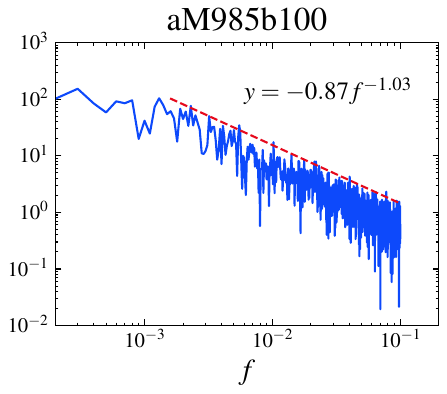}{0.31\textwidth}{}
                \fig{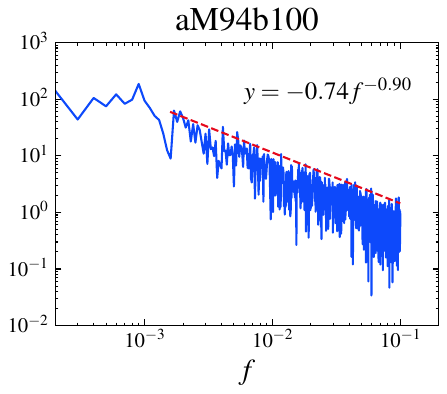}{0.31\textwidth}{}
                \fig{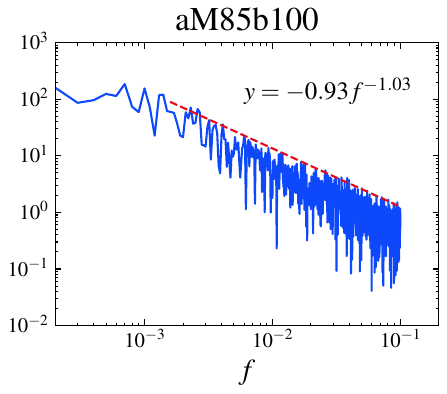}{0.31\textwidth}{}}  
    \gridline{\fig{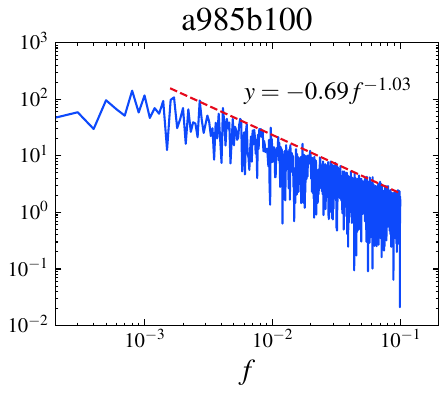}{0.31\textwidth}{}
                \fig{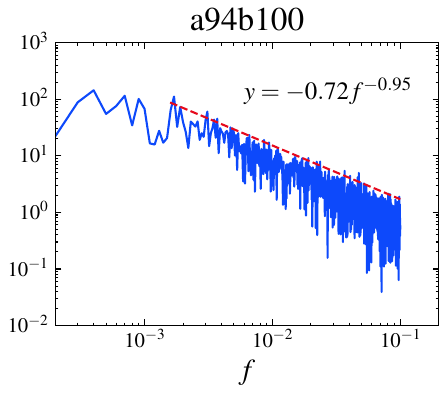}{0.31\textwidth}{}
                \fig{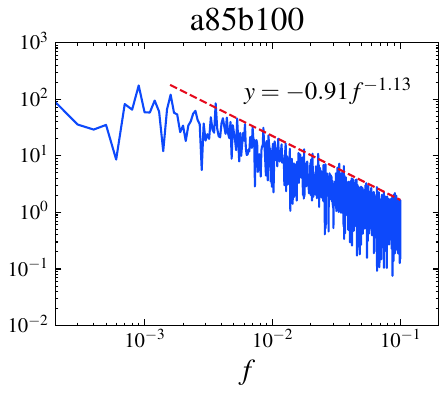}{0.31\textwidth}{}}
    \gridline{\fig{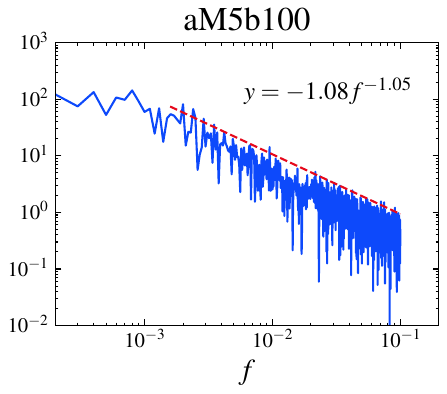}{0.31\textwidth}{} 
                \fig{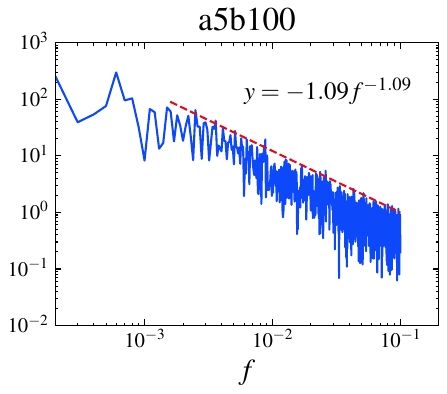}{0.31\textwidth}{}
                \fig{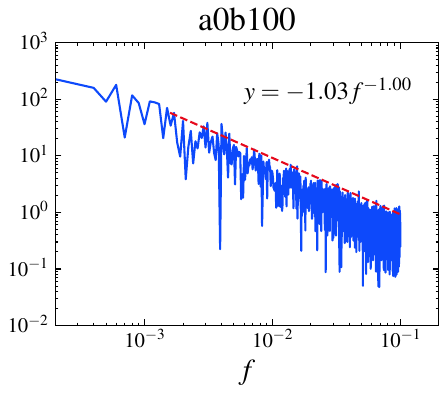}{0.31\textwidth}{}}
    \caption{Power spectrum of the MAD parameters in the time interval $10^{4} t_g < t < 2 \times 10^4 t_g$ ($1.5\times 10^4 t_g$ to $2.5 \times 10^4 t_g$ for aM94b800). In all simulations, the power is concentrated in the low frequency region, which corresponds to  periods about $500 - 2000 t_g$. In the high frequency region, the power spectrum decreases following $\sim f^{-1}$. We use a red dashed line to show this evolution.   }
    \label{fig:fft}
\end{figure}

In the MAD state,  magnetic flux eruptions play a major role in the dynamic of
the accretion: the strong magnetic pressure pushes the gas out, which leads to the fluctuations
of the accretion rate and of the magnetic flux. Moreover, these magnetic
flux eruptions have been proposed as the origin of the Sgr A$^*$ flares observed
in infrared by the GRAVITY collaboration \citep{Gra18, Gra20}. 
In particular,
\citet{DTJ20} numerically estimated the recurrence time of the flare to be between
$10^3$ and $10^4$ $r_g / c$, corresponding to 5 to 50 hours for Sgr A$^*$. The
maximum of the intensity is correlated with the sharp drop of the MAD parameter
at the onset of each eruption \citep{DTJ20}, lasting around a hundred $r_g / c$
\citep{RLC22}. The dynamic of a flux tube created in each eruptions was studied
in details by \citet{PMY21}. They found that the motion of the low-density/high
magnetization region is at first strongly radial because of the magnetic tension,
and then tends to circularize when the field is nearly vertical. The typical
life-time of these magnetic flux tubes was found numerically to be around 2 orbits,
depending on their size and magnetic energy.

\begin{figure}
    \centering
    \gridline{\fig{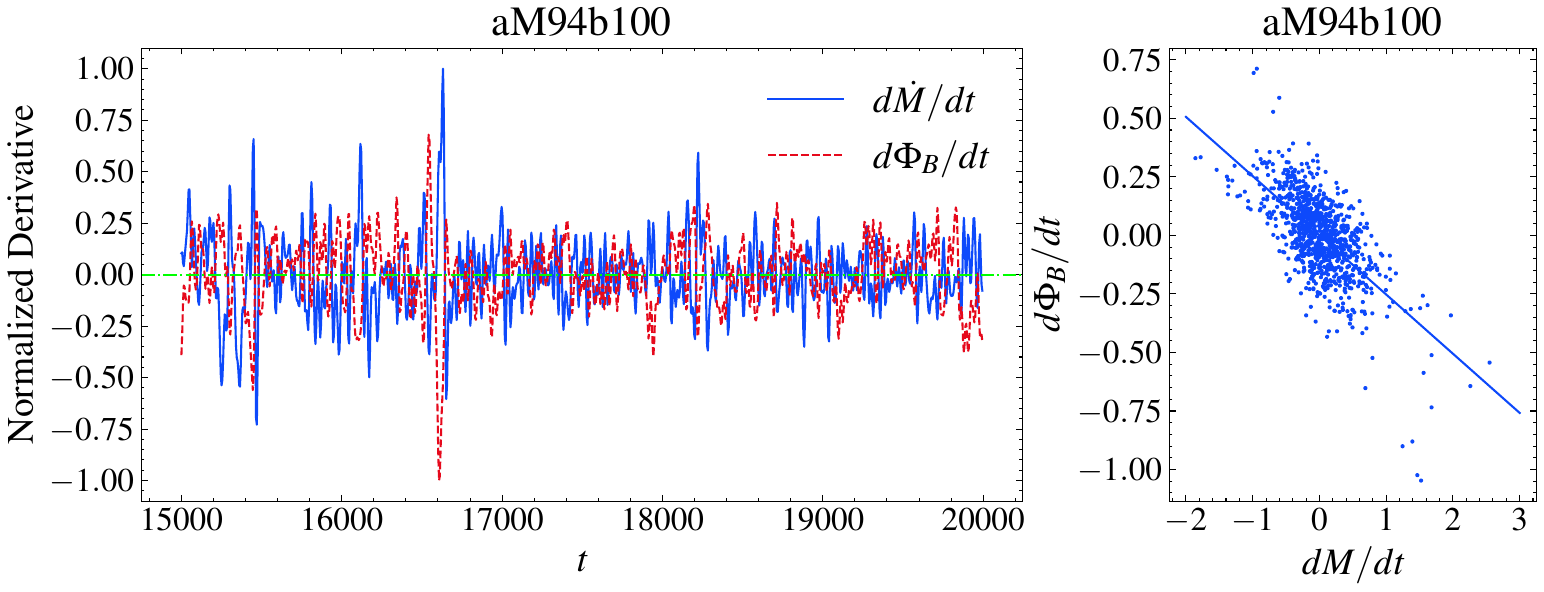}{0.9\textwidth}{}}
    \gridline{\fig{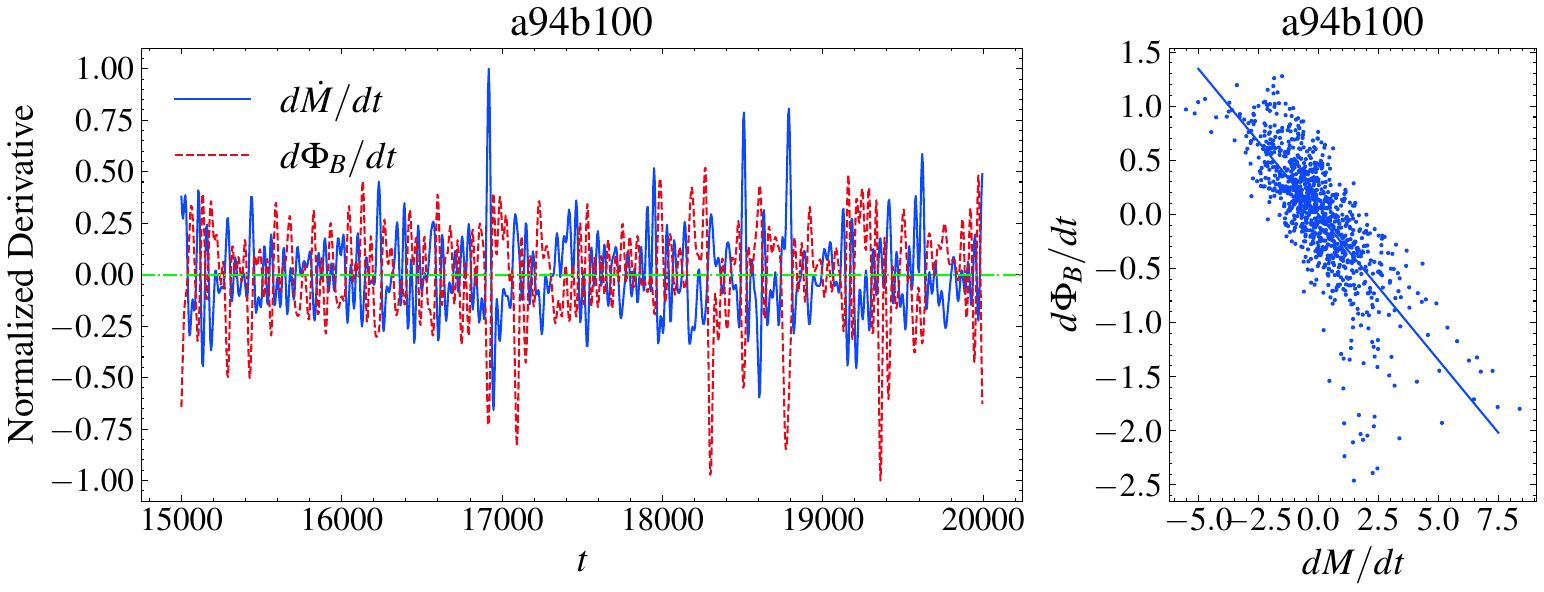}{0.9\textwidth}{}}
    \caption{Left: time evolution of the time derivatives of the mass
    accretion rate and of the MAD parameter. We show the results for aM94b100
    and a94b100, but other simulations also exhibit similar behaviors.
    These two quantities, $d\dot M/dt$ and
    $d\Phi_B / dt$ have opposite signs throughout the simulations.
    There is a clear anti-correlation between these two derivatives,
    shown in the right panels. }
    \label{fig:diff}
\end{figure}

We use the discrete Fourier transform to study the duty cycle of the MAD parameters.
The analysis is performed considering data from
$t = 10^4 t_g$ to $t = 2\times 10^4 t_g$ for which, as a preparation step,
the time series are de-trended and their average
removed, so that the fundamental Fourier coefficient is null. We show the
power spectrum of the MAD parameter $\Phi_B$ for our simulations in Figure \ref{fig:fft}.
We focus on the power spectrum from $f \simeq 4\times 10^{-3}$ to
$f \simeq 2\times 10^{-4}$, which corresponds to the period from
$T = 250 t_g$ to $T = 5000 t_g$. All our simulations show the same trend:
after a few dominant peaks, the power spectrum decays proportionally
to $f^{-1}$, characteristic of pink noise, which are shown as red dashed
lines in Figure \ref{fig:fft}. \citet{JJ22} performed a similar study, not on the MAD parameter, but rather
on the mass accretion rate. Furthermore, their
disks are smaller than ours, with $r_{\rm in} = 6$ or 12 and $r_{\rm max} = 12$ or
25 $r_g$. They found a power-spectrum of the mass accretion rate, which is
well approximated with a power-law of index $\sim 1.5$. They further report
a dependence between the index and the black hole spin. Our analysis, on the other hand,
does not show such a spin dependence on the index for the MAD parameter.

For some simulations, such as aM94b100 and a5b100,
the power spectrum shows clear peaks in the frequency window preceding
the onset of the noise. The corresponding periods are $t = 1,111 t_g $ for
aM94b100 and $ t = 1,666 t_g$ for a5b100. For the other simulations,
no dominant time scale can be unambiguously identified. Previous identifications of 
a cyclic behavior of mass accretion rate and magnetic flux were published by, e.g., \citet{CBL21}. By analysing the data of their 2D GRMHD simulations, they found a period
around $t \simeq 500 r_g / c$ \citep[see
also, e.g.,][]{Igu08,DMF23}. However, as noted by \citet{CBL21}, this pattern
is an artefact due to the nature of 2D simulations, preventing non-asymmetric
instabilities and phenomena, which are inherently responsible for the
continuous accretion in 3D, to develop.

\subsection{Relation between the MAD parameter and the mass accretion rate}
\label{subsec:phi_dotM}

The strong magnetic field during the MAD state pushes out the gas and stops, or
at least regulates the dynamics of
the in-falling of matter. Therefore, we naively expect that
an increase of the magnetic flux should result in
the decrease of the mass accretion rate, in other words, that they
are anti-correlated. An anti-correlation is also expected if accretion
proceeds by interchange instability, as matter replaces a highly magnetized
region closer to the black hole resulting in a drop in the MAD parameter
\citep{PMY21}. However, \citet{PMY21} did not find a correlation or an anti-correlation between
$\dot M $ and $\Phi_B$. We expand their results for all black hole
spins: none of our simulations show a correlation or an anti-correlation between $\dot M$ and $\Phi_B$. 

We further studied the relation between the time-derivatives
of the mass accretion rate $\dot M$ and of the MAD parameter $\Phi_B$. We first remove
their average and de-trend the time series, then we use Fourier transformation
to filter out the high (and weak) frequency components. We checked that this
procedure does not change the results presented below. Instead, it allows
for a better visualisation of the result. The processed time-derivatives are
displayed in the left column of Figure \ref{fig:diff} for aM94b100 and a94b100.
The blue line is the normalized $d\dot{M}/dt$, and the red dashed line
represents the normalized $d\Phi_B/dt$. Here, by ''normalise'' we mean that
the two time-derivatives are stretched so that the maximum of their absolute
value is 1. We show the resulting time series for aM94b100 and a94b100 as examples, but 
all other simulations exhibit a similar behavior. To the (positive)
heights of $d\dot{M}/dt$ correspond the (negative) lows of $d\Phi_B/dt$ and vice-versa.

The time-derivatives of $\dot M$ and $\Phi_B$ are clearly anti-correlated. In
the right column of Figure \ref{fig:diff}, we show this anti-correlation. 
This means that when the mass accretion rate increases, the flux threading
the horizon decreases. This can be understood as follows. Focusing in the
region close to the equator, the horizon surface is separated into two regions:
the accretion funnel from which the matter falls inward and the highly magnetized,
low density regions of the magnetic flux eruption. As the MAD parameter increases,
the magnetic pressure outside of the accretion funnel increases, and it gets compressed
resulting in a lower accretion rate. Similarly, as the magnetic flux eruption
develops, the magnetic flux at the horizon drops, thereby reducing the magnetic
pressure and allowing for a larger accretion funnel and accretion rate. The latter is
also enhanced by the interaction between the low density magnetic flux tube and
the inner region of the turbulent disk at  $r \sim 10-15 ~r_g$. Indeed, the magnetic
flux tube velocity is sub-keplerian, reducing the velocity of the matter in the
disk, which then falls ''radially'' onto the black hole.

We further perform a linear fit to this anti-correlation, and show the dependence of the slope $k\equiv (d\Phi_B/dt) / (d\dot{M}/dt)$ on the
spin $a$ in the bottom left panel of Figure \ref{fig:spin-factor}. We find that the non-spinning
black hole has the steepest slope $k$, which means that smaller variations
in mass accretion rates correspond to larger variation in the MAD parameter,
compared to other black hole spins. A smaller slope $k$ is in general a characteristic of large absolute value of spin $a$, apart for $a = 0.5$
or for $a = 0.85$, which would deserve further investigations. Finally, we point out
that the filtering performed in preparation to the data changes the
value of $k$ by a factor of up to about two, but the trend remains the same if
one accounts for this re-scaling. We further note that there seem to be a strong dependence of $k$ on the resolution, as is shown in the bottom left panel of Figure \ref{fig:spin-factor} by the two red dots, which are the slopes obtained from simulations aM94b100 and a0b100.

\subsection{Jet efficiency}

The jet efficiency, given by Equation \eqref{eq:jet-efficiency} depends on the
magnetic flux through the horizon normalised by the accretion rate $\dot M$.
In agreement with many previous publications
\citep[see, e.g.,][]{Tchekhovskoy2011MNRAS.418L..79T, Tchekhovskoy2012MNRAS.423L..55T,
MTB12, Narayan2022MNRAS.511.3795N}, we observe that the jet efficiency of
positive spin black hole with $a > 0.5$ is on average larger than $100$ \%,
meaning that the black hole is actually loosing energy: the jet is powered
by the Blandford and Znajek mechanism \citep{BZ77}.

We use the fitting formula proposed by \citet{Tchekhovskoy2010ApJ...711...50T}
and latter used by \citet{Narayan2022MNRAS.511.3795N} to interpret the
efficiency, namely
\begin{equation}
    \eta = \frac{\kappa}{4\pi}\Omega_H^2\Phi_B^2\left(1 + 1.38 \Omega_H^2 - 9.2 \Omega_H^4\right).
    \label{eq:Tchekhovskoy_jet_efficiency}
\end{equation}
Here $\kappa = 0.08 \sqrt{\pi}$ is a numerical constant and $\Omega_H = ac/r_H$
is the angular velocity at the horizon. This value of $\kappa$ is chosen such that 
$\sqrt{\pi}$ accounts for the difference in the definition of $\Phi_B$ used here and in \citet{Tchekhovskoy2010ApJ...711...50T}, and the
numerical factor 0.08 is chosen to match the normalisation of our numerical data.
The predicted efficiency, \textit{i.e.} using the measured value of the MAD parameter
$\Phi_B$ together with Equation \eqref{eq:Tchekhovskoy_jet_efficiency}, and the
simulated efficiency, namely directly measured from our simulations as defined
by Equation \eqref{eq:jet-efficiency} are shown in the right panel of Figure
\ref{fig:spin-factor} as a function spin. It is clear that they are consistent
with each other. \citet{Narayan2022MNRAS.511.3795N} reported the same behavior
but used a different value of $\kappa = 0.05$ differing from our normalisation
by a factor smaller than 2. This discrepancy could be due to the time at which
the measurements are performed. Indeed, when averaging the results of 
simulation aM94b100h at late times, $10^4 t_g < t < 5\times 10^4 t_g$, we find a lower numerical factor of $\sim 0.06$.

\subsection{Effects of the initial magnetic field strength}
\label{subsec:initial_beta0}

\begin{figure}
    \centering
    \includegraphics[width=\textwidth]{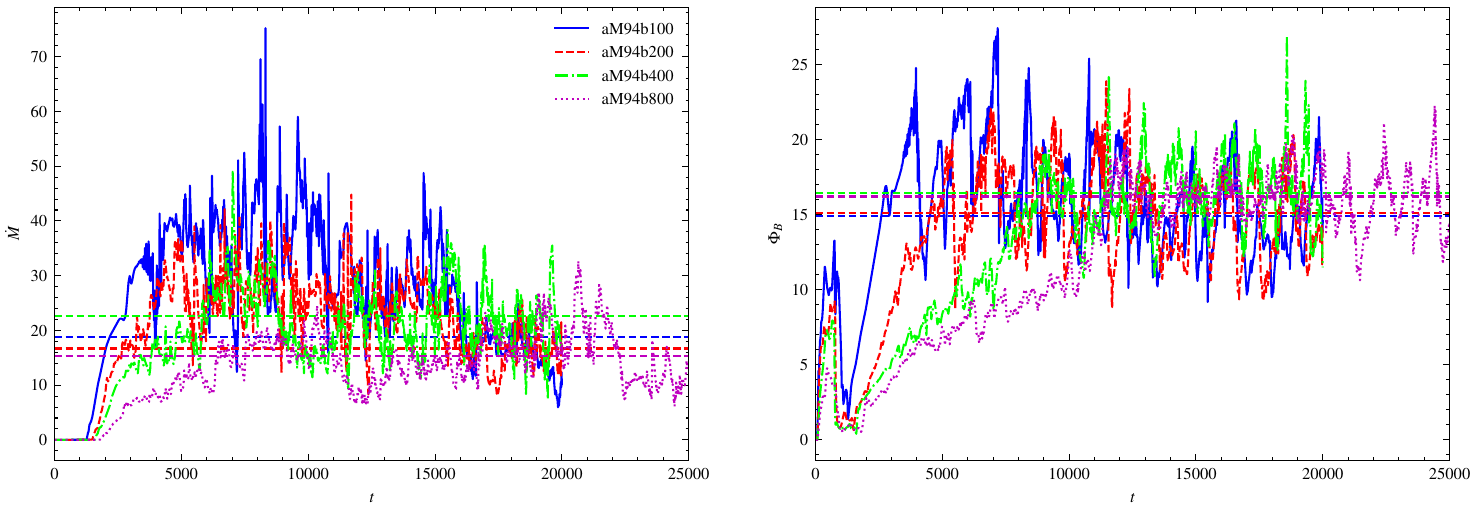}
    \caption{Mass accretion rates and MAD parameters of aM94b100, aM94b200, aM94b400, and aM94b800, which are simulations with the same initial conditions but with a different initial magnetic field strengths. In the left panel, we show the mass accretion rates. The larger the initial magnetization (the smaller $\beta_0$), the faster the mass accretion rate increases. After $t = 1.5 \times 10^4 t_g$, the mass accretion rates agree across all simulations. In the right panel, the MAD parameters of these simulations are shown to be consistent with each other after $t = 1.5 \times 10^4 t_g$. However, the time required to reach the MAD state is clearly different.}
    \label{fig:dotM-MAD-beta}
\end{figure}

\begin{figure}
    \centering
    \includegraphics{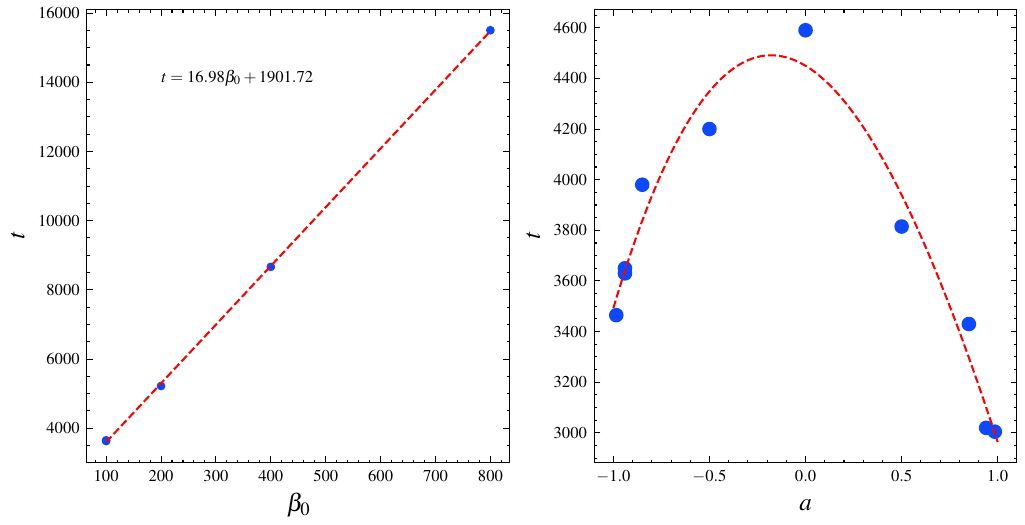}
    \caption{Left Panel: The relation between the coasting time $t_{\mathrm{MAD}}$ to reach the MAD state and the initial magnetic field. We show the results for aM94b100, aM94b200, aM94b100h, aM94b400, and aM94b800 in blue circles. The coasting time linearly depends on the initial magnetic field. We use a linear function to fit it, and the best fit result is shown by the red dashed line. Right Panel: The $t_{\mathrm{MAD}}$ as a function of the  value of the spin $a$. We use a third order polynomial function to fit this relation, and the best fitted result is shown by the dashed red line. }
    \label{fig:costt}
\end{figure}

Since the magnetic field plays a critical role in the accretion process and
its regulation, we wish to assess the solidity of our result with respect to
the initial magnetic field strength. To this end, we also
performed simulations with different values of $\beta_0$ for $a = -0.94$. In
Table \ref{tab:runs}, we list three additional simulations
with initial magnetic field strength $\beta_0 \in {100,~200,~400,~800}$. The
simulation aM94b100 has the strongest initial magnetic field strength, while
aM94b800 has the weakest. 

In Figure \ref{fig:dotM-MAD-beta}, we present the  mass accretion rates and
MAD parameters for those additional simulations with
different initial magnetic field $\beta_0$. The left panel shows that at
the beginning of the accretion process, the mass accretion rate $\dot M$
increases with the initial magnetic field strength, which can be attributed
to the fact that the transfer of angular momentum is initially driven by MRI, which development depends on the strength of the magnetic field. Therefore,
the larger the initial magnetic field, 
the higher the mass accretion rate. However, at late
time $t > 15,000 t_g$, we see that the mass accretion rates achieve similar values 
across all simulations.

In the right panel of Figure \ref{fig:dotM-MAD-beta}, it is seen that
the MAD parameters also achieve similar values across all simulations during
the MAD state, here reached after $t > 15,000 t_g$ 
specifically for aM94b800. This suggests that the
MAD state is independent of the initial magnetic field strength. However,
the time required to reach the MAD state varies. To investigate the
dependence of the required time on the initial $\beta_0$, we use the
following criteria to define $t_{\mathrm{MAD}}$, which is the time
required to reach the MAD state
\begin{enumerate}
    \item The mean MAD parameter averaged from $t = t_{\mathrm{MAD}}$ to $t = t_{\mathrm{MAD}} + 1000~t_g$ should be larger than the $1\sigma$ lower limit of the final MAD parameter.
    \item The MAD parameter in the MAD state should be highly variable. Therefore, we require the derivative of the MAD parameter to change sign several times between $t = t_{\mathrm{MAD}}$ to $t = t_{\mathrm{MAD}} + 1000~t_g$.
\end{enumerate}
According to these two criteria, the time it takes to
reach the MAD state $t_{\mathrm{MAD}}$ for
aM94b100, aM94b100h, aM94b200, aM94b400 and aM94b800 is 3650 $t_g$,
3630 $t_g$, 5220 $t_g$, 8665 $t_g$ and 15505 $t_g$, respectively.
In the left panel of Figure \ref{fig:costt}, we display the time required
to reach the MAD state as a function of the initial magnetic field
strength $\beta_0$. We find that $t_{\mathrm{MAD}}$
increases linearly with $\beta_0$, namely it increases as the initial
magnetic field strength weakens. The best fitting result is a linear function,
$t_{\rm MAD} = 17\beta_0 +
1.9\times 10^3 $. This result suggests that the accumulation of the magnetic
flux linearly depends on the initial magnetic field strength. %\deletedGQ{The MRI amplifies the magnetic field, and the amplified magnetic field leads to the MAD state. This amplification process is proportional to the time.}
We also study the dependence of $t_\mathrm{MAD}$ 
on the spin for the other simulations with the same initial $\beta_0$.
The right panel of Figure \ref{fig:costt} displays the
dependence of $t_\mathrm{MAD}$ on the spin $a$. As the spin increases from
$a = -1$ to $a = 0$, $t_{\rm MAD}$ increases. It then decreases as $a$ increases
from 0 to 1. This dependence is fitted with a third order
polynomial which results in $t_{\rm MAD} = 1.9\times 10^2 a^3 -1.2\times 10^3 a^2 - 
4.6\times 10^2 a + 4.4 \times 10^3$.

\subsection{Evolved disk and jet structures}
\label{subsec:disk-jet}

The strong magnetic field at the center of the accretion system
shapes the disk and launches the bipolar jet at low radii. Therefore, the
disk structure  is an important characteristics
of the MAD state. In the bottom right panel of Figure \ref{fig:spin-factor},
we show the time-averaged disk height $h/r$, which is averaged from
$t = 10^4 t_g$ to $t = 2 \times 10^4 t_g$. 
We find that retrograde disks are thicker than prograde disks. In the inner
region ($5 r_g < r < 15 r_g$), the disk height increases as the absolute value
of the spin increases, which is consistent with the results of 
\citet{Narayan2022MNRAS.511.3795N}. However, contrary to the findings in this
aforementioned paper, we find that the non-spinning black hole has the thinnest
disks of all, while they argued that the
disk of the non-spinning black hole is thicker than that of prograde disks. A possible explanation to this discrepancy is the fact that our analysis is carried at earlier times.

\begin{figure}
    \centering
    \begin{tabular}{c}
    \includegraphics[width=0.75\linewidth]{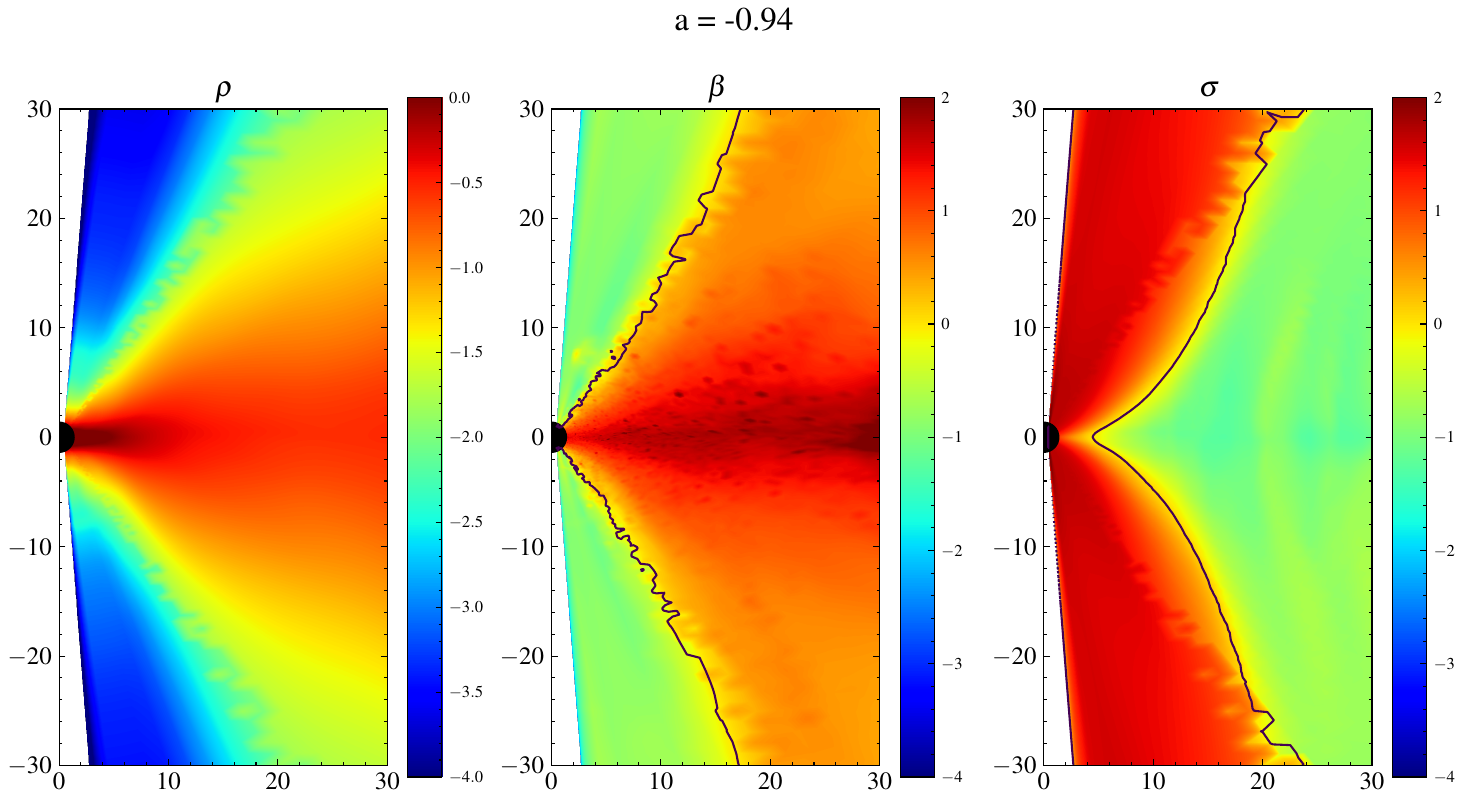} \\
    \includegraphics[width=0.75\linewidth]{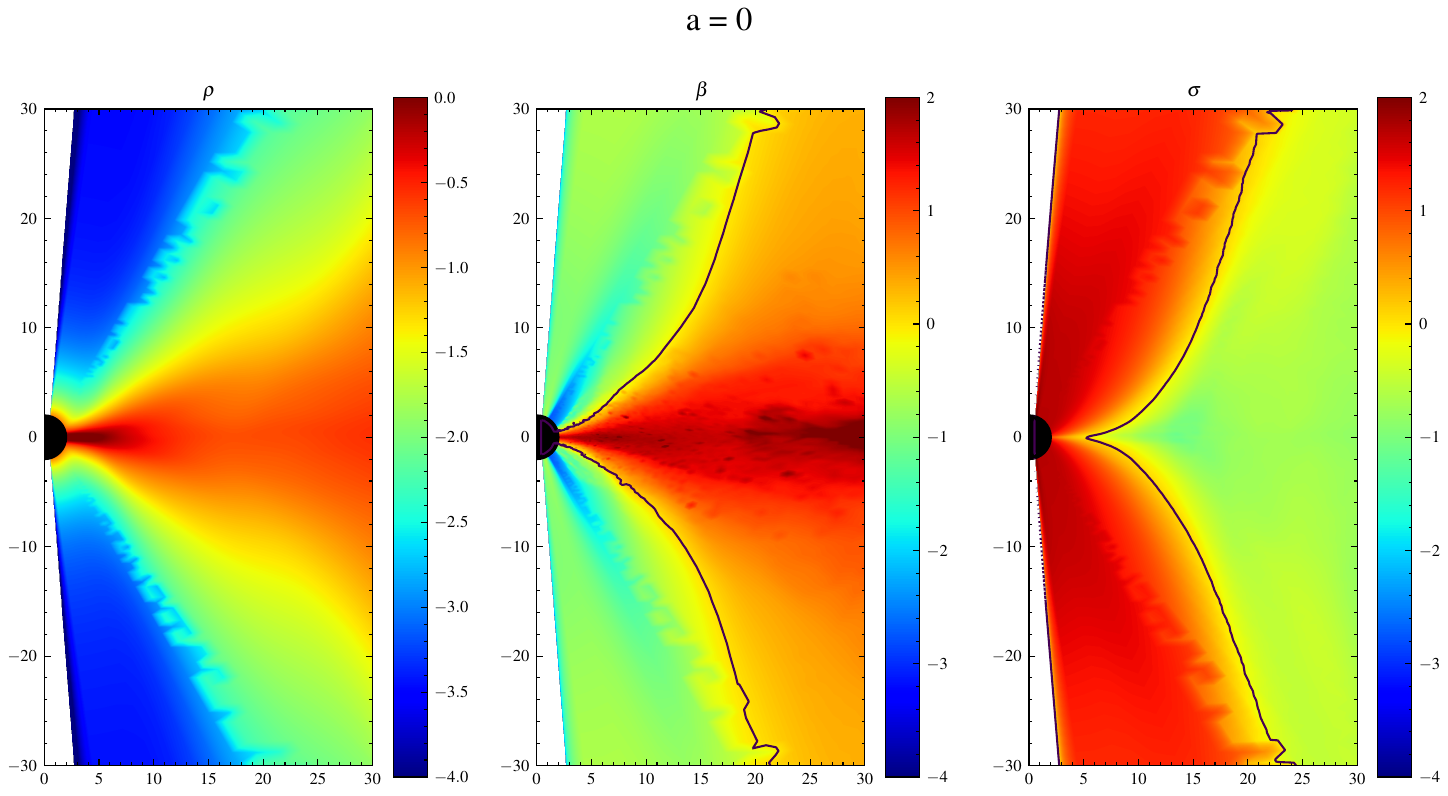} \\ 
    \includegraphics[width=0.75\linewidth]{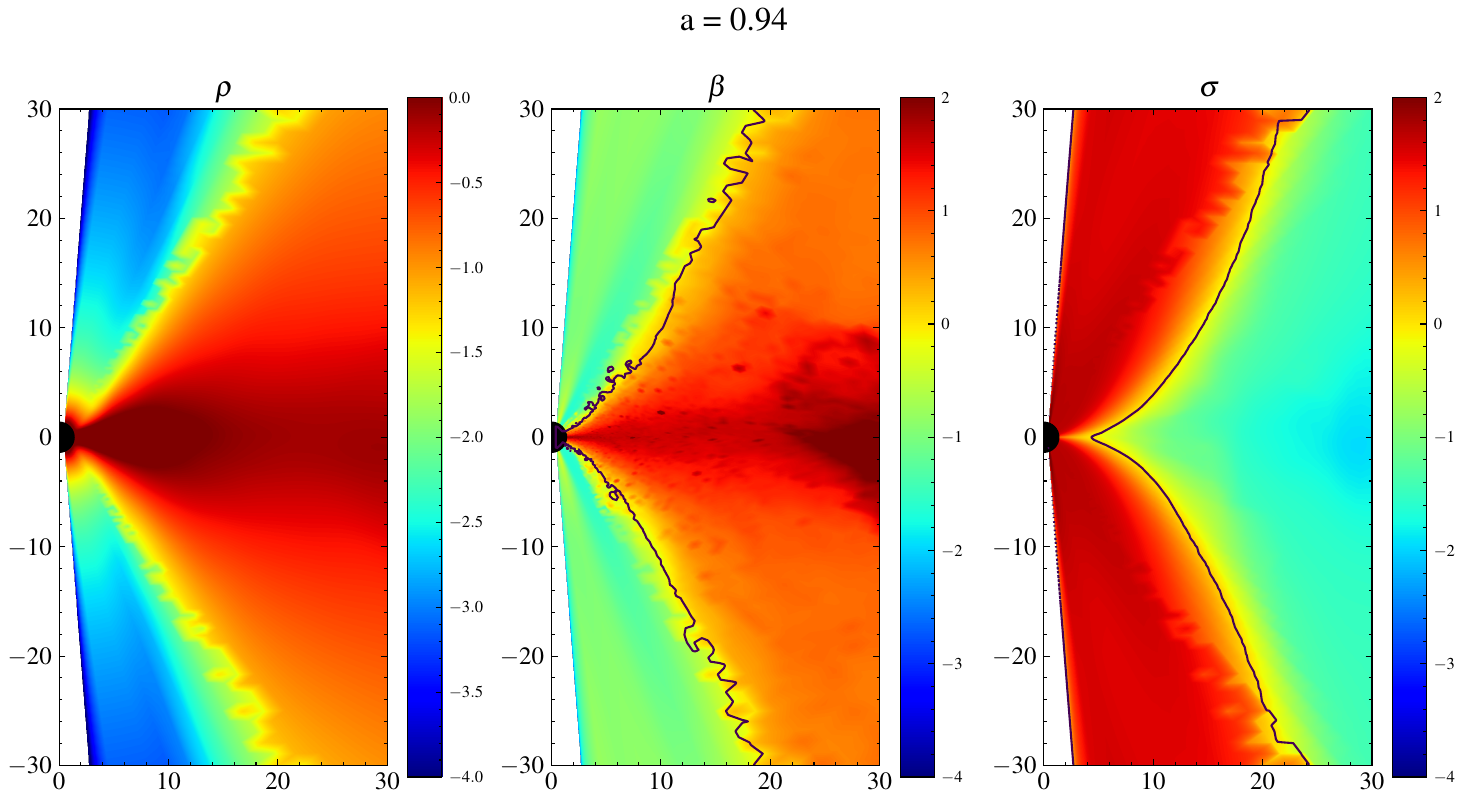}
    \end{tabular}
    \caption{Time and azimuthal averaged density $\rho$, ratio of gas to magnetic pressure $\beta$ and magnetization $\sigma$ for aM94b100 (top), a0b100 (middle) and a94b100 (bottom). The black lines in the middle and right panels represent the $\beta = 1$ and $\sigma = 1$ conditions, respectively.}
    \label{fig:aM94Map}
\end{figure}

We show the time and azimuthal averages of the density $\rho$, of the
plasma parameter $\beta$ and of the magnetization $\sigma$ for the
simulations aM94b100, a94b100 and a0b100 in Figure \ref{fig:aM94Map}.
These values are obtained using Equation \ref{eq:phi-average}, and are then
averaged from $t = 10^4 t_g$ to $t = 2\times 10^4 t_g$. The difference between
these three spins is clear. In the density $\rho$ plot, the non-spinning
black hole has the thinnest disk, while the prograde disk is the thickest,
which is consistent with the bottom right panel of Figure \ref{fig:spin-factor}.
The $\beta$ plots show similar behaviors as the density. There are clear
vacuum regions inside the jet, especially for a0b100, which are attributed to
the numerical flooring rather than physical effects.

\begin{figure}
    \centering
    \includegraphics{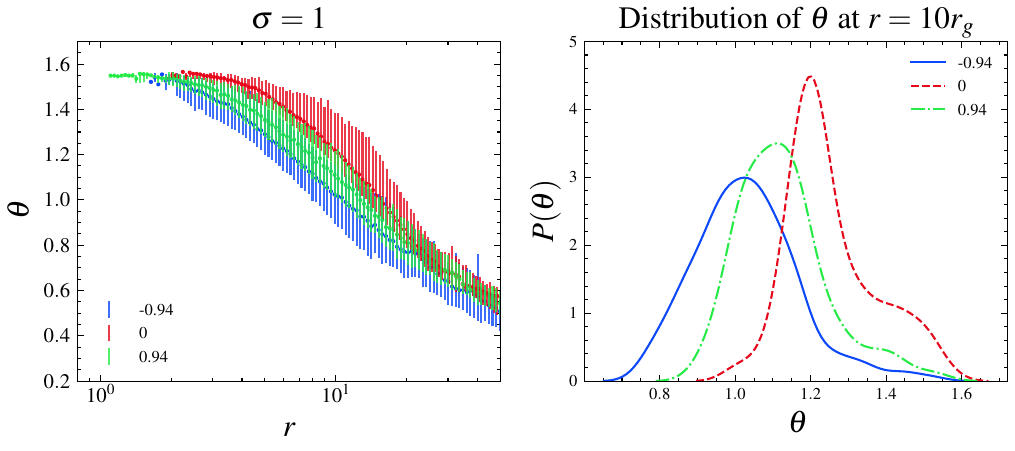}
    \caption{Left Panel: The 1$\sigma$ region of jet boundary for $a = -0.94,~ 0$ and
    $0.94$. We use the magnetization parameter $\sigma = 1$ condition as the jet-disk
    boundary and derive the $1\sigma$ variations and median
    of the polar angle $\theta$ for $\sigma = 1$ at a given radius $r$. The vertical lines
    are $1\sigma$ variations and the dots
    are the median values. Right Panel:
    The kernel density estimation (KDE) of the polar angle $\theta$
    such that $\sigma= 1$  at radius
    $r = 10 r_g$. The non-spinning black hole has the widest jet,
    corresponding to the largest $\theta$, while the prograde disk with $a = 0.94$
    has a wider jet than the retrograde disk with $a = -0.94$. In
    addition, the non-spinning black hole has the narrowest
    peak, which suggests that the jet boundary of a non-spinning black hole is
    the most stable.}
    \label{fig:sigma-three} 
\end{figure}

We show the polar angle of the jet boundary, which we associate to $\sigma = 1$, as a function of radius for
aM94b100, a94b100 and a0b100 in the left panel of Figure \ref{fig:sigma-three}.
The vertical lines correspond to $1\sigma$ variations and
the points represent the median values of the polar angle $\theta$ for
a specific radius $r$. These values are obtained for $ 10^4 t_g < t <  2 \times 10^4 t_g$.
As shown in Figure \ref{fig:mad-factor}, all three runs remain in the MAD state during
this time period, and maintain inflow-outflow equilibrium. It is 
seen that the prograde disk has a wider jet than the retrograde at $r < 20 r_g$, which
is consistent with the findings of \citet{Narayan2022MNRAS.511.3795N}. It is also
clearly seen that the non-spinning black hole has the widest jet. This result
is consistent with Figure \ref{fig:aM94Map}. 

In order to quantify the degree of variation compared to their mean, we
use the standard error $\sigma_\theta/ \bar{\theta}$, where $\sigma_\theta$
is the $1\sigma$ variations of $\theta$ and $\bar{\theta}$ is the median.
The temporal and radial (between $2 r_g $ and $ 30 r_g$ ) averages of $\sigma_\theta/ \bar{\theta}$
for aM94b100, a94b100 and a0b100 are 0.11, 0.08 and 0.07,
respectively. We further show the kernel density estimation of $\theta_{\sigma = 1}$ at
$r = 10 r_g$ in the right panel of Figure \ref{fig:sigma-three}.
The boundary of the retrograde disk has the largest variations, as
expected from large shear stresses at the boundary between the jet and the disk, which
have opposite toroidal velocity. Our results are, in this sense, consistent with those
of \citet{Wong2021ApJ...914...55W}, who argued both analytically and numerically
that retrograde disks have stronger shear across the jet-disk boundary than prograde ones.

%\clearpage

\subsection{Pressure}
\label{subsec:pressure}

\begin{figure}
    \centering
    \includegraphics[width=0.9\textwidth]{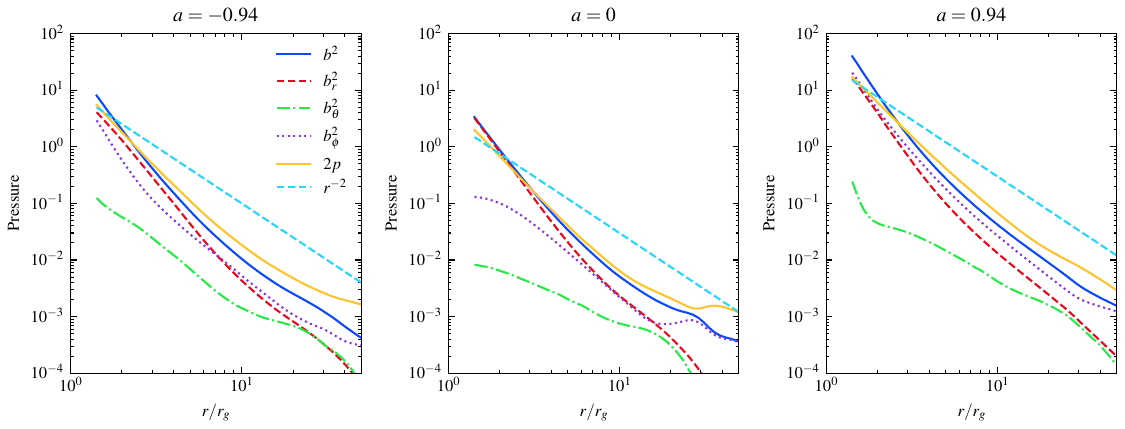}
    \caption{The angular- and time-averaged gas pressure $p$ (yellow) and magnetic
    pressure $b^2$ (blue) for aM94b100, a0b100 and a94b100. We further show
    each component of the magnetic pressure, $b_r^2$ (red), $b_\theta^2$ (green)
    and $b_\phi^2$ (purple). These values are calculated by Equation
    \ref{eq:pressure-average} and then time-averaged. The dashed (light blue) curve represents an $r^{-2}$ dependence.}
    \label{fig:pressureDiffComp}
\end{figure}

In section \ref{subsec:disk-jet}, we demonstrated that prograde disks are thinner and have
wider jets. We attribute these to the pressure distribution
inside the disk and jets. In Figure \ref{fig:pressureDiffComp}, we show the
different components of the magnetic pressure and of the gas pressure
for aM94b100, a0b100 and a94b100. These components are calculated using Equation
\eqref{eq:pressure-average} and are further time-averaged. This implies that they are biased towards large density regions. We find that the overall total pressure $p + b^2 /2$ is the lowest for
non-rotating black hole and the largest for prograde disk.
At small radii, the magnetic pressure $b^2$, represented by the blue line, dominates over the gas pressure. As the
radius increases, the gas pressure gradually becomes dominant. The
transition radius for $a = 0.94, \left. 0, \right.-0.94$ are $r_{\mathrm{eq}} =
2.71 r_g, \left. 3.13 r_g, \right. 2.35 r_g$, respectively. 
% \deletedD{It seems that
% the transition radius is the smallest for the largest MAD parameters.} \authorcommentD{This is not true}

\citet{Begelman2022MNRAS.511.2040B} analyzed a prograde disk, which is somewhat smaller than the ones we analyze here. They reported that the gas
pressure evolves $\propto r^{-2}$ while the magnetic pressure was found to evolve faster
closest to the black hole. Here we find that the gas pressure evolution is
somewhat steeper than $r^{-2}$. This could  be due to the different disk structure,
different resolution we are using or the shorter time interval over which the averaged is performed. We note that a similar radial evolution of the gas pressure, namely $p_g \propto r^{-2}$ was reported by \citet{Tchekhovskoy2011MNRAS.418L..79T} and \citet{MTB12} who found that $p_g \propto r^{-1.9}$ for their "thinner" disk models.

We further show in Figure \ref{fig:pressureDiffComp} the radial dependence of all individual magnetic pressure components. 
The total magnetic pressure has a similar radial evolution in our work and in the work of \citet{Begelman2022MNRAS.511.2040B}, namely its radial evolution is steeper than $r^{-2}$ close to the black hole. 
As is seen from Figure \ref{fig:pressureDiffComp}, the polar component $b_\theta^2$ never dominates
for any of the spins and is much lower than the radial and toroidal components.
Second, at small radii inside the ISCO, the main difference between rotating and
non-rotating black holes is the contribution of the toroidal field $b_\phi^2$: it
is negligible compared to the radial component for a non-rotating black hole,
while both components are of the same order for rapidly rotating black holes. The toroidal
field even dominates for the prograde disk very close to the black hole. This
explains the difference between the conclusions of \citet{Begelman2022MNRAS.511.2040B}, who studied disks around rapidly rotating black holes,
and \citet{Chatterjee2022ApJ...941...30C} who studied non-spinning black hole systems, which is the importance of the toroidal
component in a MAD accretion process. Similar radial evolution of the magnetic
field components were reported by \citet{Tchekhovskoy2011MNRAS.418L..79T} and \citet{MTB12} 
who found that $b_r \propto r^{-1.5}$ and $b_\phi \propto r^{-1}$ for their
"thinner" disk models. Clearly, the radial dependence of $b^r$ does not seem
to depend on the black hole spin, while the toroidal component does.

\begin{figure}
    \centering
    \includegraphics[width=\linewidth]{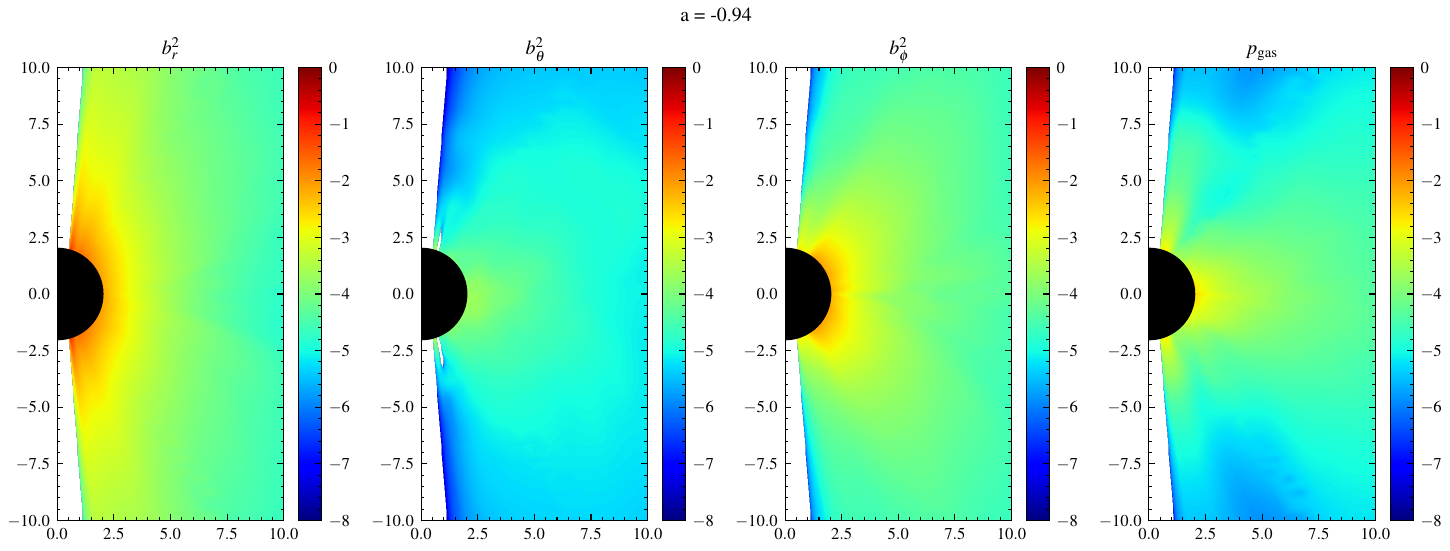}  \\
    \includegraphics[width=\linewidth]{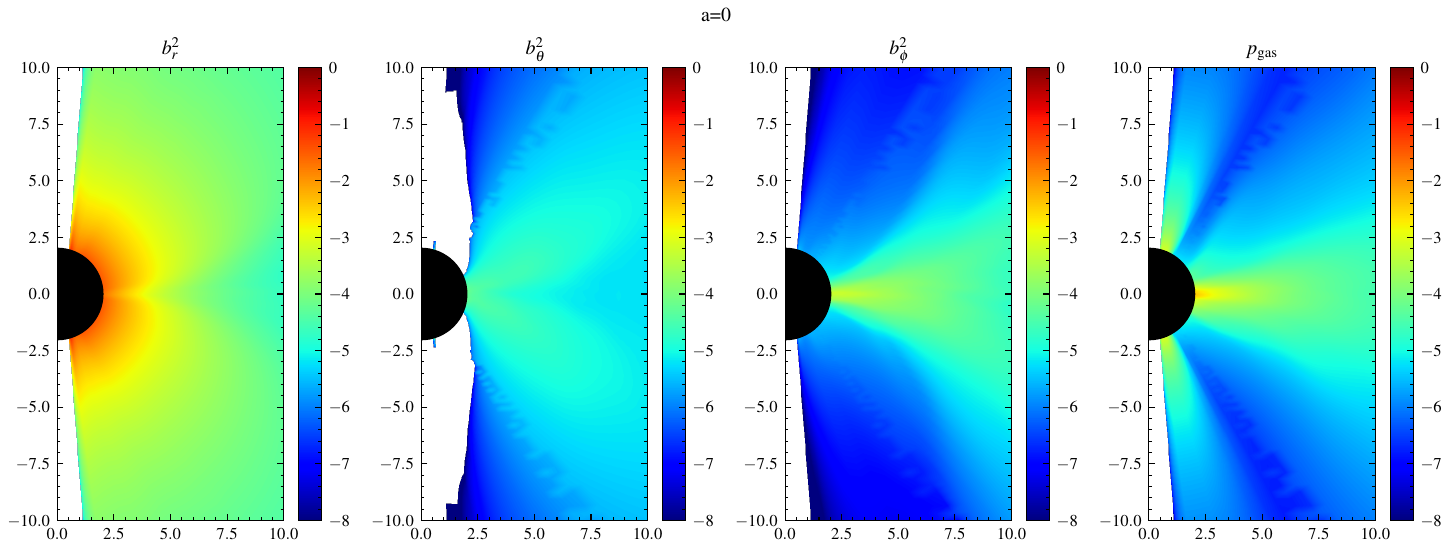} \\
    \includegraphics[width=\linewidth]{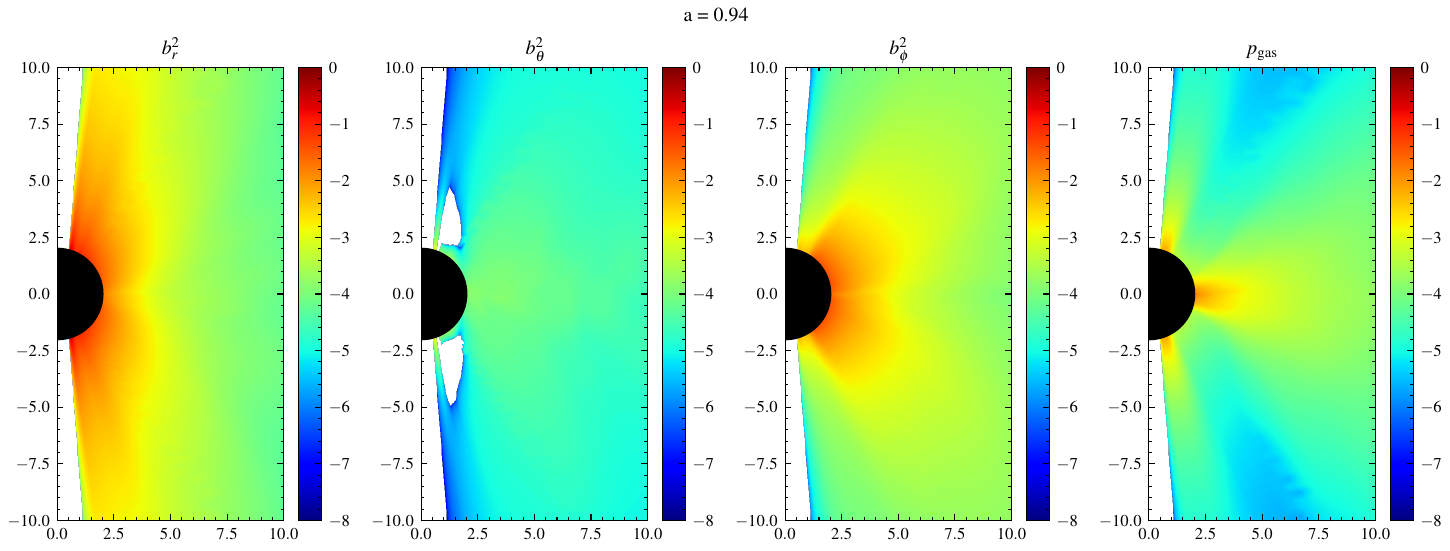} 
    \caption{The time and azimuthal averages of the magnetic pressure components, namely $b_r^2$ (first column), $b_\theta^2$ (second column), and $b_\phi^2$ (third column), and of the gas pressure (last column) for $a = -0.94$ (top), $a = 0$ (middle) and $a = 0.94$ (bottom). The color scaling is the same for all spins and all components to better display the differences. The most striking difference is the small contribution of the azimuthal component $b^\phi$ to the total magnetic pressure for the non-spinning black hole $a = 0$.}
    \label{fig:mapPressureM094}
\end{figure}

We show the $\phi-$ and time- averaged maps of the pressure components of- aM94b100,
a0b100 and a94b100 in Figure \ref{fig:mapPressureM094}.
All sub-figures use the same color scaling, allowing an easy comparison
between them. At small radii close to the black hole, the radial
component $b_r^2$ are large for all three simulations and decrease
quickly as the radius increases. For $a = \pm 0.94$, large values of $b_r^2$
are also found at large radii $r \sim 10r_g$ in the jet
region. 
The polar component, $b_\theta^2$, is negligible in all three cases.
Note that the blank regions in the color-maps of $b_\theta^2$ are
caused by numerical truncating errors: 
this component is too small relative to the other two components.
The distributions of the toroidal component $b_\phi^2$ also shows a great
variance. The non-spinning black hole has the weakest $b_\phi^2$ component,
while 
it is larger for the spining black holes.
We note the drop of $ |b^\phi|^2$ at the equator: this is because
$b^\phi$ changes sign at the equator due to the presence of a current sheet
separating the two hemispheres. This drop is  more pronounced and visible for
the spinning black holes.

The gas pressure $p_g$ of the non-spinning black hole
contains clear weak channels about $\theta \simeq \frac{\pi}{3}$ and
$\theta \simeq \frac{2\pi}{3}$. These channels are also visible for the spinning
black holes, but are less pronounced.
As we previously discussed in section \ref{subsec:disk-jet}, these channels are caused
by the numerical floors applied in the regions closest to the pole, which are
required to maintain numerical stability in our simulations.

\section{Angular Momentum Flux}
\label{sec:angular}

\begin{figure}
    \centering
    \begin{tabular}{c}
    \includegraphics[width=0.9\textwidth]{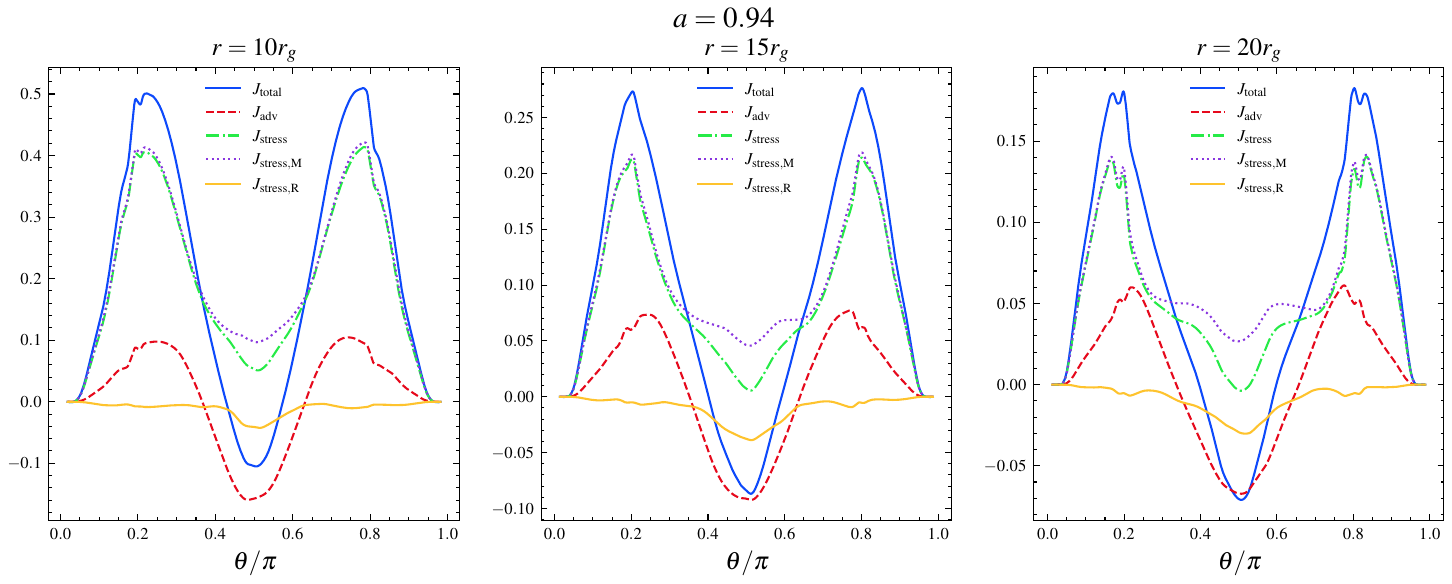} \\
    \includegraphics[width=0.9\textwidth]{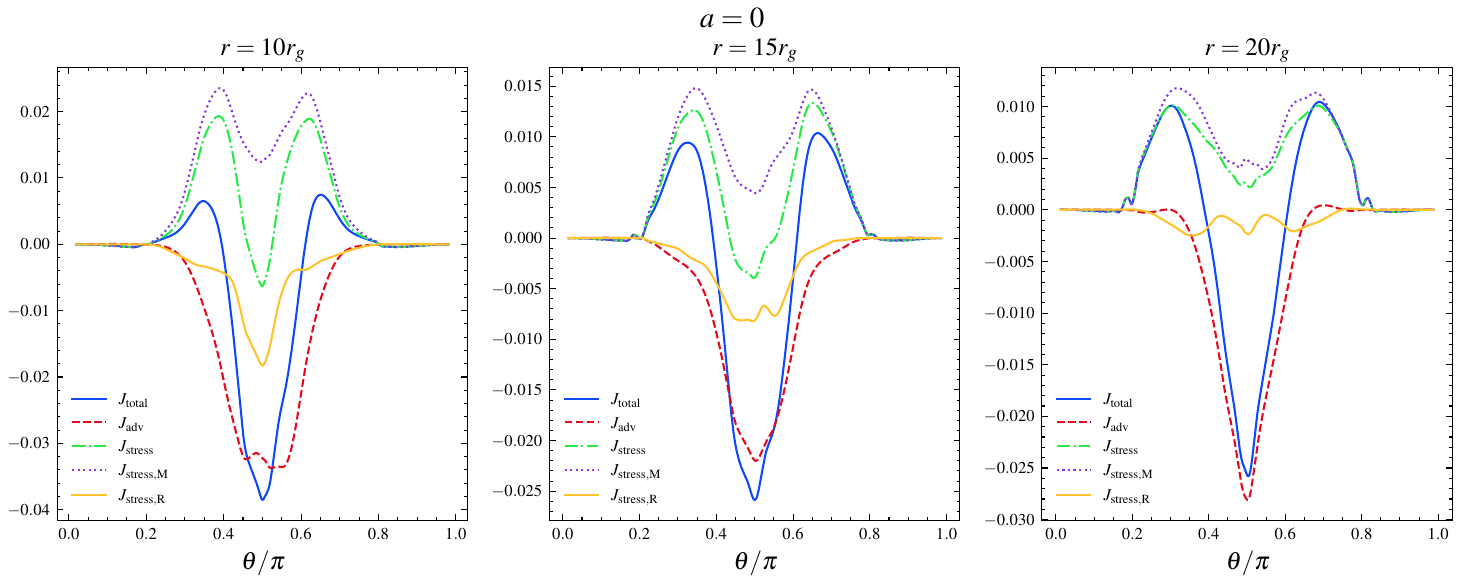} \\ 
    \includegraphics[width=0.9\textwidth]{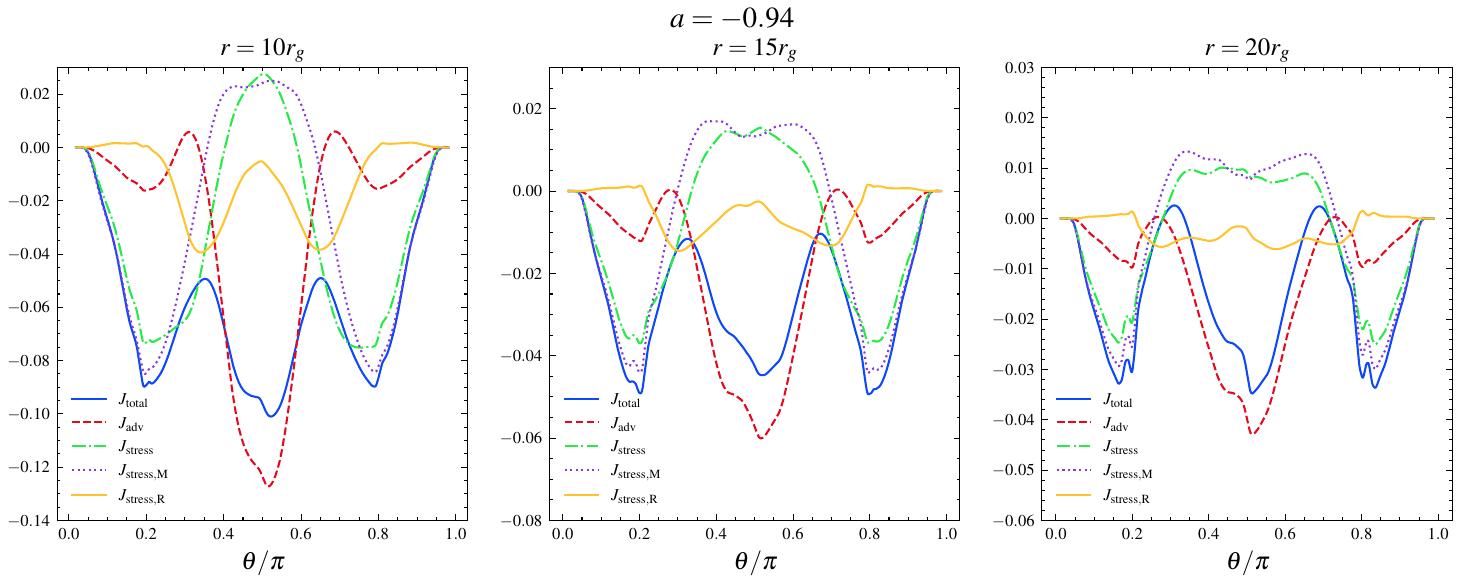}
    \end{tabular}
    \caption{The radial angular momentum flux components $\dot J^r$ as a function of the polar angle $\theta$ for a94b100 (top), a0b100 (middle) and aM94b100 (bottom) at 3 different radii, $r = 10, ~15, ~20 r_g$.}
    \label{fig:angularMomentumFlux}
\end{figure}

It has long been argued that magnetic fields drive the transfer of angular momentum during
accretion via the development of the magneto-rotational instability \citep{BH91, BH98},
but also by helping to launch a disk wind and producing a strong jet, both components
capable of significantly transporting angular momentum. In this section, we
investigate the dependence of angular momentum flux on the spin of the black hole. We
use the definitions in \citet{Chatterjee2022ApJ...941...30C} to define the total
angular momentum flux $\dot J_{\mathrm{total}}$, the  advected angular momentum
flux $\dot J_{\mathrm{adv}}$ and the  angular momentum flux due to stresses
$\dot J_{\mathrm{stress}}$, 
\begin{equation}
    \dot J^i_{\rm total} (r, \theta ) = \left \langle T^i_{~\phi} \right \rangle _{_{\phi,t}} 
\end{equation}
\begin{equation}
	\dot J^i_{\rm adv} (r, \theta ) = \left \langle \left ( \rho + u_g + \frac{b^2}{2} \right )u^i \right \rangle _{_{\phi,t}} \left \langle u_\phi \right \rangle _{_{\phi,t}} 
\end{equation}
\begin{equation}
    \dot J^i_{\rm stress} (r, \theta ) = \dot J^i_{\rm total} (r, \theta ) - \dot J^i_{\rm adv} (r, \theta )
\end{equation}
where $\langle X \rangle_{j}$ represent the average of $X$ with respect to variable
$j$. Note that for this section, there is no weighting by the density $\rho$. We
further decompose the stressed induced angular momentum flux into its
Maxwell $\dot J_{\mathrm{stress, M}}$ and its Reynolds $\dot J_{\mathrm{stress, R}}$ components:
\begin{equation}    
    \dot J^i_{\rm stress, M} (r, \theta ) = \left \langle \frac{b^2}{2} u^i u_\phi - b^i b_\phi \right \rangle _{_{\phi,t}} 
\end{equation}
\begin{equation}    
    \dot J^i_{\rm stress, R} (r, \theta ) = \left \langle \left ( \rho + u_g + \frac{b^2}{2} \right ) u^i u_\phi \right \rangle _{_{\phi,t}} - \dot J^i_{\rm adv}
\end{equation}
Following these definitions, the time- and $\phi$-averaged angular momentum flux
in the radial direction of a94b100, a0b100 and aM94b100 are shown in
Figure \ref{fig:angularMomentumFlux} at three different radii, namely
10$r_g$, 15$r_g$ and 20$r_g$.

For non-spinning black hole, depicted in the middle
line of Figure \ref{fig:angularMomentumFlux}, the results we obtained are similar to the results reported in \citet{Chatterjee2022ApJ...941...30C}. The radial component of the
total angular momentum flux $\dot J^r_{\mathrm{total}}$ is negative in the disk
region at all sampled radii, $r = 10 r_g$, $15 r_g$, and $20 r_g$. This indicates
that as the matter accretes, it brings a net angular momentum to the black hole.
However, at $\theta \simeq 0.35\pi$ and $0.65\pi$, the total angular momentum flux
changes sign to become positive. This shows that angular momentum is being transported
away from the accretion disk by the wind.
The radial angular momentum flux converges to 0 as it approaches the
poles, underlying the weakness of the jet for $a = 0$.
In the disk, the contributions of $\dot J^r_{\mathrm{adv}}$ and
$\dot J^r_{\mathrm{stress, R}}$ dominate and are always negative, suggesting
that they are responsible for the inward transport of angular momentum in the disk,
as was previously demonstrated by \citet{Chatterjee2022ApJ...941...30C}. We
note that the relative importance of the Reynolds components is large at small
radii but decreases towards large radii and is in agreement with the results
of \citet{Chatterjee2022ApJ...941...30C} at $r = 20 r_g$.
Furthermore, the Maxwell
stress component $\dot J^r_{\mathrm{stress, M}}$ is always positive with two maxima,
reached at $\theta \simeq 0.4\pi$ and $\theta \sim 0.6\pi$. At these high
latitudes, the radial component of the total angular momentum flux is dominated by
$J^r_{\mathrm{stress, M}}$ underlying the importance of magnetic fields in
the production and dynamics of the wind above the disk in the transport
of angular momentum. 

The results of the black hole with positive spin $a = 0.94$ are displayed in the
top line of Figure \ref{fig:angularMomentumFlux}. We find this case to
be qualitatively similar to the non-spinning black hole scenario: the radial component of the total angular
momentum flux is negative in the disk and positive at high latitudes, with a change
of sign at $\theta \sim 0.4 \pi$ and $\theta \sim 0.6 \pi$. Outside the disk,
we find that the Maxwell component strongly dominates and reaches its maxima 
symmetrically with respect to the equator at $\theta \sim 0.25 \pi$ and $\theta \sim 0.75 \pi$.
The radial components of the angular momentum flux remain large up to the poles.
This indicates the existence of powerful magnetized winds and jets, through which the disk
and the black hole lose a substantial fraction of their angular momentum. We further
see that the
angular momentum flux contributed by advection is negative in the disk region. However,
it is positive at high latitude, which is different from the case of the non-spinning
black hole. This may be due to the large contribution of the magnetic fields
energy density $b^2/2$ in its expression. Finally, we find that the Reynolds
stress component is much smaller than any of the other components and can be
safely ignored.

For the negative spin simulation $a = -0.94$ displayed in the bottom line of
Figure \ref{fig:angularMomentumFlux}, we find that the radial angular
momentum flux distribution with angle is significantly distinct from that of the
non-rotating and positive spin black holes. This is mostly due to the fact that
the poloidal velocity changes sign: the plasma corotate with the black hole
close to the pole while it corotates with the disk at the equator. First, we find that the radial component of the 
total angular momentum flux is always negative at all angles. As
a result, the black hole gains angular momentum and its spin decreases towards 0. This
is consistent with the result presented in Figure \ref{fig:flux-r}. The contribution
of the advection to the flux of angular momentum is similar to the cases of positive
and null spins, 
namely its contribution is maximum and negative at the equator. The Maxwell
component contribution is positive at the equator, showing that magnetic fields are
responsible for taking away radial angular momentum also in this case. However
this contribution is negative around $\theta \simeq 0.2\pi$ and $0.8 \pi$ and dominates
the radial total angular momentum flux at those angles. 

\begin{figure}
    \centering
    \gridline{\fig{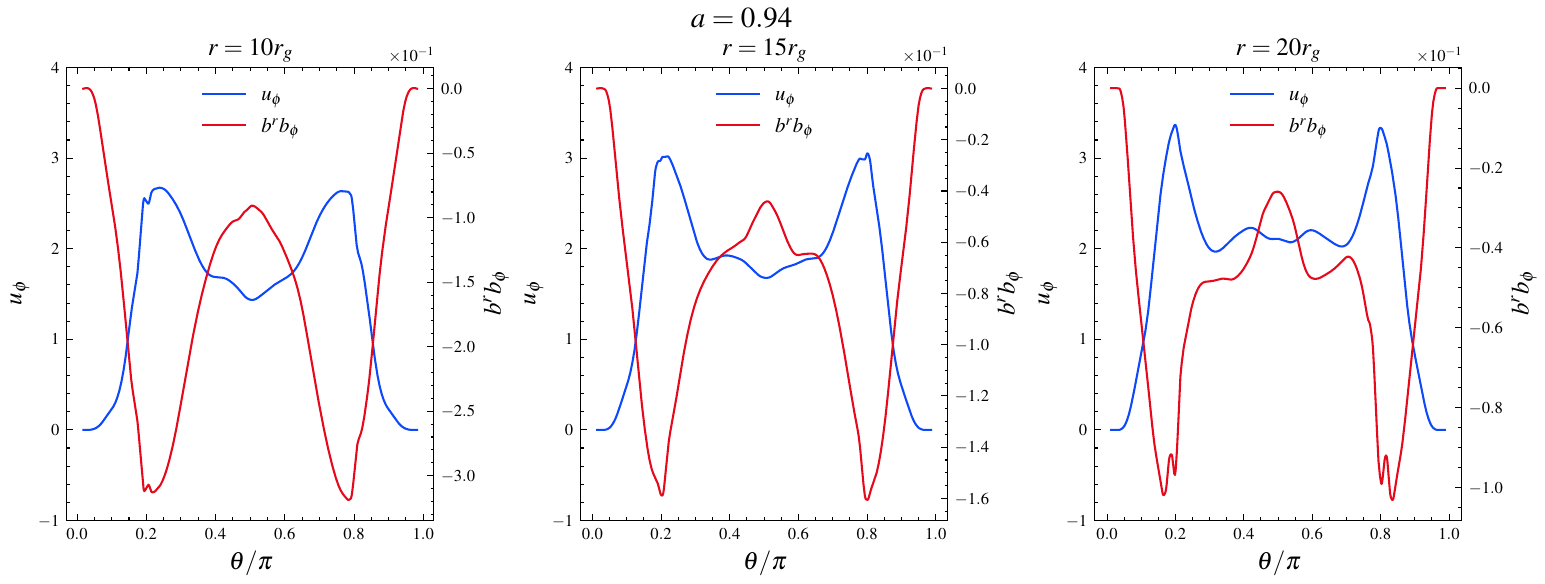}{0.8\textwidth}{}}
    \gridline{\fig{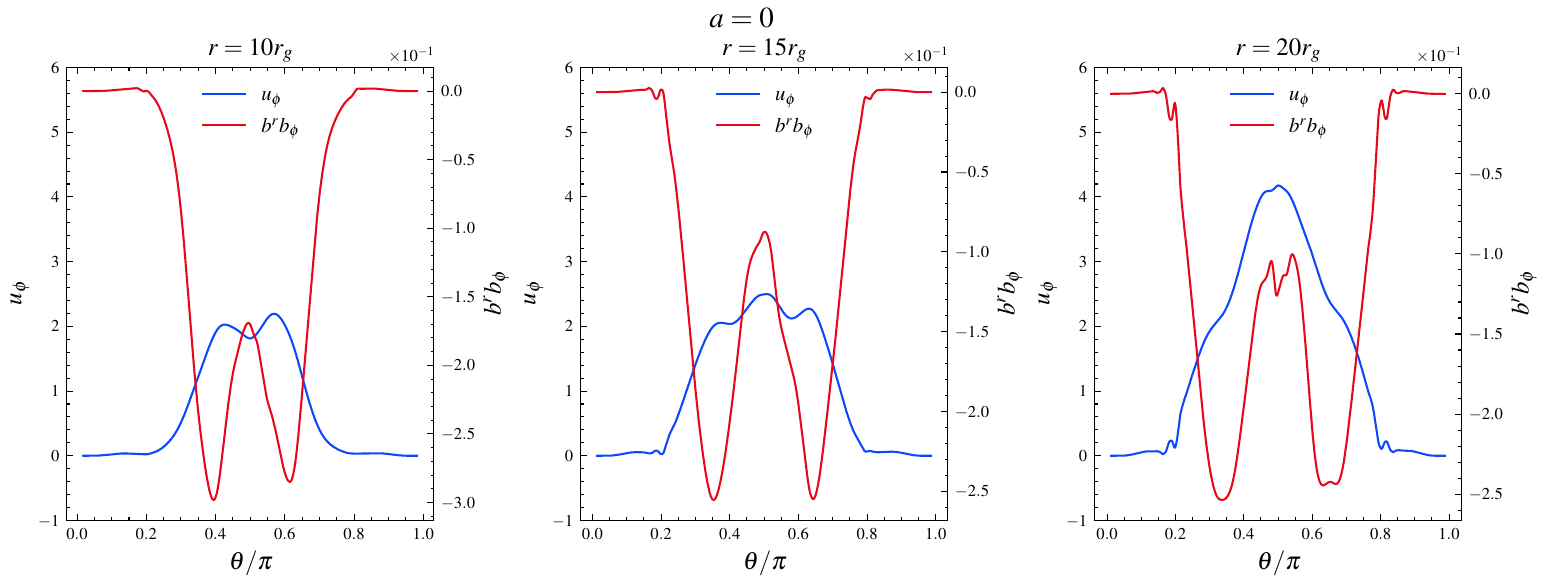}{0.8\textwidth}{}}
    \gridline{\fig{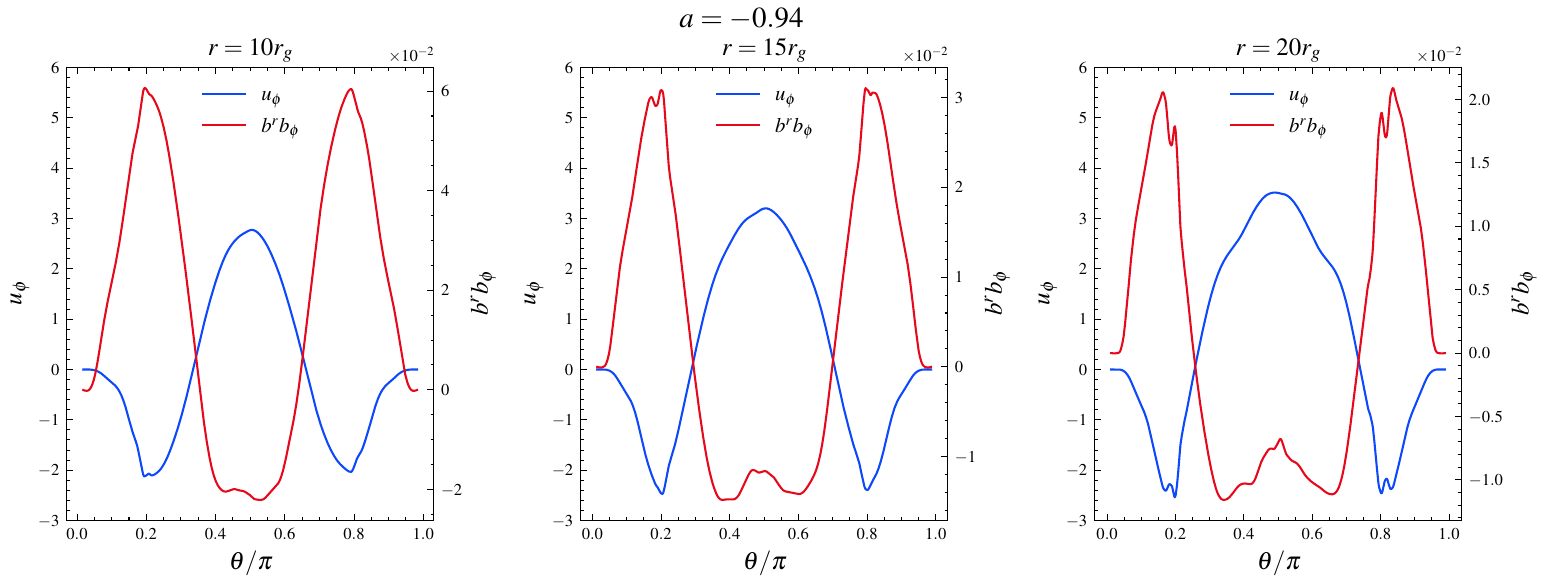}{0.8\textwidth}{}}
    \caption{The toroidal component of the 4-velocity $u^\phi$ and $b^r b_\phi$ as a function of $\theta$ for a94b100 (top), a0b100 (middle) and aM94b100 (bottom), at the radii at which the radial angular momentum flux are obtained}.
    \label{fig:uphi-brbphi}
\end{figure}

The signs of angular momentum flux are mainly determined by the sign of the azimuthal
velocity $u_\phi$ and of $b^r b_\phi$. In Figure \ref{fig:uphi-brbphi}, we show the
$\phi$- and time-averaged evolution of $u_\phi$ and of $b^r b_\phi$ with the polar
angle $\theta$ at the three radii selected for the angular momentum flux in Figure
\ref{fig:angularMomentumFlux}. For the $a = 0$ and $a = 0.94$ cases, the toroidal
velocity $u_\phi$ is positive for all angles $\theta$ as expected, therefore, the sign
of the advection contribution to the angular momentum flux $\dot J_{\mathrm{adv}}$
depends on the sign of $u^r$. On average, $u^r$ is negative in the disk region since
the matter is accreted towards the black hole. It is however positive at high
latitude, as matter is carried away in the disk wind and the jet. This change of sign
explains the pattern of $\dot J_{\mathrm{adv}}$ seen in Figure
\ref{fig:angularMomentumFlux} for $a = 0.94$. For the non-spinning black hole,
the azimuthal velocity quickly  goes to 0 at high latitude, which explains
the null contrbution of the advection to the angular momentum flux in this region.
On the other hand, the term  $b^r b_\phi$ is always negative at all angles $\theta$ for the black hole
with spin $a = 0.94$. In this case, the maxima (one in each hemisphere) of $|b^r b_\phi|$
%for the black hole with spin $a = 0.94$
is reached closer to the poles than for the non-rotating black hole, further underlying
the different disk structure: rotating black holes have wider disk than non-rotating
ones, as was already demonstrated in Section \ref{subsec:disk-jet} and in
Figures \ref{fig:aM94Map} and \ref{fig:sigma-three}. 
The polar angles $\theta$ at which the maxima of $b^r b_\phi$ are reached correspond 
to the angle at which the contribution of the Maxwell stress is maximal. 

The actual sign of the component $b^r b_\phi$ depends on our arbitrary choice of initial
disk magnetization, and specifically on the orientation of the initial magnetic field
loop. A different orientation would lead to a different sign of the $b^r b_\phi$
component. Indeed, the sign of $b_\phi$ is set by frame-dragging and it is positive in
the south hemisphere and negative in the north hemisphere, for prograde disks. However, the sign of $b^r$ would be opposite if the initial
magnetic field loop would have had a different orientation, as demonstrated in
\citet{MTB12}. We note here another difference between the $a = 0$ and
$a = 0.94$ black holes: $b^r b_\phi$ is negative close to the poles for the
rotating black hole, while it is very small for the non-rotating black hole.

The situation is different for the retrograde disk, with a different evolution of the
azimuthal velocity $u_\phi$ and of the term $b^r b_\phi$ with polar angle $\theta$,
both shown in the third line of Figure \ref{fig:uphi-brbphi} for $a = -0.94$.
The toroidal velocity $u_\phi$ is also positive in the disk region. However,
it becomes negative close to the poles around $\theta \simeq 0.2\pi$ and 
$\theta \simeq0.8 \pi$. 
The term $b^r b_\phi$ is also different from that of the
prograde disk. First, it has a different sign close to the pole: $b^r b_\phi$ is
positive for a retrograde disk since (i) $b_\phi$ is negative in the south hemisphere and
positive in the north hemisphere because of frame dragging, and (ii) $b^r$ is negative
in the south hemisphere and positive in the north hemisphere. Here, the sign of $b^r$ is
set by the orientation of the initial magnetic field loop. At the equator, however,
$b^r b_\phi$ has the same sign for both prograde and retrograde disks. This explains the
sign of the Maxwell stress contribution to the radial angular momentum flux.

\begin{figure}
    \centering
    \gridline{\fig{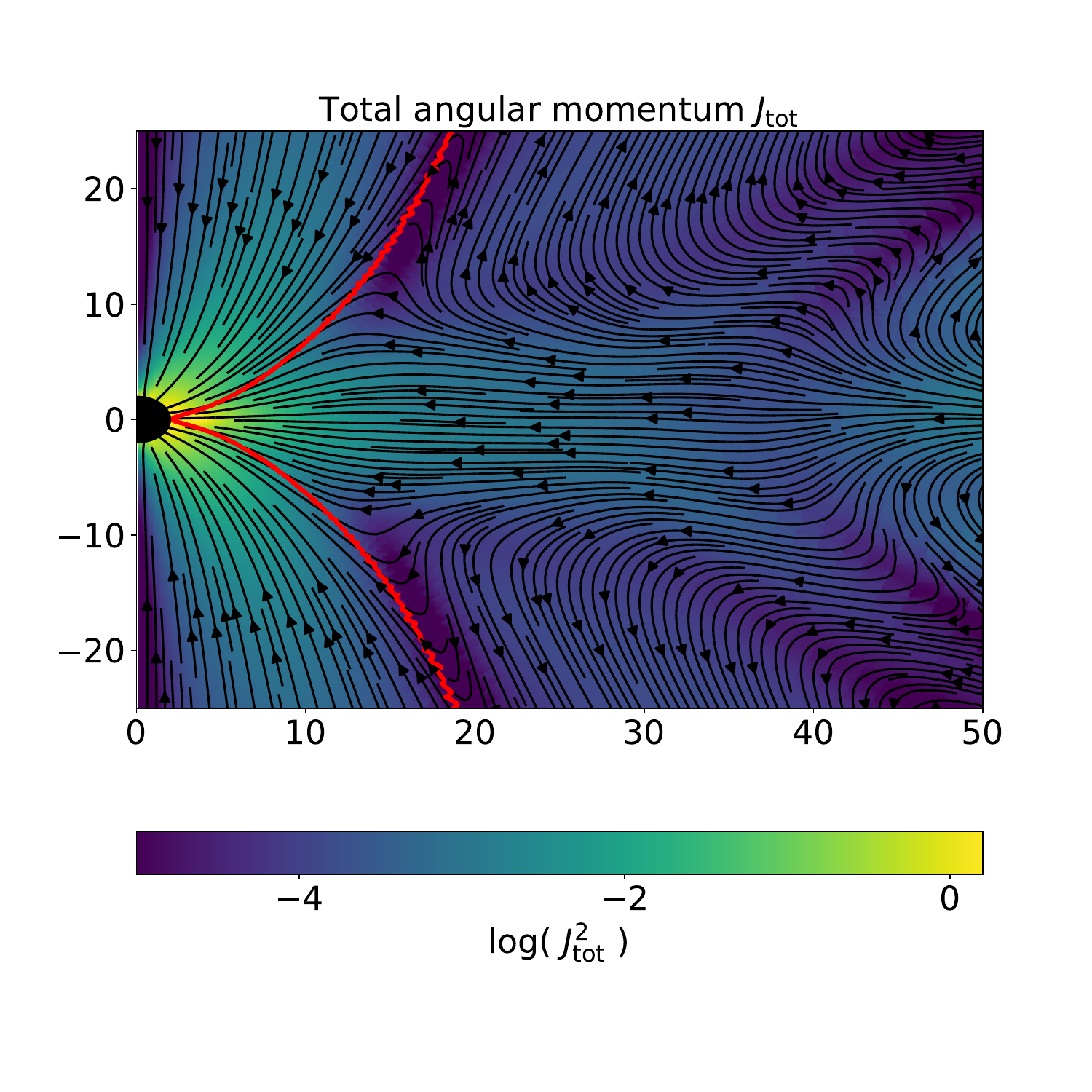}{0.32\textwidth}{}       
              \fig{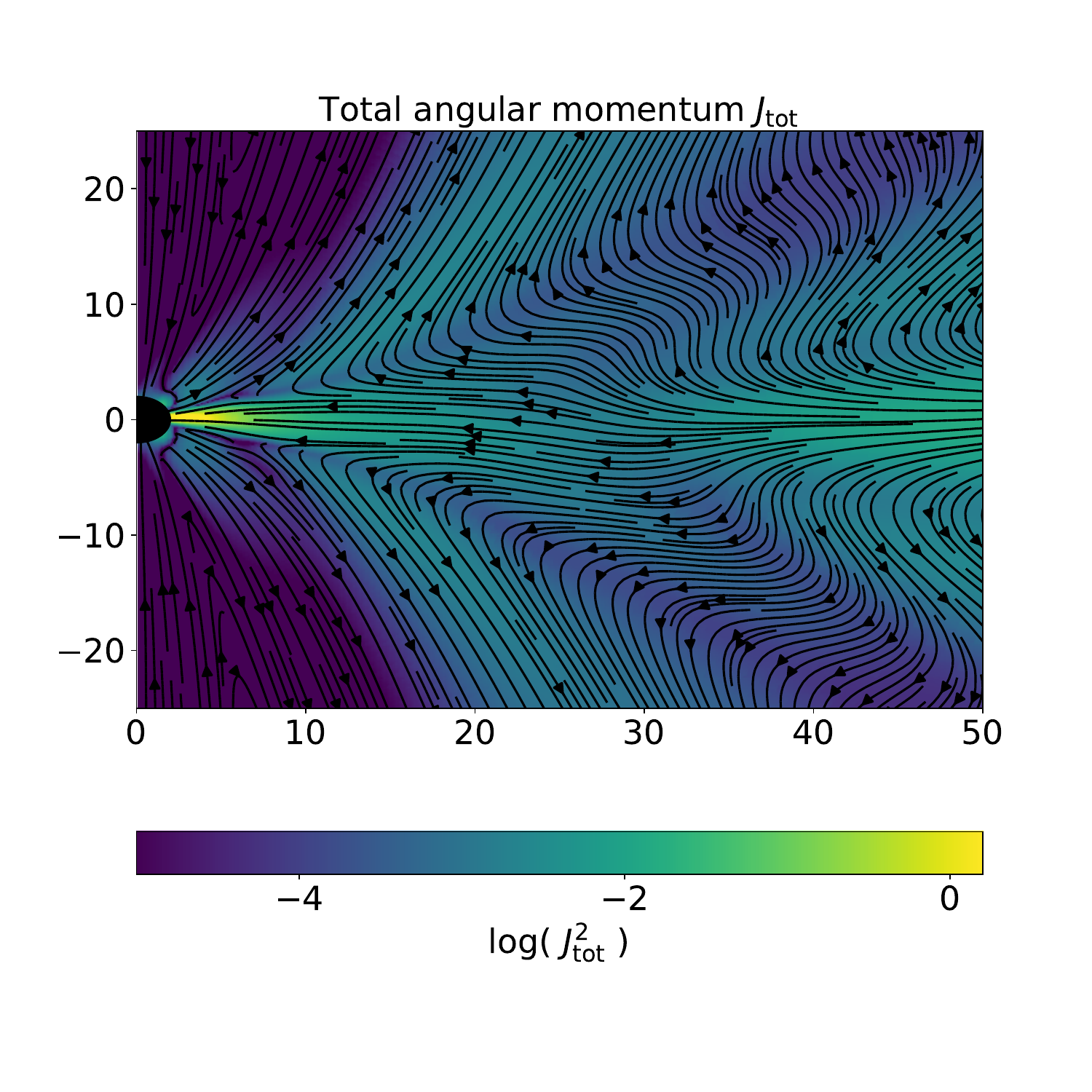}{0.32\textwidth}{} 
              \fig{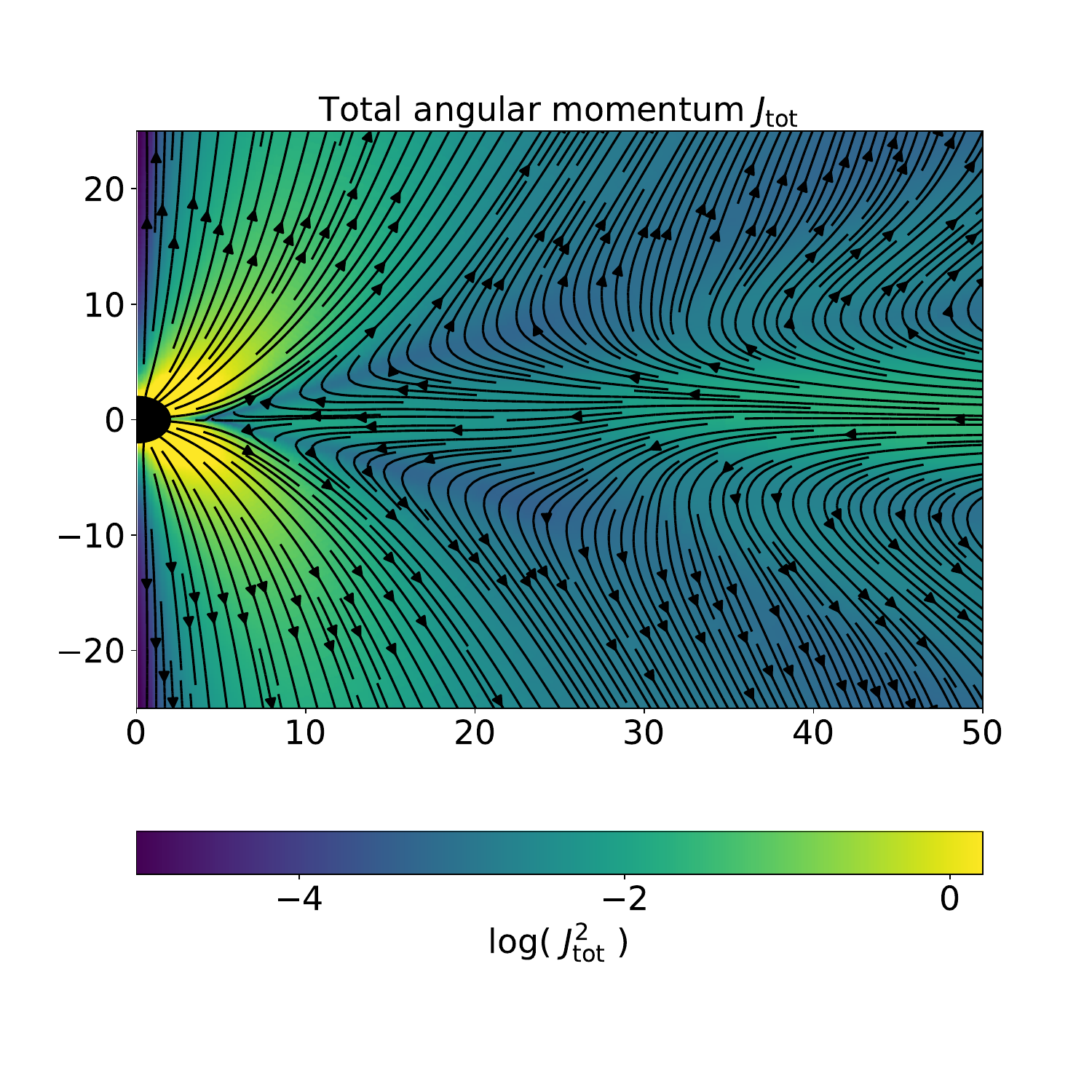}{0.32\textwidth}{}}
    \vspace*{-10mm}
    \gridline{\fig{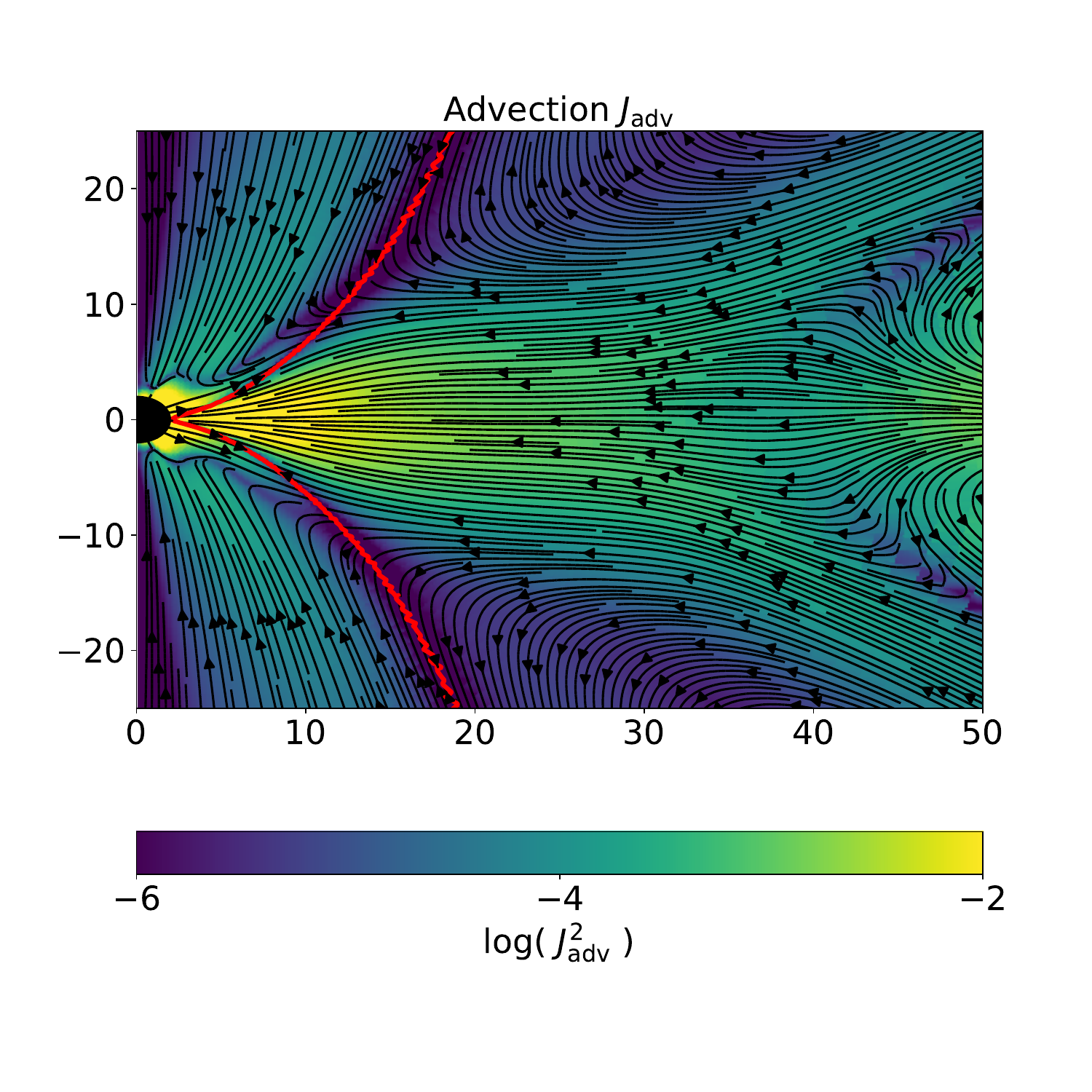}{0.32\textwidth}{}       
              \fig{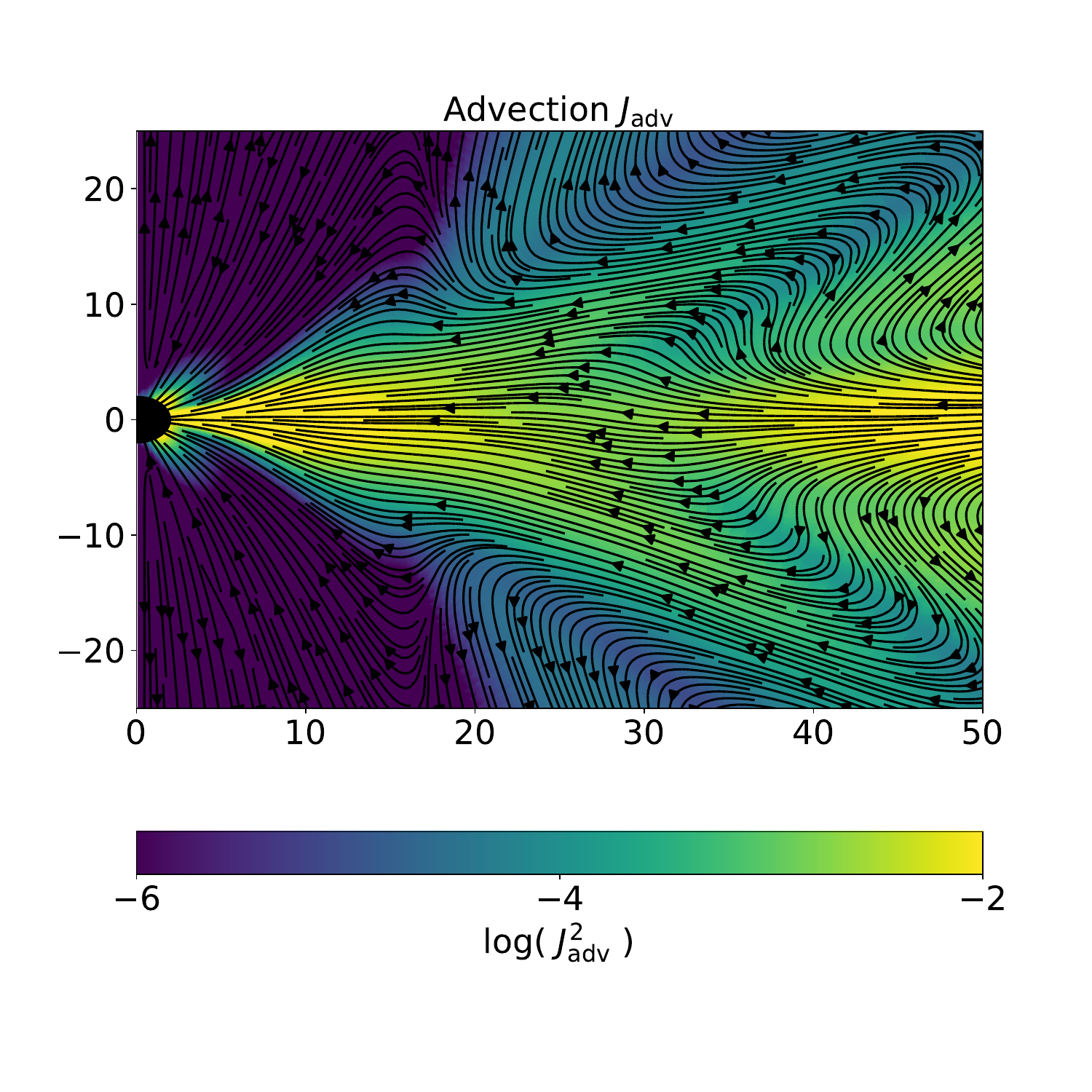}{0.32\textwidth}{} 
              \fig{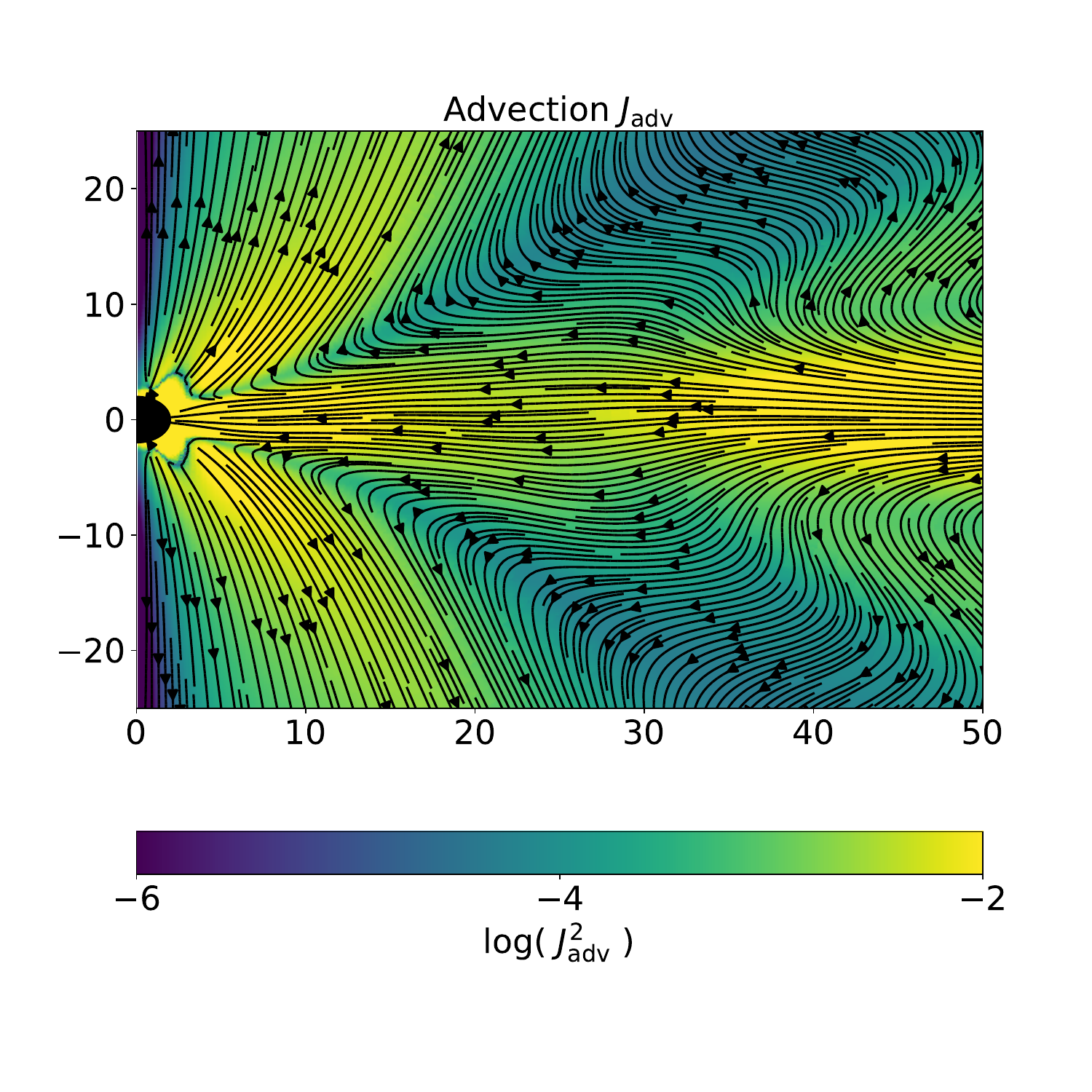}{0.32\textwidth}{}}
    \vspace*{-10mm}
    \gridline{\fig{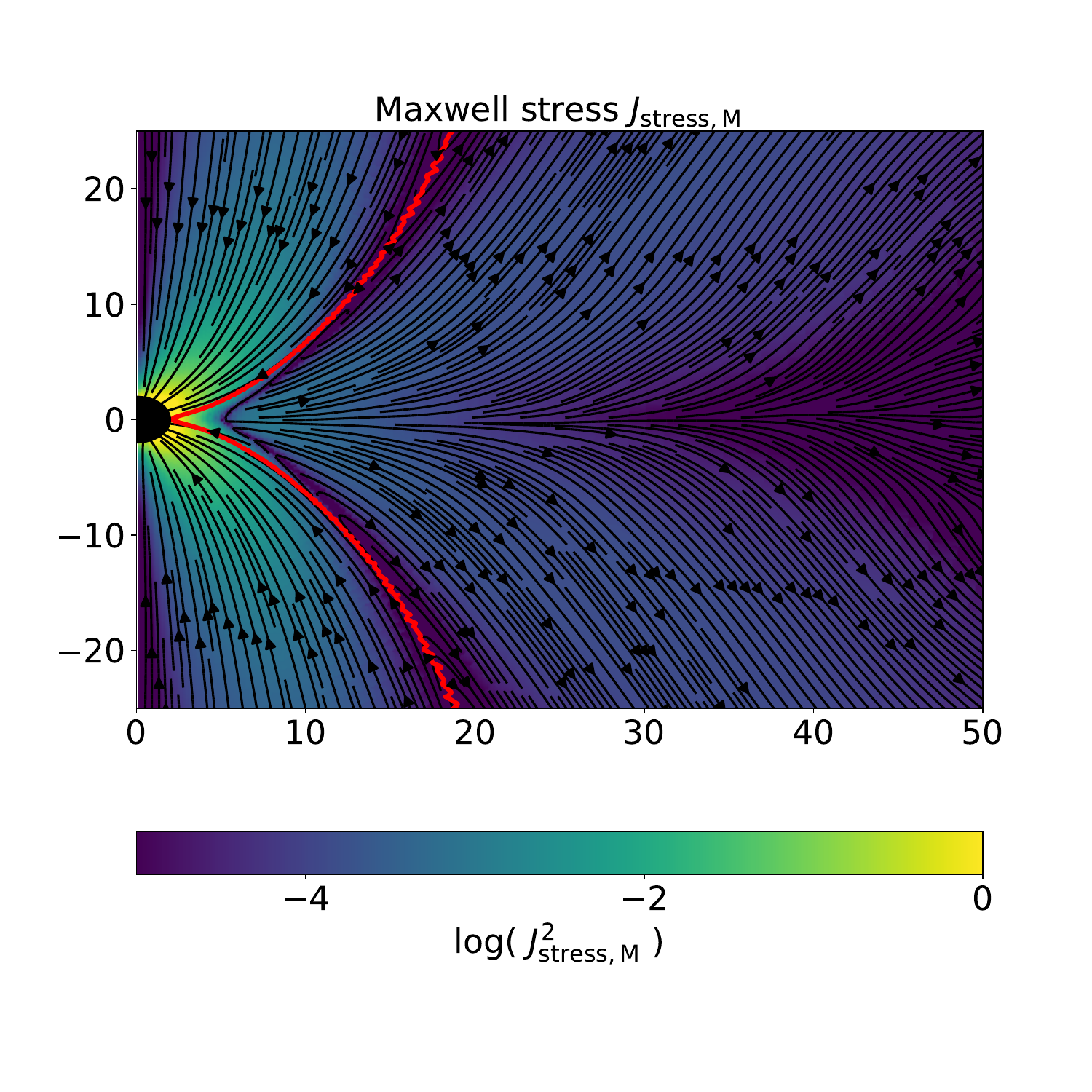}{0.32\textwidth}{}
              \fig{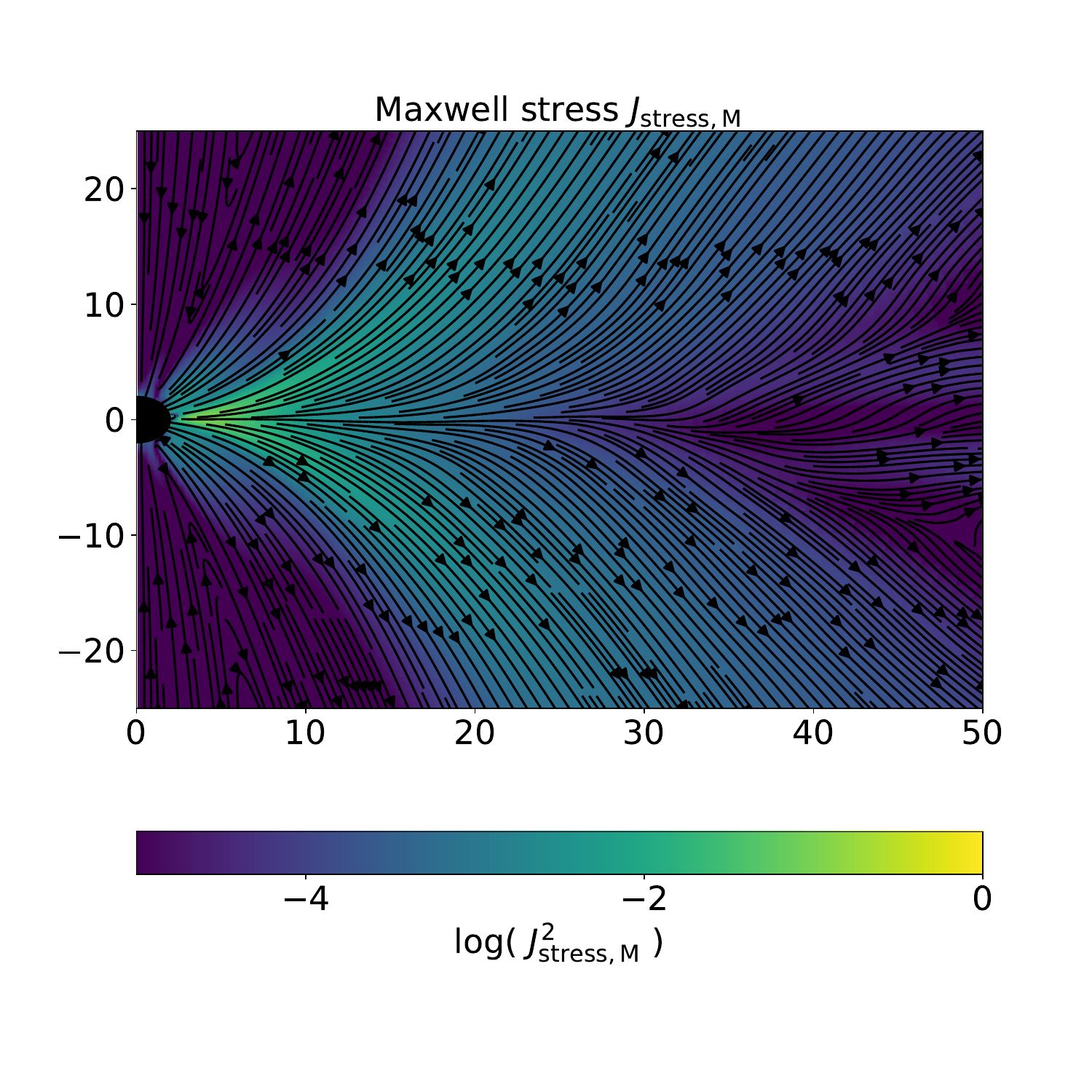}{0.32\textwidth}{}
              \fig{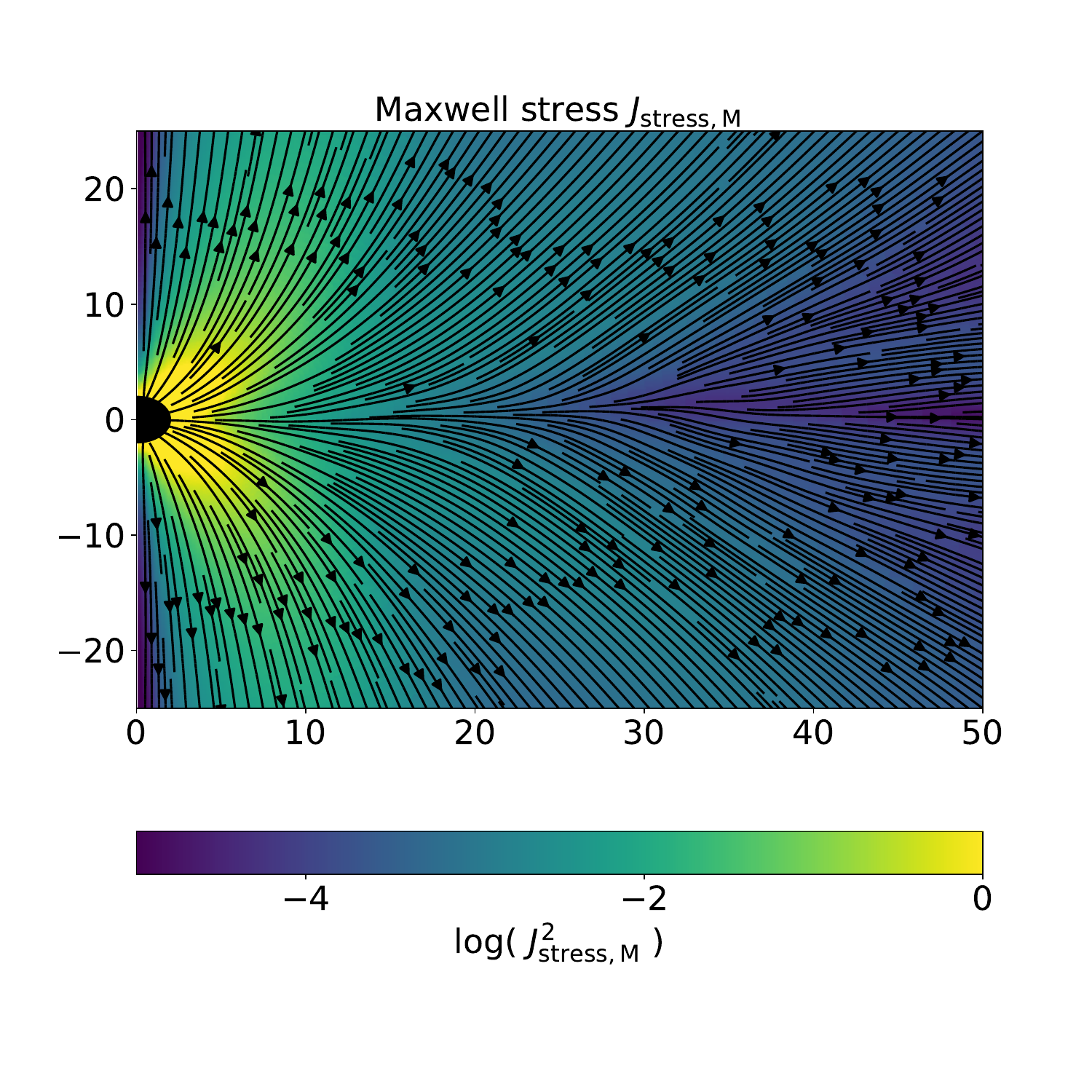}{0.32\textwidth}{}}
    \caption{Time and $\phi-$averaged maps of the angular momentum flux of the simulations with spin $a = -0.94$ (left), $a = 0$ (middle) and $a = 0.94$ (right). The color map represents the modulus of the angular momentum flux, $\sqrt{({\dot J^\theta})^2 + ({\dot J^r})^2}$. The first row shows the total angular momentum flux, while the second and third rows show the contributions of the advection and Maxwell stress components, respectively. The color coding for each row is the same to ease the comparison. For $a = -0.94$, we use the red lines to denote the transition radius where the toroidal velocity component $u^\phi$ changes sign. It is seen that its position corresponds to the region in which the radial component of the Maxwell stress changes sign as well.}
    \label{fig:Jstreamline}
\end{figure}

In order to understand how angular momentum is transported throughout the disk, we show in
Figure \ref{fig:Jstreamline} colormaps of the angular momentum flux modulus
\begin{equation}
\dot J^2 = ({\dot J}^r )^2 + ({\dot J}^\theta)^2,    
\end{equation}
for the total,  advection and Maxwell stress components and the associated streamlines.
%The total angular momentum flux is underlined with a density map.
The top line shows the total angular momentum flux,
while the middle and bottom lines show the contribution of the advection and Maxwell
stress, respectively. The left, middle and right columns are for spins $a = -0.94, ~0, ~ 0.94$
respectively. We note that the scale of the color coding is the same in each line
to ease the comparison between the different black hole spins. Further, we point out
that the equilibrium radius for these simulations is around $r = 30 r_g$, prompting
caution for larger radius. First, it is clear that each of these figures is antisymmetric
with respect to  the equator, as expected. We note that these figures
are obtained directly from the data from our simulation without imposing the symmetry
as done by \citet{Chatterjee2022ApJ...941...30C}. From the top line, we further see
that for all three spins, angular momentum is lost by the disk through the disk
wind.

There is one more difference between those figures: the angular momentum flux
at the pole is negative for $a = -0.94$ and $a = 0$, while it is positive for
$a = 0.94$. Also, the contribution to the total angular momentum flux in this
region appears to be small compared to the flux at the equator as seen in Figure \ref{fig:angularMomentumFlux}. In addition, Figure
\ref{fig:Jstreamline} shows
the presence of a transition layer in the simulation with a negative and a null spin. In this transition layer, the angular momentum flux changes direction, being oriented outwards in the wind and inward in the jet.
The position of this transition layer is at larger angle from the pole
for $a = -0.94$ than for $a = 0$. This transition layer also appears in the 
Maxwell stress contribution.

The streamlines of the Maxwell stress contribution to the total momentum flux are shown 
in the last line of Figure \ref{fig:Jstreamline}. We find two interesting features.
First both the prograde and retrograde disks display a large Maxwell stress
contribution to the angular momentum flux in the jet region contrary to the
non-spinning black hole. This underlines that the jets of a non-rotating
black hole in the MAD regime are weak and do 
not transport a substantial amount of angular momentum (and in fact energy) to
their surrounding environments. The largest outward contribution in
the non-rotating black hole actually comes from the magnetized wind.
It is also clear that the jet of the prograde disk is stronger than that
of the retrograde black hole and deposit angular momentum faster into its environment. 
We further see a substantial Maxwell stress contribution into the wind, which 
shares the same characteristics as the total angular momentum flux, meaning it
is weaker for retrograde disks. Finally, we find that for the retrograde
disk, a region of nearly zero Maxwell stress contribution separates the
wind region from the disk region. This transition is associated with the
change of sign of the toroidal velocity $u^\phi$, shown by the
red line in Figure \ref{fig:Jstreamline} for $a = -0.94$. It is clear that the
red line follows the transition region.

\section{Conclusion}

\label{sec:conclusion}

In this work, we performed several GRMHD simulations of thick accretion disks
in the MAD regime around black holes characterized by different spin $a$ with cuHARM. Our
key results can be summarized as follows: 

\begin{enumerate}

\item We studied the angular momentum flux for our simulations with
$a = -0.94, ~0, ~0.94$ and underlined the differences. We found that i) a substantial
amount of angular momentum is transported away in the magnetized wind and that
the Maxwell stresses are larger for rotating black hole than for non-rotating
black hole as expected because of frame-dragging. In fact, the amount of angular
momentum transported by the "jet" for the non-rotating black hole is small and
negligible. These results were provided in Section \ref{sec:angular} and are clearly displayed in Figure \ref{fig:Jstreamline}.

\item We did not find any correlation between the mass accretion rate and the MAD parameter. However, we did find an anti-correlation between their time derivatives, displayed for our simulations aM94b100 and a94b100 on Figure \ref{fig:diff}. We provided a heuristic explanation for this result in Section \ref{subsec:phi_dotM}.

\item We underlined the difference in the magnetic field component strengths
for spinning and non-spinning black holes in Section \ref{subsec:pressure}. From Figures \ref{fig:pressureDiffComp} and \ref{fig:mapPressureM094}, it is seen that the $\theta$ component is always
subdominant, while the relative importance of the $\phi$ component depend on the
spin of the black hole. In the non-rotating case, the toroidal component is
negligible compared to the radial component close to the horizon, while these
components are comparable for the rotating black hole. The toroidal component even
dominates very close to the horizon for $a = 0.94$. We therefore recover both results from
\citet{Begelman2022MNRAS.511.2040B} and \citet{Chatterjee2022ApJ...941...30C} on the 
importance of the toroidal component in regulating the accretion, and confirmed
that the difference has its origin in the black hole spin.

\item We underlined the differences in the structure of the disks and the jets in the MAD state as a function of spin in Section \ref{subsec:disk-jet}. In particular, we found that retrograde
disks are wider than the corresponding prograde disks, in agreement with the findings
of \citet{Narayan2022MNRAS.511.3795N}. Correspondingly, the jets of prograde disk are
narrower than the jets of retrograde disks. These results are displayed in Figure \ref{fig:sigma-three}.

\item We studied the MAD parameters and investigated their dependence on the spin and the
initial magnetic field strength in Sections \ref{subsec:mass_accretion} and \ref{subsec:initial_beta0}. We found that our numerical results are in very
good agreement with those of \citet{Narayan2022MNRAS.511.3795N}, namely the MAD
factor increases with spin until $a = 0.5$ after which it decreases. The fitting formula
from \citet{Narayan2022MNRAS.511.3795N}, which we renormalized to account for the
difference in definitions, accurately describes the distribution of $\Phi_B$ with spin. 
Therefore, this result holds across resolution, simulation duration and numerical method. The results of this analysis are summarized in the top raw of Figure \ref{fig:spin-factor}.
We find that the MAD parameter does not have a strong dependence on the initial
magnetic field strength parameter $\beta_0$. It only takes longer for the simulation
to reach the MAD state, namely for the magnetic field to saturate to its final
value. Once the MAD state is achieved, all simulations with different $\beta_0$ share the same characteristics, as shown in Figure \ref{fig:dotM-MAD-beta}.

\item We attempted to identify a characteristic variability time in the MAD regime by
studying the temporal variation of $\Phi_B$ via the Fourier transform in Section \ref{subsec:time_characteristics}. Contrary to the 
2D case \citep[see e.g.][]{CBL21}, no clear period could be identified unambiguously. 
This is because in 3D simulation, accretion proceed via non-axisymmetric instabilities,
such as the interchange instability \citep{SSP95, MTB12, Begelman2022MNRAS.511.2040B}.
Yet, the typical variability time we find is around a few hundreds $t_g$ to a thousands, 
comparable to the estimates from \citet{LDF16} and \citet{JJN22}. This time scale is also comparable
to the time scale inferred by \citet{Wong2021ApJ...914...55W} who studied the layer
between the disk and the jet.

\end{enumerate}

Our results shed light on the differences in the accretion dynamics of disks in
the MAD state across spins, prompting for more detailed analysis of the transport
of angular momentum by jets and winds, as well as the the role of the toroidal
magnetic field in (i) shaping the disk and jet and (ii) providing the norm of
the MAD parameter, and via the norm understanding the efficiency of conversion
between accretion luminosity and black hole spin energy deposited into the Poynting jets.

\begin{acknowledgments}
 G.-Q. Zhang acknowledges support by the China Scholarship Council for 1 year study at Bar-Ilan University. DB and AP acknowledge support from the European Research Council via the ERC consolidating grant $\sharp$773062 (acronym O.M.J.). B.-B.Z. acknowledges the support by the National Key Research and Development Programs of China (2022YFF0711404, 2022SKA0130102), the National SKA Program of China (2022SKA0130100), the National Natural Science Foundation of China (Grant Nos. 11833003, U2038105, U1831135, 12121003), the science research grants from the China Manned Space Project with NO.CMS-CSST-2021-B11, the Fundamental Research Funds for the Central Universities, the Program for Innovative Talents and Entrepreneur in Jiangsu.   This work is performed on a HPC server equipped with 8 Nvidia DGX-V100 GPU modules at Nanjing University. We acknowledge the IT support from the computer lab of the School of Astronomy and Space Science at Nanjing University.
\end{acknowledgments}

\bibliography{main}
\bibliographystyle{aasjournal}

\end{document}